# Survey on the Teacher-Learner Collaboration

N. M. Chase[a]

School of Arts and Sciences, MCPHS University, 179 Longwood Avenue, Boston, MA 02115 USA[b]

*Abstract: We present a scantron-based survey which examines the teacher-learner collaboration. The survey includes questions which probe student preparedness, behaviors, attitudes, and expectations – as well as questions addressing key measures of excellence in teaching and course structure[1]. By presenting two novel approaches to displaying the survey data, we demonstrate that examination of the data in its entirety can provide information of great value to future students, as well as faculty and administrators [2-3].*

## I. Introduction

Many end-of-semester course surveys provide feedback which (for the most part) focuses on the student-consumer's impression of the instructor's performance; the resulting survey data is typically formatted as a table of counts and proportions (of students entering each possible response to each survey question). This format leaves the instructor (and administrators) to puzzle about which aspects of the student experience and the instructor's teaching methods might have motivated each response [4-19]. With the goal of obtaining greater insight, we have created a 20 question scantron-based survey which examines the teacher-learner collaboration; the survey includes questions which probe student preparedness, behaviors, attitudes, and expectations, as well as questions addressing key measures of excellence in teaching and course structure. [1] This paper is intended to demonstrate that the most interesting and valuable information is revealed when survey results are formatted to show every response entered by each individual student; it is there that we find information which we might use to improve our guidance of student learning and cognitive development. This detailed view of survey data consists of a large table containing one row for each individual respondent; the table is 20 columns wide, so that within each row one can read that individual's verbal responses to each of the 20 survey questions. [2] Color-coding is used to facilitate apprehension of key results within the huge table (printed on large A3 paper); blue font is used to flag the most desirable outcomes (suggesting a successful collaboration) while red font is used to flag the least desirable outcomes (suggesting areas in need of improvement). We refer to this tabulation of the survey data as the "Colorful Individual Student" view. Now examination of the Colorful Individual Student view of survey data for numerous semesters leads one to see apparent "correlations" between responses to pairs of survey questions. In search of any statistically significant correlations between responses to pairs of survey questions, we have examined the pooled survey data (650 students) from 12 semesters of PHY 270 and PHY 274, for Sp2014 through Sp2018 (all taught by a single instructor). Bar graphs are easily created to graphically illustrate correlations and, embedded within each bar graph, one can show the results of hypothesis tests for independence - using both Spearman Rank correlation and Kendall Tau-beta tests for independence. [3] (Survey data was processed and color-coded via Mathematica.)

This paper is organized as follows. Section II presents the survey instrument. Section III comments on some of the many important insights one might glean from the Individual Student View of the pooled survey data (650 students). Due to the space limitations of journal format, the Colorful Individuals view is presented in online supplementary materials [4]. For completeness and by way of contrast, Section IV comments very briefly on the more "standard" summative view of the data [5]; the Counts and Proportions table is also presented in supplementary materials presented online [4]. Section V presents the methodology and outcomes of our search for statistically significant correlations between student responses to pairs of survey questions. Of the 380 illustrative bar graphs, we present only a selection of some of the most interesting bar graphs here, with the results of Spearman Rank and Kendall-Tau tests for independence embedded in each. Additional Bar Graphs are presented in online supplementary materials [4]. In Section VI, we summarize our conclusions and comment on future work.

## II. The Survey

The Physics Faculty Group plans to use the results of this questionnaire to improve the quality of its courses. We hope that all students will participate and that their answers will be candid and serious. Questions on which you have no opinion may be left blank.

Please respond to questions on the scantron form provided. Do not put your name or your ID number on the scantron form! FILL IN ONLY YOUR ANSWERS TO THE QUESTIONS and the TEST FORM "A" BUBBLE. Submit your scantron form by placing it face down in the white box on the front table, as you leave the lecture hall.

1. My GPA is
   a. 3.50 – 4.00    b. 3.00 – 3.49    c. 2.50 – 2.99.    d. 2.00 – 2.49.    e. less than 2.00.

2. I have attended
   a. all of the class periods.    b. all but 1-2 class periods.    c. all but 3-4 classes.
   d. most of the class periods.    e. few of the class periods.

3. I have studied
   a. all of the required reading material.    b. most of the required reading material.
   c. some of the required material.    d. none of the required material.

4. Of the homework problems assigned, I have done
   a. between 95% and 100%.    b. between 85% and 94%.    c. between 70% and 84%.
   d. less than 70%.    e. No homework problems were assigned.

5. I feel that I deserve a grade of
   a. A.    b. B.    c. C.    d. D.    e. F.

6. I expect to receive a grade of
   a. A.    b. B.    c. C.    d. D.    e. F.

7. My work with the math required for this course was
   a. very quick and accurate.    b. quick and reasonably accurate.
   c. moderately paced and mostly accurate.
   d. less quick and/or accurate than I would have liked.

8. The material covered in this course is
   a. too difficult.    b. difficult but reasonable.    c. fairly simple.    d. too simple.

9. The material covered in this course was
 a. very exciting.    b. interesting.    c. of some interest.    d. of no interest.

10. I learned
 a. a great deal.    b. a moderate amount.    c. almost nothing.

11. I tended to begin work on assigned homework problems
 a. within a day of the problems being assigned.
 b. within 2 - 3 days of the problems being assigned.
 c. within 2 weeks of the problems being assigned.    d. within the week before an exam.
 e. No homework problems were assigned.

12. The textbook used in this course was
 a. excellent.    b. good.    c. adequate.    d. poor.    e. No textbook was used.

13. The pace of the course was
 a. too rapid for proper understanding.    b. reasonably fast.    c. moderate.
    d. somewhat slow.    e. much too slow.

14. Difficult and/or subtle points were explained
 a. very clearly.    b. adequately.    c. poorly.    d. not at all.

15. Outside of class, the professor was
 a. accessible for questions.    b. could rarely be found.    c. I did not look for any extra help.

16. The professor
 a. encourages questions and answers them seriously.
 b. does not encourage questions but answers them seriously.
 c. does not encourage questions and is reluctant to answer them.
 d. openly discourages questions in class.

17. Examinations in this course are
 a. too difficult.    b. difficult but reasonable.    c. moderate.
    d. fairly easy.    e. much too easy.

18. Supplementary materials which the instructor posted in Blackboard were
 a. very helpful.    b. somewhat helpful.    c. not helpful.
     d. I never consulted posted course supplements.

19. I found the use of Mastering Physics
 a. very helpful.   b. somewhat helpful.   c. not at all helpful.
 d. I have not tried much Mastering Physics homework.
 e. I have not tried any Mastering Physics homework.

20. What is your overall evaluation of the course?
 a. The best or very nearly the best.   b. Excellent.   c. Good.   d. Fair.   e. Poor.

# III.  *The Colorful Individuals View of the Data*

Each row of the Colorful Individuals view of the data [2] shows the responses of an individual student to each of the survey's twenty questions.   Any insights which may be gained from this data must, of course, be informed by an understanding of the level, structure, and learning goals of PHY 270 and PHY 274 - as well as the demographics of the survey population.  We now provide a brief summary of that requisite information. The courses PHY 270 and PHY 274 comprise a 2-semester calculus based introductory physics sequence. Both courses emphasize guiding students towards gaining an in-depth understanding of the foundations of physics; students are expected to demonstrate their understanding by analysing and solving fairly complicated problems, by using multi-step reasoning from first physics principles. For a great many students, these courses appear to provide some of their first exposures to being required to interpret and figure out solutions to complicated problems.   The courses are required of students in many diverse degree programs: Pharm.D., Pre-Med (Physical Therapy, Occupational Therapy, Physicians Assistant), Pharmaceutical Sciences, and a relatively small number of majors in Chemistry and Biology; there are no Physics, Mathematics, or Engineering majors.  Students range from those having no prior course work in physics to those who have successfully completed 2 to 4 years of introductory physics course work in high school (including advanced placement courses) .  All students are required to have completed a 2 semester calculus prerequisite; however, facility with mathematics varies widely over the course population.  In addition, students range from year 1 to year 4 in their respective degree programs - including many "repeaters" of the course. The data clearly reflects the diversity of the course population - and, to some degree, the interests indicated by the student's choice of degree program.

In seeking insights into how instruction might be improved, one might first examine responses of students who reported that "Difficult &/or Subtle Points" were explained "Poorly".  (There were not many of these, but one tries to reach everyone!)  This group of students tends to report various combinations of the following traits: non-ideal attendance, a low completion rate for HW, weaker math skills, starting a month's worth of HW assignments within 1 week of an exam, finding the course material too difficult, finding exams too difficult, reporting that NO HW was assigned, reporting that no textbook was used, feeling that pace of the course was too fast, no interest in the material covered, finding that the professor's Blackboard Supplements were not helpful, feeling that he/she deserves a grade of "A" but expects to receive a lower grade.

One might then focus on responses of students who reported that "the pace of the course" was "too fast for proper comprehension".  (Again, there were not many of these, but the goal is to do better!)   Among this group of students, one finds various combinations of the same traits listed above for students responding that difficult &/or subtle points were explained "Poorly".

Of course, given access to all of the survey data, many (if not most) instructors would also want to know the "characteristics" of students who report:  learning a great deal or learning almost nothing, finding the material very exciting or finding the material "of no interest".  Both Faculty and Administrators might learn much by examining the characteristics (preparedness, behaviors, attitudes, and expectations) of students who rate the course as "the best or very nearly the best" or "excellent - as well as the characteristics of students who rate the Course Overall as poor.   Understanding the traits, behaviors, and attitudes of the individuals in each group might enable us to be more effective in guiding our students' learning!

    Overall, the Colorful Individuals view of the survey data provides student-generated data which strongly supports the advice which most instructors would give to their students: attend class, do all the assigned readings and problem assignments, do homework in a "timely manner", work to improve any weakness in mathematics skills, etc. (The analysis presented in the Section V shows that the correlations "suggested" by examining several semesters of the Colorful Individuals view are indeed statistically significant.) After completing the statistical analysis, the author undertook an experiment - appending brief remarks on the 12 semesters of survey data to the standard in-class advice on how to learn physics (and succeed in the course) without unnecessary stress. Those brief remarks seemed to strongly pique student interest! (They stood in stark contradiction to the advice many students report receiving from their friends in the upper classes!)

## IV. The Counts and Proportions Representation of the Survey Data

Although the counts and proportions view of the data [4,5] in our survey does provide some insights into aspects of the teacher-learner collaboration which need improvement, it leaves the instructor to wonder about exactly how to make the necessary adjustments. Clearly, the design of improved "course evaluation" instruments must encompass the design of optimally informative and useful representations of the survey data!

## V. Bar Graphs - Searching for Correlations Between Responses to Pairs of Survey Questions

In search of any interdependence between responses to pairs of survey questions, we first constructed contingency tables containing the count data for responses to each pair of survey questions; an example of one such contingency table is shown in Table 1. For each contingency table, we then calculated proportions (fractions) - the proportion of students responding "A" to Question X while responding "B" to Question Y. Bar graphs were then constructed to display those proportions. Not surprisingly, the bar graphs displayed several striking trends; however, one wondered whether the apparent interdependencies were statistically significant. Thus, we used both Spearman Rank correlation and Kendall Tau-beta to perform hypothesis tests [3] [20]. The null hypothesis was taken to be that the populations are independent; we performed the tests at the alpha = 0.05 level of significance. The P-values (embedded in the bar graphs) give the probability of rejecting the null hypothesis when the null hypothesis is actually True. This statistical analysis provided a mechanism for using "student generated" data in a way which might inform future students - as well as faculty and administrators.

*A detailed view of the "meanings" of the bar graphs:*

As an example, let us consider Question 20 (Course Overall) and Question 17 (Difficulty of Exams). A contingency table was constructed, as shown below. From the contingency table, we see that 94 respondents rated the course as "The best or very nearly the best". Of these, the proportion reporting that "the exams were too difficult" was 4/94 = 0.043; the proportion saying exams were "difficult but reasonable" was 33/94 = 0.35. On the other hand, of the 9 respondents rating the course overall as "Poor", the proportion reporting that "the exams were too difficult" was 4/9 = 0.44 and the proportion reporting that exams were difficult but reasonable was 2/9 = 0.22.

The bar graph array appearing in Figure 1 displays these proportions (and several others).

| X | too difficult | difficult but reasonable | moderate | fairly easy | much too easy | | CHECKSUM |
|---|---|---|---|---|---|---|---|
| The best or very nearly the best | 4 | 33 | 37 | 19 | 1 | 1 | 95 |
| Excellent | 7 | 96 | 76 | 27 | 1 | 0 | 207 |
| Good | 15 | 106 | 88 | 21 | 3 | 0 | 233 |
| Fair | 11 | 32 | 26 | 7 | 2 | 0 | 78 |
| Poor | 4 | 2 | 2 | 1 | 0 | 0 | 9 |
| | 2 | 3 | 2 | 0 | 0 | 9 | 16 |
| CHECKSUM | 43 | 272 | 231 | 75 | 7 | 10 | 638 |

Table 1: Contingency table for Question 20 (Course Overall) and Question 17 (Difficulty of Exams)

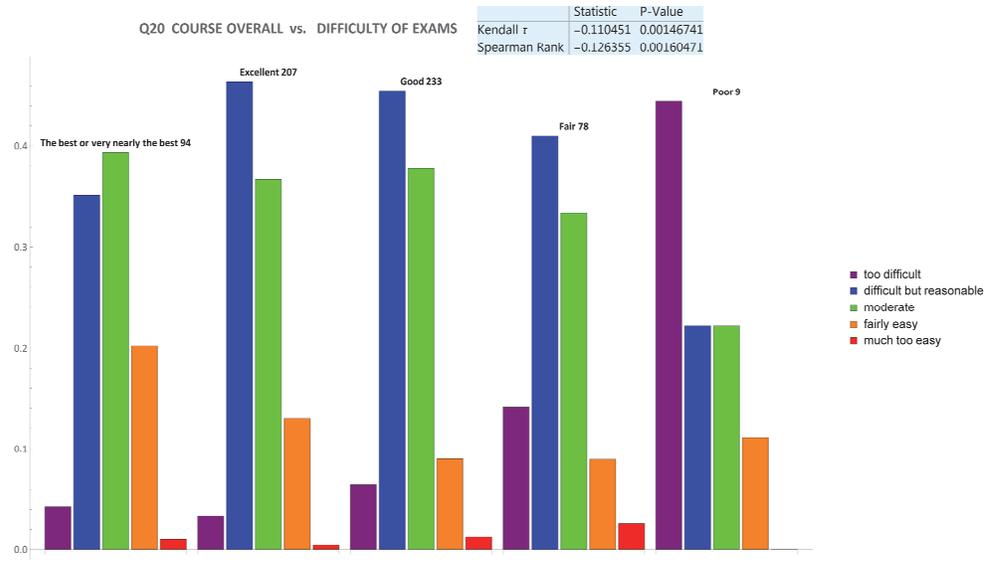

Figure 1 : Course Overall and Difficulty of Exams

1. **Amount Learned:** Selected results appear in Figures 2 through 5. Students who reported that they learned the most from the course tended to report: better class attendance, completion of more reading and problem assignments, feelings that they deserved higher grades and expected to receive higher grades, better mathematics preparation, starting problem assignments in a more timely manner, finding the material more interesting, finding greater clarity in the instructor's explanations of difficult and subtle points, finding the instructor's Blackboard Supplements more helpful, and higher ratings of the course overall. One must note that the relationships between the variables are not invariably monotonic and that the often very strong correlations between the variables do not (of course) indicate "causation". Nevertheless, one might (for example) conclude that if one is interested in learning more from the course, attending class, doing homework, building math skills, striving to developing an interest in the material (etc) might help!

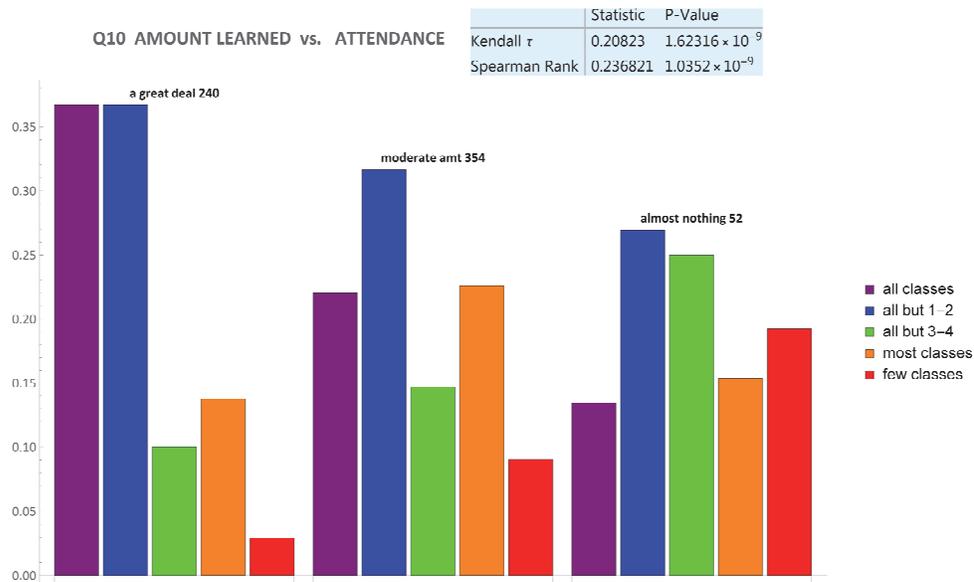

Figure 2 : Amount Learned versus Attendance

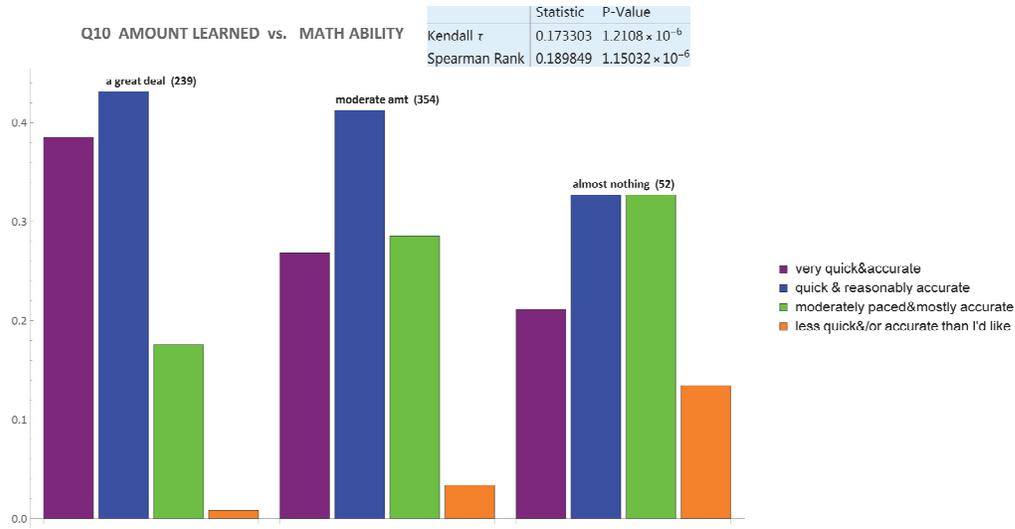

Figure 3 :  Amount Learned versus Math Ability

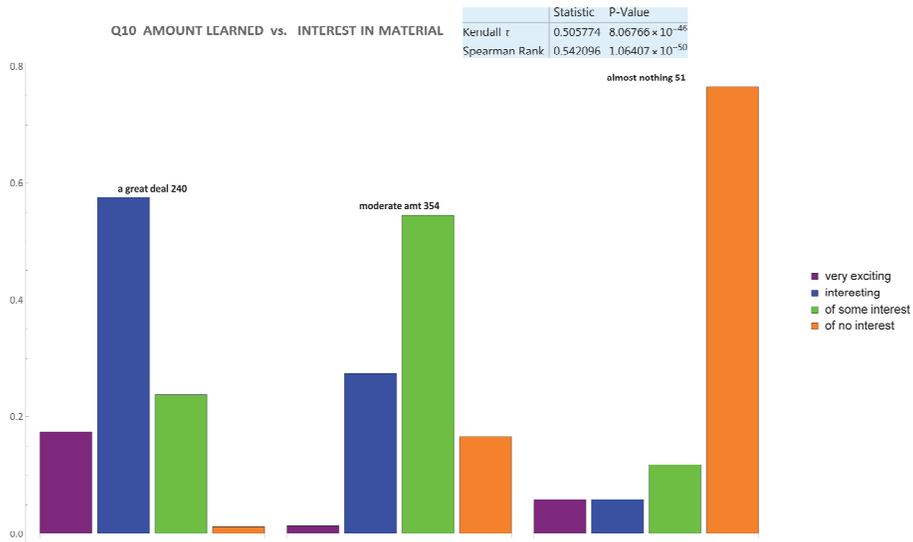

Figure 4 :  Amount Learned versus Interest in Material

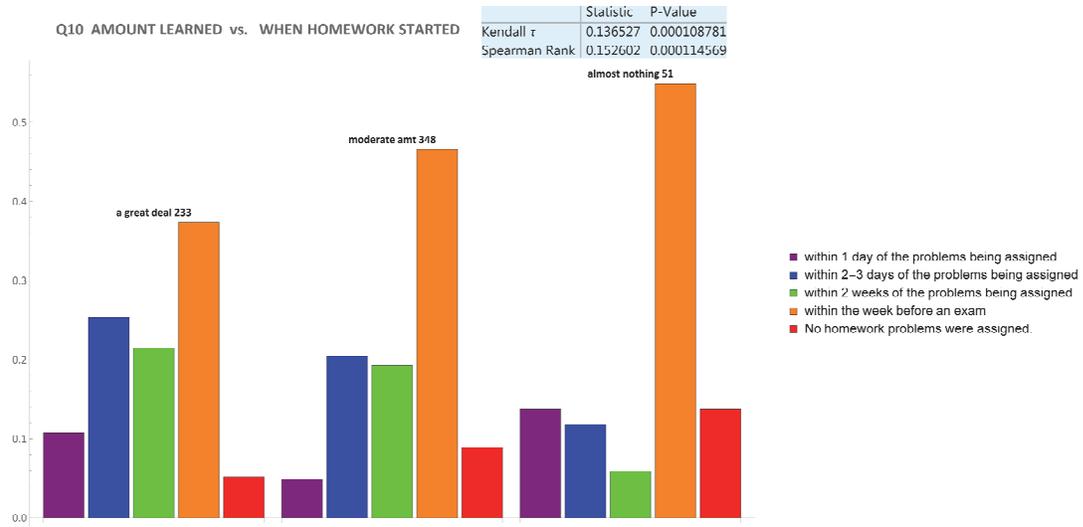

Figure 5 : Amount Learned versus When Homework Started

ADDED NOTE: Hypothesis tests indicated that the reported Amount Learned appeared to be independent of reported GPA, Difficulty of Material, and Difficulty of Exams.

**2. Difficulty of Exams:** Selected results appear in Figures 6 through 8. Students who reported that they found exams too difficult (only 7.3%) tended to find the material too difficult. They tended to report not feeling encouraged to ask questions in class - while the vast majority of respondents (92%) reported that the instructor encouraged questions and answered them seriously. Students who reported that exams were too difficult tended to report finding that the pace of the course too rapid for proper understanding - while the vast majority of respondents (94%) did not. It is not surprising that those who responded that exams were too difficult also tended to report weaker math preparedness and, not surprisingly, were less likely to feel that difficult &/or subtle points were explained very clearly or adequately. They tended to report having lower GPAs and, not coincidentally, tended to assign a lower rating to the course overall.

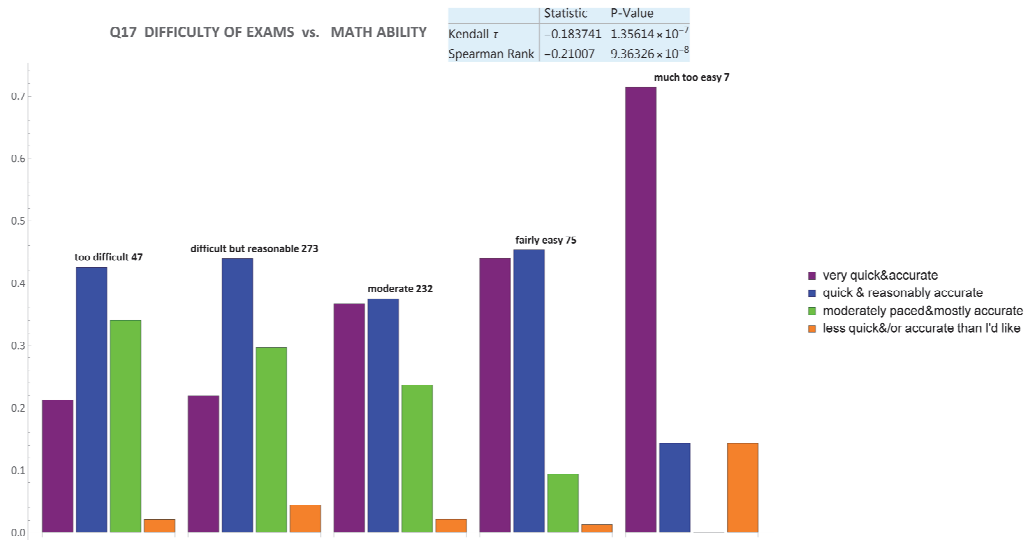

Figure 6 : Difficulty of Exams versus Math Ability

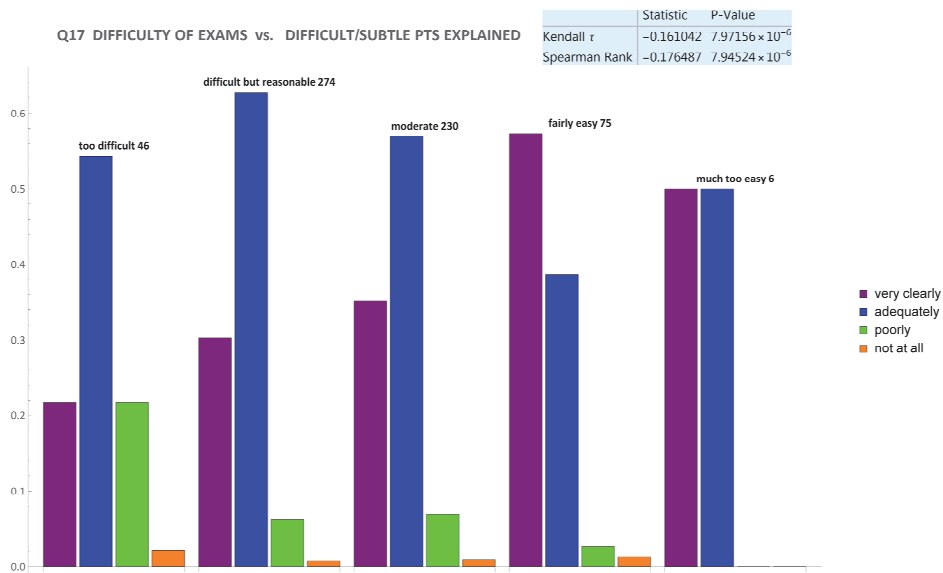

Figure 7 : Difficulty of Exams versus Explanations of Difficult/Subtle Points

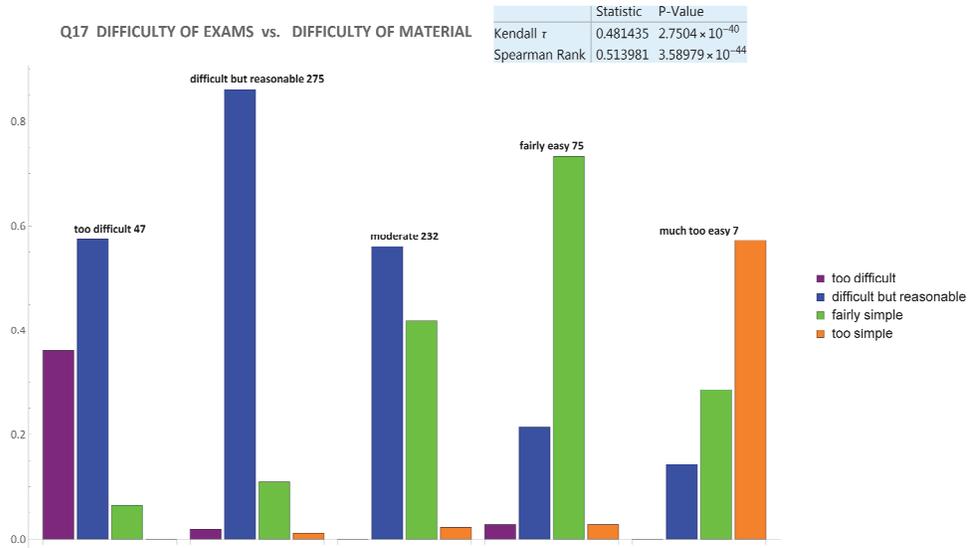

Figure 8 : Difficulty of Exams versus Difficulty of Material

**3. Math Ability:** Selected results appear in Figures 9 through 13. Students who reported feeling less adept with the mathematics required by the course tended to report: lower GPA's, fewer reading assignments completed, feeling they deserved and expected to receive lower course grades. They tended to feel the material was more difficult and less interesting. As mentioned previously, this group was more likely to feel that difficult &/or subtle points were explained poorly or not at all and they were more likely to assign a lower rating to the course overall.

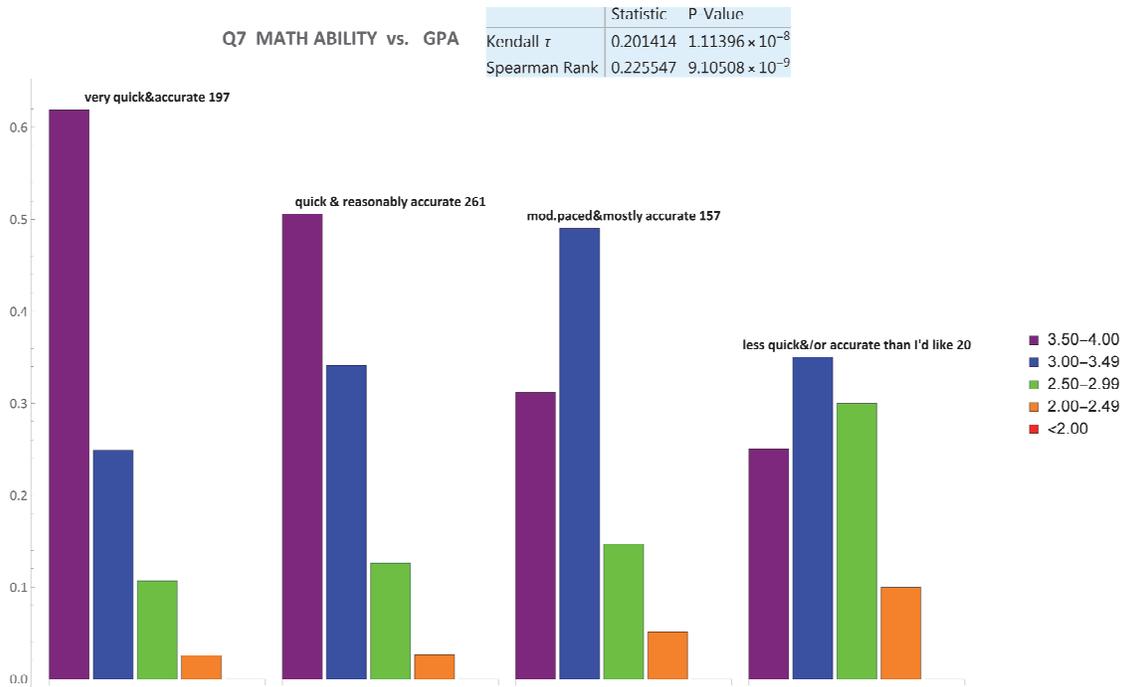

Figure 9 : Math Ability versus GPA

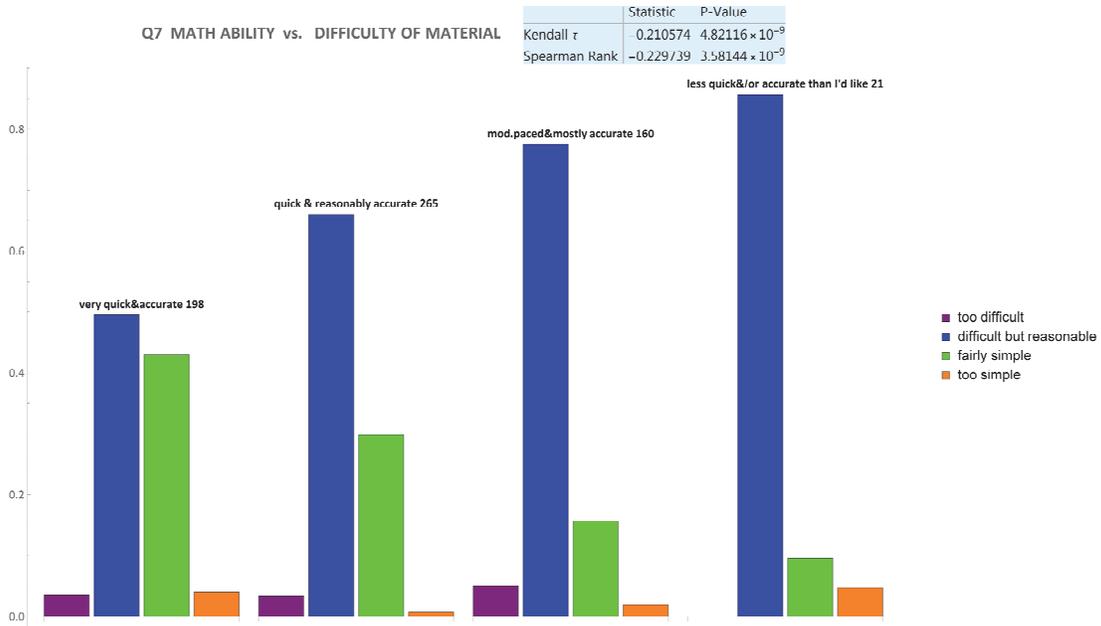

Figure 10 : Math Ability versus Difficulty of Material

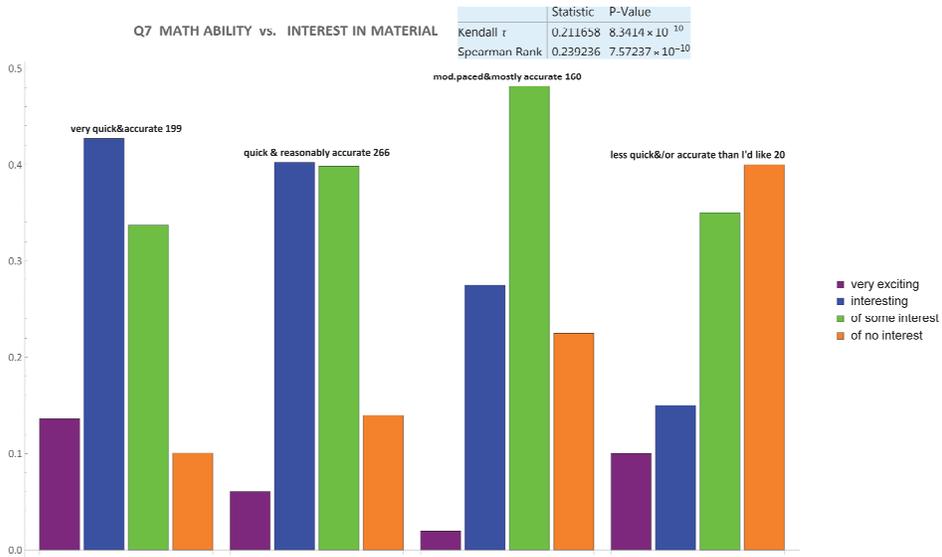

Figure 11 : Math Ability versus Interest in Material

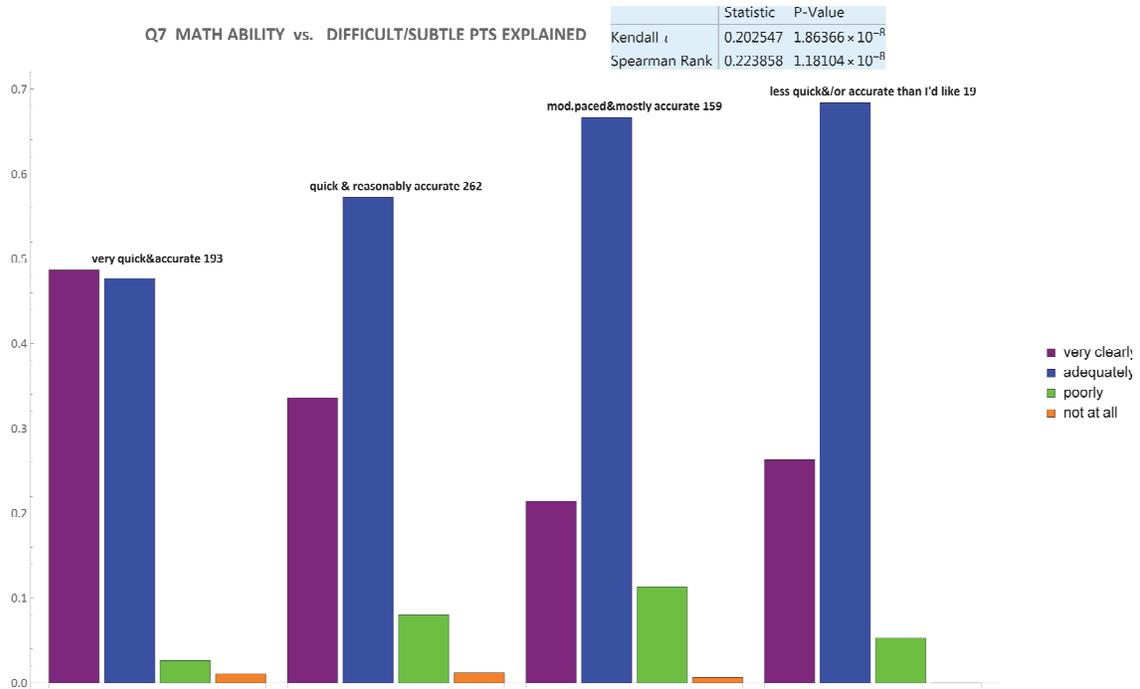

Figure 12 : Math Ability versus Explanations of Difficult/Subtle Points

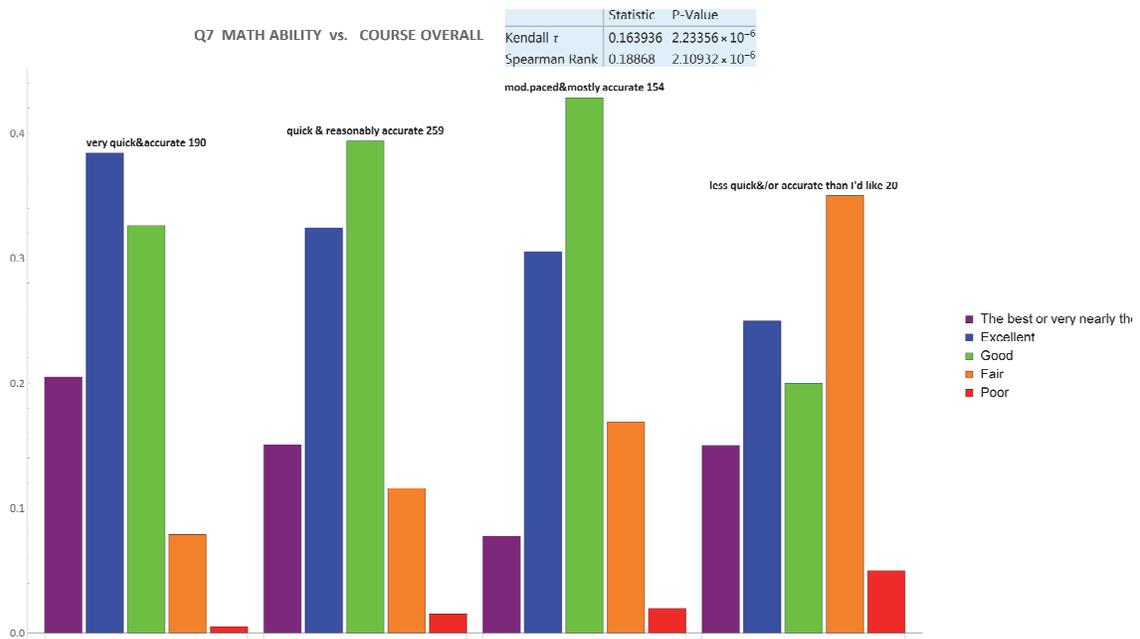

Figure 13 : Math Ability versus Course Overall

**4. The Course Overall :** Selected results appear in Figures 14 through 19.  *Higher Overall Course ratings* tended to come from students reporting better math preparation, better records of attendance in class, a better record of completing assigned readings, and stronger interest in the course material. This group also tended to report starting homework in a more timely manner and completing a higher percentage of it. They tended to report:  that difficult and subtle points were explained more clearly, that exams were less difficult, and they tended to feel more encouraged to ask questions in class.  This group tended to feel that they deserved higher grades and expected to receive them.  *Lower Overall Course ratings* tended to come from students reporting weaker math skills, worse attendance records, a lower completion rate for assigned readings, and less interest in the course material. This group also tended to report starting homework "at the last minute" and completing less of it. They tended to report that difficult and subtle points were explained less clearly, that exams were more difficult, and they tended to feel less encouraged to ask questions in class.  This group tended to report expecting lower course grades. Hypothesis tests indicated that reported ratings of the Course Overall appeared to be independent of reported GPA and only weakly correlated with the perceived Pace of the Course.

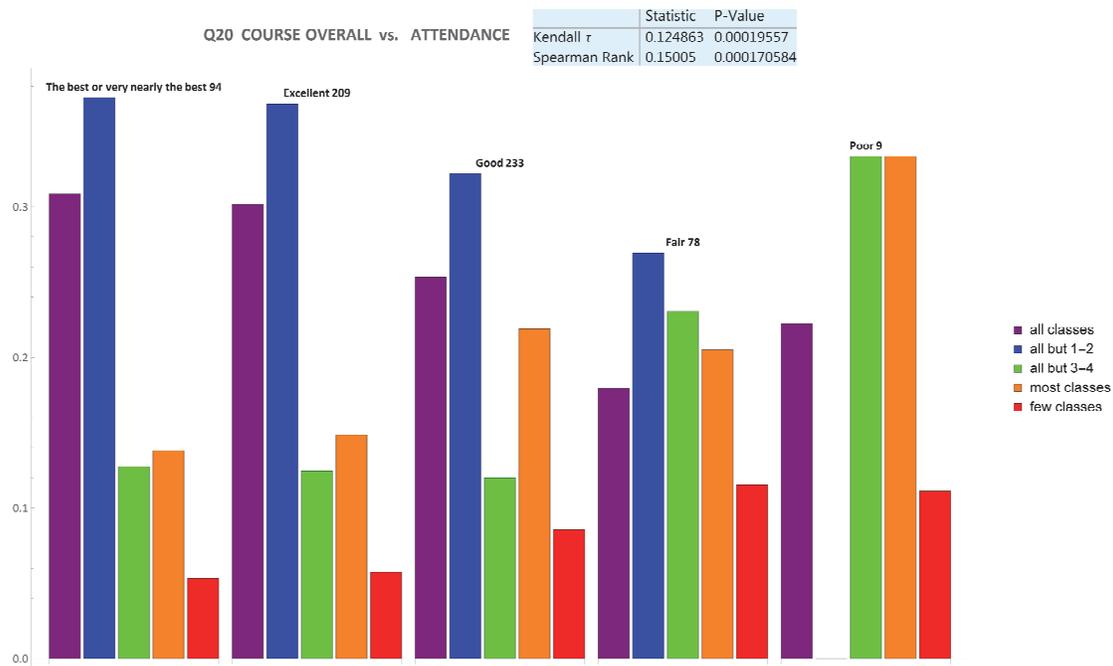

Figure 14 :  Course Overall versus Attendance

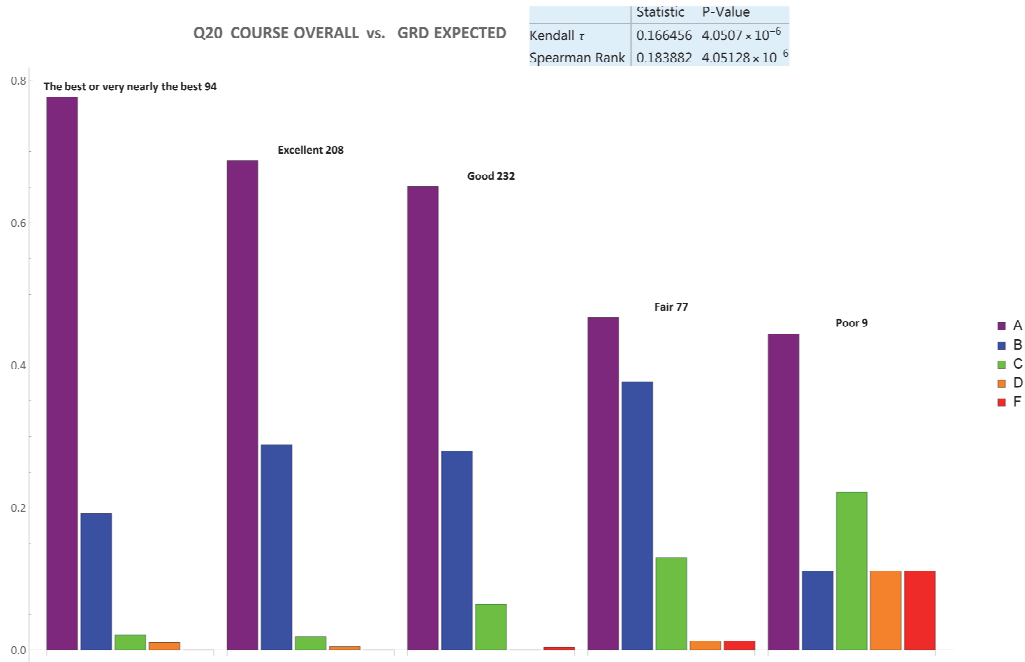

Figure 15 : Course Overall versus Grade Expected

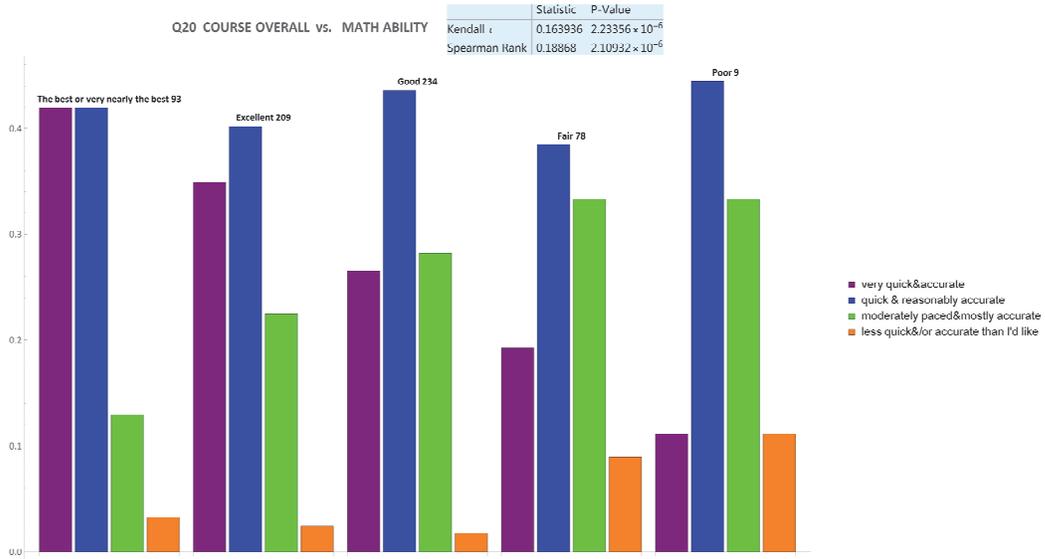

Figure 16 : Course Overall versus Math Ability

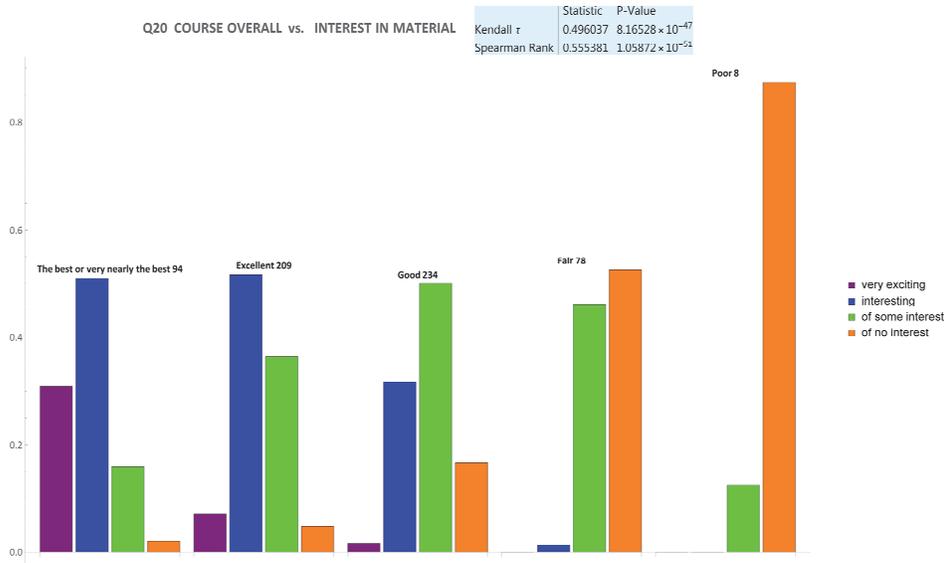

Figure 17 : Course Overall versus Interest in Material

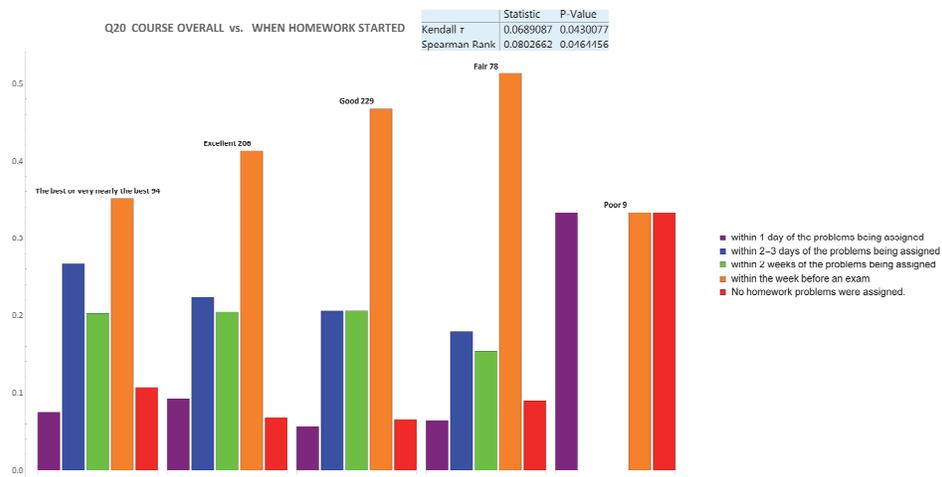

Figure 18 : Course Overall versus When Homework Started

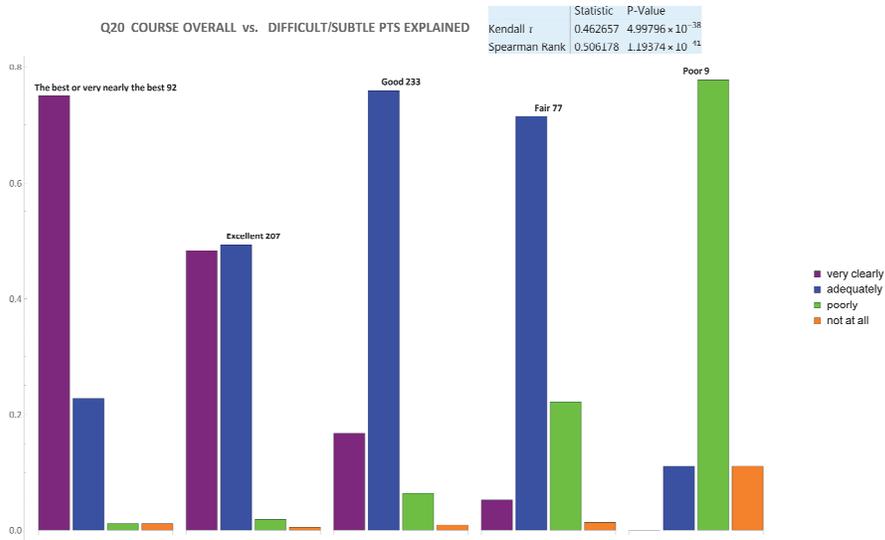

Figure 19 : Course Overall versus Explanations of Difficult/Subtle Points

It is important to emphasize the (already noted) vast academic diversity of the course populations, which included a significant number of students reporting weaker mathematics preparation and little or no prior course work in physics - along with a significant number reporting strong mathematics skills and quite substantial prior physics course work. Based on this analysis of the survey data, one sees that some of the lower ratings of the course overall may have come from students for whom a more challenging course would be preferable! This is an important lesson!

# *VI. Conclusion*

We have advocated for an end-of-semester survey on the Teacher-Learner Collaboration, with data processed to show a) each individual student's response to each of the survey questions and b) a statistical analysis to find any statistically significant correlations between responses to pair of survey questions. To the extent that student responses were candid and serious, the colorful individuals view of the survey data provides considerable (sometimes astonishing) insights into student expectations and their experiences with the course! The (often far less than ideal) behaviors reported by students suggest that respondents were remarkably candid - for the most part. Examination of the bar graphs and their embedded statistical analysis permits one to better understand the course population (on a global scale). Insights gained from the survey can inform the instructor's teaching methods, permit the instructor to better advise students, and also inform administrators about what is going on in a course. Specifically, the survey can assist administrators with evaluation of teaching and learning (viewed as a collaborative effort) and also document any additional administrative support which may be required. While the correlations we found between responses to certain pairs of survey questions would not be the slightest bit surprising to any seasoned physics instructor, in this work we have shown that student-generated data supports our expectations. We note that examination of *individual semesters* of PHY 270 or PHY 274 suggests similar correlations between responses to pairs of survey questions.

When an almost identical survey was used in a large lecture course in Organic Chemistry (taught by a colleague of the author)[3], the resulting bar graphs showed correlations similar to many of those presented here for physics; some minor differences might be attributed to the greater emphasis which physics courses place on problem-solving and the greater emphasis which organic chemistry courses place on factual recall. In another recent extension of this work, the author examined merged Physics Lab Survey data for Sp 2015 through F 2019 - for two different Lab Instructors (with each instructor's surveys analyzed separately). Sample sizes for the two lab instructors' surveys were, unfortunately, only roughly similar (134 and 194). The interdependencies between responses to pairs of survey questions were, for the most part, similar to each other - and to those shown in this paper. One notable (and intriguing) exception: for one lab instructor, the "Lab Overall" rating was significantly correlated with both Mathematics Ability and GPA; for the second instructor, there was only a weak correlation between the same pairs of variables. It seems that differences in teaching styles may be the root cause of the disparity. However, further study of this remains for a future work.

## *End Notes*

**a.** norma.chase@mcphs.edu   Semi-retired, May 2023, Current email:  nmchase18@comcast.net

**b.** formerly known as Massachusetts College of Pharmacy & Allied Health Sciences

**1.** Some of the survey questions were taken from the end-of-semester survey used by the Physics Department at Northeastern University in the late 1970's - mid-1980's.  At Northeastern, only counts and proportions/percents tables were generated from the data gathered.  The scantron-based survey is administered in the last class of the semester in order to eliminated the biases which may be introduced by student collaborations in completing the survey online.  Numbered tickets are distributed for entry in the scantron form, in place of student ID numbers.

**2.**  Blank spaces in the Colorful Individuals Table occur where students chose to provide no responses to particular questions.

**3.**   Spearman rank correlation is likely familiar to most readers; however, according to some authors [20], Kendall-Tau-beta is a more reliable statistic for data which includes many repeated ranks.

**4.**   Link to Online Supplementary Materials:  
https://1drv.ms/f/s!AqiUghAiepIRrHRPxuhUwrL5vGOC?e=L8EamO

**5.**  Some respondents bubbled two or more responses to a single question;  some  bubbled "d" or "e" in response to a question providing only three possible responses.  In the Counts and Proportions Table, for each response to a question, we cite the proportion (of all students replying to the question, legitimately or not) who submitted that response.  That is why, for some of the questions, the proportions do not add up to one as expected.

**April 24, 2020**

# 2 COLORFUL INDIVIDUALS DATA

April 24, 2020

| GPA | ATTENDANCE | RDGS COMPLETED | HW COMPLETED | GRD DESERVED | GRD EXPECTED | MATH ABILITY | MATERIAL COVERED | MATERIAL COVERED | I LEARNED | HW STARTED | TEXTBK | PACE OF LECTS | DIFF/SUBTLE PTS | PROFESSOR | PROFESSOR | EXAMS WERE | BLKBD SUPPLEMENTS | MAST.PHYSICS | COURSE OVERALL |
|---|---|---|---|---|---|---|---|---|---|---|---|---|---|---|---|---|---|---|---|
| 3.50–4.00 | all but 1–2 | most readings | 70–84% | A | A | quick & reasonably accurate | difficult but reasonable | of some interest | | within 1 day of assignment | good | reasonably fast | very clearly | accessible for questions | encourages questions & answers them seriously | difficult but reasonable | somewhat helpful | | |
| 3.50–4.00 | few classes | some readings | 95–100% | A | A | very quick&accurate | fairly simple | of some interest | almost nothing | within 2 weeks | good | moderate | very clearly | accessible for questions | | | | very helpful | |
| 3.50–4.00,2.50–2.99 | all classes | some readings | 95–100% | B | B | less quick&/or accurate than I'd like | difficult but reasonable | of some interest | | | moderate | | | | | | | | |
| 3.50–4.00 | most classes | some readings | 85–94% | A | A | very exciting | too difficult,too simple | a great deal | | within week before exam | excellent | moderate | very clearly | | I did not seek extra help | | too difficult,difficult but reasonable,much too easy | very helpful | not at all helpful | Excellent |
| 3.50–4.00 | all but 1–2 | all readings | 95–100% | A | A | very quick&accurate | fairly simple | of some interest | moderate amt | within 2–3 days | NO txtbk used | moderate | adequately | accessible for questions | encourages questions & answers them seriously | | very helpful | not at all helpful | Fair |
| 3.50–4.00 | all but 1–2 | some readings | 85–94% | A | A | very quick&accurate | difficult but reasonable | interesting | a great deal | within week before exam | adequate | moderate | very clearly | accessible for questions | encourages questions & answers them seriously | | very helpful | not at all helpful | Excellent |
| 3.00–3.49 | all but 3–4 | most readings | 95–100% | A | B | mod. paced&mostly accurate | difficult but reasonable | of some interest | moderate amt | within 2 weeks | adequate | too rapid for proper understanding | adequately | accessible for questions | encourages questions & answers them seriously | | difficult but reasonable | somewhat helpful | not at all helpful | Good |
| 3.50–4.00 | all but 1–2 | all readings | 95–100% | A | A | very quick&accurate | fairly simple | of some interest | a great deal | within 2–3 days | excellent | somewhat slow | very clearly | accessible for questions | encourages questions & answers them seriously | | fairly easy | very helpful | not at all helpful | The best or very nearly the best |
| 2.50–2.99 | all classes | most readings | 85–94% | A | B | | difficult but reasonable | very exciting | a great deal | within 2–3 days | excellent | moderate | very clearly | accessible for questions | encourages questions & answers them seriously | | difficult but reasonable | somewhat helpful | | The best or very nearly the best |
| 3.50–4.00 | all but 1–2 | all readings | 95–100% | A | A | quick & reasonably accurate | difficult but reasonable | interesting | a great deal | within 2–3 days | poor | somewhat slow | very clearly | accessible for questions | encourages questions & answers them seriously | | | somewhat helpful | | The best or very nearly the best |
| 3.50–4.00 | all classes | some readings | 85–94% | B | B | mod. paced&mostly accurate | difficult but reasonable | interesting | moderate amt | within week before exam | NO txtbk used | moderate | adequately | | E | encourages questions & answers them seriously | difficult but reasonable | somewhat helpful | very helpful | Good |
| 3.50–4.00 | all classes | some readings | 70–84% | A | B | quick & reasonably accurate | fairly simple | of some interest | moderate amt | within 2 weeks | NO txtbk used | reasonably fast | very clearly | | I did not seek extra help | encourages questions & answers them seriously | | moderate | not at all helpful | Excellent |
| 3.50–4.00 | all classes | all readings | 95–100% | A | A | very quick&accurate | fairly simple | very exciting | a great deal | within 2–3 days | excellent | reasonably fast | very clearly | accessible for questions | encourages questions & answers them seriously | | fairly easy | very helpful | very helpful | The best or very nearly the best |
| 2.50–2.99 | all classes | all readings | 95–100% | A | B | quick & reasonably accurate | difficult but reasonable | interesting | moderate amt | within 2–3 days | | adequately | | could rarely be found | encourages questions & answers them seriously | | difficult but reasonable | somewhat helpful | not at all helpful | Fair |
| 3.50–4.00 | all classes | all readings | 95–100% | A | A | mod. paced&mostly accurate | too difficult | too rapid | of no interest | within 2 weeks | NO txtbk used | | adequately | poorly | | not helpful | | not at all helpful | | Fair |
| 3.00–3.49 | all but 1–2 | most readings | <70% | B | B | mod. paced&mostly accurate | fairly simple | of no interest | a great deal | within week before exam | NO txtbk used | moderate | adequately | | I did not seek extra help | encourages questions & answers them seriously | difficult but reasonable | somewhat helpful | not at all helpful | Good |
| 3.00–3.49 | all but 1–2 | fairly easy | 85–94% | A | B | quick & reasonably accurate | difficult but reasonable | of some interest | moderate amt | within week before exam | NO txtbk used | moderate | adequately | | I did not seek extra help | encourages questions & answers them seriously | difficult but reasonable | somewhat helpful | somewhat helpful | Excellent |
| 3.50–4.00 | all classes | most readings | 95–100% | A | A | very quick&accurate | difficult but reasonable | very exciting | a great deal | within 2 weeks | good | reasonably fast | adequately | accessible for questions | does not encourage questions but answers them seriously | | difficult but reasonable | somewhat helpful | very helpful | Excellent |
| 3.00–3.49 | all but 1–2 | most readings | 95–100% | A | B | very quick&accurate | too simple | of some interest | moderate amt | within week before exam | NO txtbk used | much too slow | adequately | accessible for questions | does not encourage questions but answers them seriously | | much too easy | very helpful | somewhat helpful | Good |
| 2.00–2.49 | all but 1–2 | most readings | 95–100% | B | B | very quick&accurate | difficult but reasonable | very exciting | a great deal | within 2–3 days | adequate | reasonably fast | very clearly | could rarely be found | encourages questions & answers them seriously | | difficult but reasonable | very helpful | very helpful | The best or very nearly the best |
| 3.50–4.00 | most classes | all readings | 95–100% | A | A | mod. paced&mostly accurate | difficult but reasonable | interesting | a great deal | within 2–3 days | adequate | reasonably fast | very clearly | accessible for questions | encourages questions & answers them seriously | | difficult but reasonable | somewhat helpful | | Good |
| 2.50–2.99 | all but 1–2 | all readings | 95–100% | B | B | mod. paced&mostly accurate | difficult but reasonable | of some interest | moderate amt | within 2–3 days | excellent | moderate | very clearly | accessible for questions | encourages questions & answers them seriously | | moderate | very helpful | very helpful | Excellent |
| 3.50–4.00 | all but 3–4 | all readings | 95–100% | A | A | quick & reasonably accurate | too difficult | interesting | almost nothing | within 1 day of assignment | good | too rapid for proper understanding | poorly | accessible for questions | encourages questions & answers them seriously | | too difficult | very helpful | very helpful | Excellent |
| 3.00–3.49 | all classes | all readings | 85–94% | A | A | mod. paced&mostly accurate | difficult but reasonable | interesting | moderate amt | within 1 day of assignment | NO txtbk used | moderate | adequately | accessible for questions | encourages questions & answers them seriously | | difficult but reasonable | very helpful | not at all helpful | Excellent |
| 3.50–4.00 | all classes | some readings | 85–94% | A | A | quick & reasonably accurate | fairly simple | interesting | a great deal | within 2–3 days | poor | moderate | adequately | accessible for questions | encourages questions & answers them seriously | | difficult but reasonable | very helpful | not at all helpful | Excellent |
| 3.50–4.00 | all but 1–2 | most readings | 95–100% | A | A | very quick&accurate | difficult but reasonable | interesting | a great deal | within week before exam | poor | moderate | adequately | accessible for questions | I did not seek extra help | | difficult but reasonable | very helpful | not at all helpful | Good |
| 3.00–3.49 | all classes | most readings | 85–94% | B | B | mod. paced&mostly accurate | fairly simple | of some interest | moderate amt | within 2 weeks | NO txtbk used | moderate | adequately | very clearly | encourages questions & answers them seriously | | moderate | somewhat helpful | not at all helpful | Good |
| 3.50–4.00 | few classes | all readings | 95–100% | A | A | less quick&/or accurate than I'd like | difficult but reasonable | of no interest | almost nothing | within week before exam | poor | moderate | adequately | accessible for questions | encourages questions & answers them seriously | | much too easy | very helpful | not at all helpful | Excellent |
| 3.50–4.00 | all but 3–4 | most readings | 95–100% | A | A | very quick&accurate | fairly simple | of no interest | a great deal | within week before exam | poor | moderate | adequately | very clearly | accessible for questions | encourages questions & answers them seriously | moderate | very helpful | | Good |
| 3.00–3.49 | most classes | most readings | 95–100% | B | B | mod. paced&mostly accurate | fairly simple | of some interest | moderate amt | within 2 weeks | poor | moderate | adequately | | I did not seek extra help | encourages questions & answers them seriously | difficult but reasonable | somewhat helpful | not at all helpful | Good |
| 3.50–4.00 | all classes | all readings | 95–100% | A | A | quick & reasonably accurate | difficult but reasonable | of some interest | moderate amt | within 2–3 days | NO txtbk used | moderate | very clearly | accessible for questions | encourages questions & answers them seriously | | moderate | very helpful | very helpful | |
| 2.50–2.99 | all but 1–2 | all readings | 95–100% | A | B | quick & reasonably accurate | difficult but reasonable | interesting | a great deal | within 2 weeks | adequate | moderate | very clearly | accessible for questions | encourages questions & answers them seriously | | moderate | very helpful | somewhat helpful | The best or very nearly the best |
| 3.50–4.00 | all but 1–2 | all readings | 70–84% | A | A | quick & reasonably accurate | fairly simple | of some interest | moderate amt | within 2–3 days | adequate | much too slow | adequately | accessible for questions | encourages questions & answers them seriously | | fairly easy | very helpful | somewhat helpful | |
| 3.50–4.00 | all but 1–2 | most readings | <70% | B | C | very quick&accurate | difficult but reasonable | of some interest | moderate amt | within 2 weeks | adequate | somewhat slow | adequately | accessible for questions | encourages questions & answers them seriously | | moderate | very helpful | not at all helpful | Excellent |
| 3.50–4.00 | most classes | most readings | 85–94% | A | A | quick & reasonably accurate | difficult but reasonable | interesting | a great deal | within 2–3 days | adequate | reasonably fast | adequately | accessible for questions | encourages questions & answers them seriously | | difficult but reasonable | somewhat helpful | not at all helpful | Excellent |
| 2.50–2.99 | all but 1–2 | all readings | 95–100% | C | D | quick & reasonably accurate | difficult but reasonable | of no interest | almost nothing | within 2–3 days | good | reasonably fast | adequately | accessible for questions | encourages questions & answers them seriously | | too difficult | somewhat helpful | not at all helpful | Fair |
| 3.50–4.00 | all but 1–2 | most readings | <70% | A | A | very quick&accurate | fairly simple | of some interest | moderate amt | within week before exam | NO txtbk used | reasonably fast | very clearly | accessible for questions | encourages questions & answers them seriously | | moderate | not helpful | somewhat helpful | Excellent |
| 3.00–3.49 | most classes | all readings | 85–94% | A | B | fairly simple | of some interest | moderate amt | within 2–3 days | good | moderate | adequately | accessible for questions | encourages questions & answers them seriously | | moderate | moderate | very helpful | Excellent | |
| 2.50–2.99 | all but 1–2 | most readings | 85–94% | B | C | very quick&accurate | difficult but reasonable | interesting | moderate amt | within 2 weeks | excellent | moderate | adequately | | I did not seek extra help | encourages questions & answers them seriously | difficult but reasonable | very helpful | very helpful | Good |
| 3.50–4.00 | all classes | most readings | 95–100% | A | A | quick & reasonably accurate | difficult but reasonable | interesting | moderate amt | within 2 weeks | good | reasonably fast | very clearly | accessible for questions | encourages questions & answers them seriously | | difficult but reasonable | very helpful | | Good |
| 2.50–2.99 | all but 1–2 | all readings | 95–100% | A | A | very quick&accurate | difficult but reasonable | very exciting | moderate amt | within 1 day of assignment | adequate | moderate | very clearly | | I did not seek extra help | encourages questions & answers them seriously | difficult but reasonable | somewhat helpful | not at all helpful | The best or very nearly the best |
| 3.50–4.00 | all but 1–2 | most readings | 95–100% | A | B | mod. paced&mostly accurate | difficult but reasonable | interesting | a great deal | within 2 weeks | poor | moderate | very clearly | | I did not seek extra help | encourages questions & answers them seriously | too difficult | very helpful | somewhat helpful | Excellent |
| 3.50–4.00 | all but 1–2 | none of rdgs | <70% | B | B | quick & reasonably accurate | fairly simple | of some interest | moderate amt | within week before exam | NO txtbk used | reasonably fast | very clearly | | I did not seek extra help | encourages questions & answers them seriously | moderate | somewhat helpful | haven't tried much Mastering HW | Excellent |
| 3.50–4.00 | all classes | all readings | 95–100% | A | A | quick & reasonably accurate | difficult but reasonable | interesting | a great deal | within 2–3 days | NO txtbk used | reasonably fast | very clearly | accessible for questions | encourages questions & answers them seriously | | difficult but reasonable | somewhat helpful | | The best or very nearly the best |
| 2.00–2.49 | all but 1–2 | all readings | NO HW assigned | B | B | very quick&accurate | difficult but reasonable | very exciting | a great deal | NO HW assigned | NO txtbk used | very clearly | accessible for questions | encourages questions & answers them seriously | | | moderate | very helpful | haven't tried any Mastering HW | The best or very nearly the best |
| 3.50–4.00 | most classes | all readings | <70% | B | B | quick & reasonably accurate | fairly simple | interesting | a great deal | within week before exam | NO txtbk used | somewhat slow | adequately | accessible for questions | encourages questions & answers them seriously | | difficult but reasonable | somewhat helpful | haven't tried any Mastering HW | Excellent |
| 3.50–4.00 | all but 1–2 | none of rdgs | 85–94% | A | A | very quick&accurate | fairly simple | interesting | moderate amt | within 2–3 days | poor | moderate | adequately | accessible for questions | encourages questions & answers them seriously | | moderate | somewhat helpful | haven't tried any Mastering HW | Good |
| 2.50–2.99 | all classes | all readings | 95–100% | A | A | mod. paced&mostly accurate | difficult but reasonable | interesting | a great deal | within 1 day of assignment | adequate | moderate | very clearly | accessible for questions | encourages questions & answers them seriously | | moderate | somewhat helpful | somewhat helpful | Fair |
| 3.50–4.00 | all but 1–2 | most readings | 95–100% | A | A | very quick&accurate | difficult but reasonable | interesting | a great deal | within week before exam | adequate | moderate | very clearly | accessible for questions | encourages questions & answers them seriously | | difficult but reasonable | very helpful | very helpful | Excellent |
| 3.00–3.49 | all but 1–2 | all readings | 95–100% | A | B | very quick&accurate | fairly simple | interesting | a great deal | within 1 day of assignment | excellent | reasonably fast | very clearly | accessible for questions | encourages questions & answers them seriously | | difficult but reasonable | very helpful | | Excellent |
| 3.50–4.00 | all classes | all readings | <70% | A | A | quick & reasonably accurate | difficult but reasonable | interesting | a great deal | within 2 weeks | good | too rapid for proper understanding | adequately | accessible for questions | encourages questions & answers them seriously | | too difficult | very helpful | very helpful | Excellent |
| 3.50–4.00 | few classes | some readings | 85–94% | A | A | very quick&accurate | difficult but reasonable | of no interest | moderate amt | within 2–3 days | adequate | moderate | poorly | | I did not seek extra help | encourages questions & answers them seriously | moderate | very helpful | not at all helpful | Good |
| 3.50–4.00 | all but 1–2 | all readings | 95–100% | A | A | quick & reasonably accurate | fairly simple | interesting | moderate amt | within week before exam | NO txtbk used | somewhat slow | adequately | could rarely be found | does not encourage questions but answers them seriously | | difficult but reasonable | somewhat helpful | not at all helpful | Fair |
| 2.50–2.99 | all but 1–2 | most readings | <70% | B | B | mod. paced&mostly accurate | difficult but reasonable | interesting | moderate amt | within week before exam | NO txtbk used | somewhat slow | very clearly | accessible for questions | encourages questions & answers them seriously | | fairly easy | very helpful | haven't tried any Mastering HW | The best or very nearly the best |
| 3.00–3.49 | all but 1–2 | all readings | 95–100% | A | B | mod. paced&mostly accurate | difficult but reasonable | of some interest | moderate amt | within 2 weeks | adequate | moderate | poorly | accessible for questions | encourages questions & answers them seriously | | very helpful | very helpful | Good | |
| 3.00–3.49 | all but 1–2 | all readings | 95–100% | A | B | mod. paced&mostly accurate | too difficult | very exciting | moderate amt | within week before exam | excellent | moderate | very clearly | accessible for questions | encourages questions & answers them seriously | | too difficult | very helpful | very helpful | Good |
| 3.50–4.00 | all classes | all readings | 95–100% | A | A | quick & reasonably accurate | difficult but reasonable | interesting | a great deal | within 2–3 days | NO txtbk used | moderate | very clearly | accessible for questions | encourages questions & answers them seriously | | too difficult | very helpful | very helpful | The best or very nearly the best |
| 3.50–4.00 | all but 3–4 | none of rdgs | <70% | A | A | quick & reasonably accurate | fairly simple | of some interest | moderate amt | NO HW assigned | NO txtbk used | somewhat slow | adequately | | I did not seek extra help | encourages questions & answers them seriously | moderate | somewhat helpful | haven't tried any Mastering HW | Good |
| 3.50–4.00 | most classes | some readings | <70% | A | A | very quick&accurate | difficult but reasonable | of some interest | moderate amt | NO HW assigned | NO txtbk used | somewhat slow | poorly | accessible for questions | encourages questions & answers them seriously | | moderate | moderate | very helpful | Good |
| 3.50–4.00 | all but 1–2 | some readings | <70% | A | A | very quick&accurate | difficult but reasonable | of some interest | moderate amt | within week before exam | NO txtbk used | moderate | not at all | | I did not seek extra help | encourages questions & answers them seriously | moderate | very helpful | haven't tried much Mastering HW | Good |
| 3.50–4.00 | most classes | none of rdgs | <70% | B | C | mod. paced&mostly accurate | difficult but reasonable | of no interest | almost nothing | within week before exam | NO txtbk used | moderate | very clearly | accessible for questions | encourages questions & answers them seriously | | difficult but reasonable | too difficult | somewhat helpful | |
| 3.50–4.00 | all but 1–2 | none of rdgs | <70% | A | A | very quick&accurate | fairly simple | interesting | a great deal | NO HW assigned | NO txtbk used | moderate | very clearly | | I did not seek extra help | encourages questions & answers them seriously | moderate | I never consulted Blkbd supplmts | somewhat helpful | Excellent |
| 2.50–2.99 | most classes | most readings | 95–100% | B | B | very quick&accurate | difficult but reasonable | of no interest | moderate amt | within 2 weeks | poor | too rapid for proper understanding | adequately | accessible for questions | encourages questions & answers them seriously | | difficult but reasonable | | somewhat helpful | Fair |

| GPA | ATTENDANCE | RDGS COMPLETED | HW COMPLETED | GRD DESERVED | GRD EXPECTED | MATH ABILITY | MATERIAL COVERED | MATERIAL COVERED | I LEARNED | HW STARTED | TEXTBK | PACE OF LECTS | DIFF/SUBTLE PTS | PROFESSOR | PROFESSOR | EXAMS WERE | BLKBD SUPPLEMNTS | MAST.PHYSICS | COURSE OVERALL |
|---|---|---|---|---|---|---|---|---|---|---|---|---|---|---|---|---|---|---|---|
| 3.00–3.49 | all but 1–2 | all readings | | A | A | quick & reasonably accurate | fairly simple | | interesting | a great deal | | moderate | | very clearly | encourages questions & answers them seriously | fairly easy | | haven't tried any Mastering HW | Excellent |
| 3.50–4.00 | all classes | all readings | 95–100% | A | A | very quick&accurate | fairly simple | | interesting | within 1 day of assignment | excellent | too rapid for proper understanding | very clearly | accessible for questions | does not encourage questions but answers them seriously | too difficult | | | Excellent |
| 3.50–4.00 | all but 1–2 | most readings | 95–100% | A | A | very quick&accurate | fairly simple | of some interest | a great deal | within 2 weeks | NO txtbk used | moderate | adequately | accessible for questions | encourages questions & answers them seriously | moderate | | not at all helpful | Excellent |
| 3.50–4.00 | all but 1–2 | most readings | 95–100% | A | B | quick & reasonably accurate | fairly simple | interesting | moderate amt | within 2–3 days | good | moderate | adequately | accessible for questions | encourages questions & answers them seriously | moderate | | very helpful | Excellent |
| 3.50–4.00 | all classes | some readings | 95–100% | A | A | very quick&accurate | fairly simple | interesting | moderate amt | within week before exam | NO txtbk used | reasonably fast | adequately | accessible for questions | encourages questions & answers them seriously | moderate | | very helpful | The best or very nearly the best |
| 3.50–4.00 | all but 1–2 | some readings | <70% | A | B | mod. paced&mostly accurate | difficult but reasonable | of no interest | moderate amt | NO HW assigned | poor | too rapid for proper understanding | poorly | I did not seek help | encourages questions & answers them seriously | difficult but reasonable | I never consulted Blkbd supplemts | not at all helpful | Good |
| 2.50–2.99 | all but 1–2 | all readings | <70% | C | F | very quick&accurate | difficult but reasonable | interesting | moderate amt | within week before exam | adequate | somewhat slow | adequately | very clearly | encourages questions & answers them seriously | too difficult | | not helpful | Good |
| 3.50–4.00 | all but 1–2 | none of rdgs | <70% | A | A | mod. paced&mostly accurate | difficult but reasonable | of some interest | a great deal | within week before exam | adequate | moderate | adequately | very clearly | I did not seek help | moderate | | not at all helpful | Excellent |
| 3.00–3.49 | most classes | some readings | 70–84% | A | A | quick & reasonably accurate | difficult but reasonable | of some interest | moderate amt | within week before exam | NO txtbk used | too rapid for proper understanding | adequately | accessible for questions | encourages questions & answers them seriously | moderate | somewhat helpful | not at all helpful | Good |
| 3.50–4.00 | most classes | none of rdgs | 95–100% | A | B | quick & reasonably accurate | difficult but reasonable | of some interest | a great deal | within 2–3 days | NO txtbk used | moderate | adequately | accessible for questions | I did not seek help | moderate | | very helpful | Good |
| 3.50–4.00 | most classes | none of rdgs | 95–100% | A | A | very quick&accurate | difficult but reasonable | interesting | moderate amt | within week before exam | NO txtbk used | moderate | poorly | accessible for questions | encourages questions & answers them seriously | difficult but reasonable | somewhat helpful | very helpful | Good |
| 3.00–3.49 | all but 1–2 | | 95–100% | A | A | quick & reasonably accurate | difficult but reasonable | interesting | a great deal | within 2–3 days | poor | moderate | poorly | accessible for questions | encourages questions & answers them seriously | moderate | | not at all helpful | The best or very nearly the best |
| 2.50–2.99 | all but 1–2 | | 85–94% | A | A | mod. paced&mostly accurate | difficult but reasonable | interesting | moderate amt | within week before exam | adequate | moderate | very clearly | very clearly | encourages questions & answers them seriously | moderate | | very helpful | The best or very nearly the best |
| 3.50–4.00 | all classes | all readings | NO HW assigned | A | A | quick & reasonably accurate | fairly simple | of some interest | moderate amt | NO HW assigned | adequate | moderate | adequately | accessible for questions | encourages questions & answers them seriously | moderate | | haven't tried any Mastering HW | Excellent |
| 3.50–4.00 | all but 1–2 | all readings | 95–100% | A | A | quick & reasonably accurate | fairly simple | of some interest | moderate amt | within 2 weeks | adequate | moderate | adequately | accessible for questions | encourages questions & answers them seriously | moderate | | very helpful | Excellent |
| 3.50–4.00 | all classes | all readings | 95–100% | A | A | quick & reasonably accurate | too difficult | very exciting | a great deal | within 1 day of assignment | good | too rapid for proper understanding | poorly | could rarely be found | encourages questions & answers them seriously | too difficult | | very helpful | Good |
| 3.00–3.49 | all but 1–2 | all readings | <70% | B | B | very quick&accurate | difficult but reasonable | of some interest | moderate amt | within week before exam | poor | somewhat slow | adequately | accessible for questions | encourages questions & answers them seriously | difficult but reasonable | | very helpful | Good |
| 3.50–4.00 | all but 1–2 | most readings | 95–100% | A | A | quick & reasonably accurate | fairly simple | of no interest | moderate amt | within 2–3 days | poor | somewhat slow | adequately | | | | | very helpful | Fair |
| 3.00–3.49 | all classes | most readings | 95–100% | A | B | quick & reasonably accurate | difficult but reasonable | interesting | a great deal | within 2–3 days | adequate | moderate | adequately | very clearly | encourages questions & answers them seriously | moderate | | very helpful | Excellent |
| 3.00–3.49 | most classes | most readings | 95–100% | B | C | mod. paced&mostly accurate | difficult but reasonable | of no interest | moderate amt | within week before exam | excellent | reasonably fast | adequately | accessible for questions | encourages questions & answers them seriously | difficult but reasonable | somewhat helpful | very helpful | Good |
| 3.00–3.49 | all but 1–2 | most readings | 85–94% | B | C | mod. paced&mostly accurate | difficult but reasonable | of no interest | moderate amt | within 2–3 days | adequate | somewhat slow | adequately | | accessible for questions | moderate | | not at all helpful | Fair |
| 3.50–4.00 | most classes | some readings | <70% | A | A | quick & reasonably accurate | difficult but reasonable | interesting | moderate amt | within week before exam | NO txtbk used | too rapid for proper understanding | adequately | | I did not seek help | difficult but reasonable | | very helpful | Good |
| 3.00–3.49 | all but 1–2 | all readings | 95–100% | A | A | mod. paced&mostly accurate | difficult but reasonable | interesting | moderate amt | within 2–3 days | adequate | moderate | adequately | accessible for questions | encourages questions & answers them seriously | moderate | | very helpful | Excellent |
| 3.00–3.49 | all but 1–2 | most readings | 95–100% | A | A | quick & reasonably accurate | difficult but reasonable | of some interest | moderate amt | within 2–3 days | adequate | somewhat slow | adequately | accessible for questions | I did not seek help | moderate | | very helpful | Good |
| 3.00–3.49 | all classes | all readings | 95–100% | A | A | very quick&accurate | fairly simple | very exciting | a great deal | within 2 weeks | NO txtbk used | moderate | adequately | very clearly | encourages questions & answers them seriously | moderate | | moderate | Good |
| 2.50–2.99 | all but 1–2 | all readings | 95–100% | A | A | very quick&accurate | difficult but reasonable | interesting | a great deal | within 2 weeks | good | moderate | adequately | very clearly | accessible for questions | moderate | | very helpful | The best or very nearly the best |
| 3.00–3.49 | all but 1–2 | most readings | 95–100% | A | B | quick & reasonably accurate | difficult but reasonable | of some interest | moderate amt | within week before exam | NO txtbk used | somewhat slow | adequately | accessible for questions | I did not seek help | difficult but reasonable | | very helpful | Good |
| 3.50–4.00 | all but 1–2 | some readings | 95–100% | B | B | quick & reasonably accurate | difficult but reasonable | interesting | moderate amt | within 2–3 days | NO txtbk used | reasonably fast | adequately | accessible for questions | encourages questions & answers them seriously | difficult but reasonable | | very helpful | Good |
| 3.50–4.00 | all but 1–2 | | 95–100% | A | A | quick & reasonably accurate | fairly simple | interesting | a great deal | within 2–3 days | good | moderate | adequately | very clearly | I did not seek help | fairly easy | | not at all helpful | Excellent |
| 2.50–2.99 | | all readings | 95–100% | A | A | mod. paced&mostly accurate | difficult but reasonable | interesting | moderate amt | within 2–3 days | NO txtbk used | moderate | adequately | accessible for questions | encourages questions & answers them seriously | difficult but reasonable | somewhat helpful | very helpful | Good |
| 3.50–4.00 | all but 1–2 | all readings | 95–100% | B | B | quick & reasonably accurate | difficult but reasonable | interesting | a great deal | within 2–3 days | adequate | moderate | adequately | accessible for questions | encourages questions & answers them seriously | difficult but reasonable | | very helpful | Excellent |
| 3.00–3.49 | all but 1–2 | most readings | 95–100% | A | A | very quick&accurate | difficult but reasonable | interesting | moderate amt | within 2–3 days | NO txtbk used | reasonably fast | adequately | accessible for questions | encourages questions & answers them seriously | difficult but reasonable | | very helpful | Good |
| 3.50–4.00 | few classes | some readings | <70% | A | A | very quick&accurate | fairly simple | of some interest | moderate amt | | | | | | | | | | |
| 3.50–4.00 | all but 3–4 | some readings | <70% | B | B | mod. paced&mostly accurate | fairly simple | of some interest | moderate amt | within 2 weeks | adequate | reasonably fast | very clearly | very clearly | I did not seek help | moderate | | not helpful | |
| 3.50–4.00 | all but 1–2 | most readings | 95–100% | A | A | very quick&accurate | fairly simple | of some interest | moderate amt | within 2–3 days | NO txtbk used | moderate | adequately | very clearly | accessible for questions | fairly easy | | very helpful | Excellent |
| 3.50–4.00 | all classes | most readings | 95–100% | A | A | mod. paced&mostly accurate | difficult but reasonable | of some interest | moderate amt | within 2–3 days | adequate | reasonably fast | adequately | very clearly | accessible for questions | difficult but reasonable | | somewhat helpful | Good |
| 2.50–2.99 | all classes | most readings | <70% | B | B | very quick&accurate | fairly simple | interesting | a great deal | within 2–3 days | good | reasonably fast | adequately | accessible for questions | encourages questions & answers them seriously | moderate | | haven't tried any Mastering HW | Excellent |
| 3.00–3.49 | all but 1–2 | most readings | <70% | A | B | mod. paced&mostly accurate | too difficult | of some interest | moderate amt | within 2 weeks | NO txtbk used | moderate | adequately | very clearly | I did not seek help | too difficult | | not at all helpful | Good |
| | all but 1–2 | most readings | | | | | | very exciting | a great deal | within 2 weeks | | good | reasonably fast | | accessible for questions | | | very helpful | The best or very nearly the best |
| 2.50–2.99 | few classes | most readings | <70% | B | B | quick & reasonably accurate | difficult but reasonable | interesting | a great deal | | NO txtbk used | moderate | adequately | | encourages questions & answers them seriously | difficult but reasonable | somewhat helpful | haven't tried much Mastering HW | Good |
| 3.50–4.00 | all but 1–2 | | 85–94% | B | C | quick & reasonably accurate | difficult but reasonable | interesting | moderate amt | within 2 weeks | NO txtbk used | moderate | adequately | very clearly | I did not seek help | difficult but reasonable | somewhat helpful | very helpful | Excellent |
| 3.50–4.00 | all but 3–4 | some readings | NO HW assigned | A | A | very quick&accurate | difficult but reasonable | of some interest | moderate amt | NO HW assigned | NO txtbk used | moderate | very clearly | very clearly | I did not seek help | difficult but reasonable | | very helpful | Excellent |
| 2.50–2.99 | all classes | all readings | 85–94% | A | A | quick & reasonably accurate | difficult but reasonable | interesting | a great deal | within 1 day of assignment | good | moderate | adequately | accessible for questions | encourages questions & answers them seriously | moderate | | very helpful | Good |
| 3.00–3.49 | all but 3–4 | all readings | <70% | A | A | quick & reasonably accurate | difficult but reasonable | of no interest | almost nothing | | NO txtbk used | moderate | adequately | very clearly | could rarely be found | moderate | | haven't tried much Mastering HW | Excellent |
| 3.50–4.00 | all but 1–2 | some readings | <70% | B | B | quick & reasonably accurate | difficult but reasonable | interesting | a great deal | within week before exam | adequate | moderate | adequately | very clearly | accessible for questions | difficult but reasonable | somewhat helpful | | Excellent |
| 3.50–4.00 | all but 1–2 | none of rdgs | 95–100% | A | A | very quick&accurate | difficult but reasonable | interesting | moderate amt | within 2–3 days | NO txtbk used | moderate | adequately | very clearly | accessible for questions | difficult but reasonable | | very helpful | Excellent |
| 3.00–3.49 | all but 3–4 | most readings | <70% | B | B | very quick&accurate | difficult but reasonable | of some interest | a great deal | within week before exam | adequate | moderate | adequately | very clearly | I did not seek help | difficult but reasonable | | haven't tried much Mastering HW | Excellent |
| 3.00–3.49 | all but 1–2 | none of rdgs | NO HW assigned | A | B | less quick&/or accurate than I'd like | difficult but reasonable | of no interest | almost nothing | NO HW assigned | NO txtbk used | moderate | adequately | | encourages questions & answers them seriously | too difficult | somewhat helpful | haven't tried any Mastering HW | Fair |
| 3.50–4.00 | all classes | some readings | <70% | A | A | very quick&accurate | difficult but reasonable | interesting | a great deal | within week before exam | NO txtbk used | moderate | adequately | very clearly | I did not seek help | difficult but reasonable | | not at all helpful | Good |
| 2.50–2.99 | all but 1–2 | all readings | <70% | A | A | quick & reasonably accurate | difficult but reasonable | of some interest | moderate amt | | | somewhat slow | adequately | | accessible for questions | difficult but reasonable | | very helpful | Good |
| 3.50–4.00 | all but 1–2 | most readings | 70–84% | A | A | quick & reasonably accurate | difficult but reasonable | of some interest | moderate amt | within week before exam | poor | too rapid for proper understanding | adequately | very clearly | I did not seek help | difficult but reasonable | | not at all helpful | Good |
| 3.50–4.00 | most classes | all readings | 95–100% | A | A | very quick&accurate | fairly simple | of some interest | moderate amt | within 2–3 days | NO txtbk used | much too slow | very clearly | very clearly | I did not seek help | much too easy | | not at all helpful | The best or very nearly the best |
| 2.00–2.49 | few classes | some readings | NO HW assigned | C | B | quick & reasonably accurate | difficult but reasonable | of some interest | moderate amt | within week before exam | NO txtbk used | reasonably fast | adequately | | encourages questions & answers them seriously | difficult but reasonable | somewhat helpful | haven't tried any Mastering HW | Good |
| 3.50–4.00 | few classes | most readings | 70–84% | A | A | quick & reasonably accurate | difficult but reasonable | of no interest | almost nothing | within week before exam | NO txtbk used | much too slow | very clearly | very clearly | accessible for questions | fairly easy | I never consulted Blkbd supplemts | not at all helpful | Good |
| 3.50–4.00 | all but 1–2 | some readings | <70% | C | B | quick & reasonably accurate | difficult but reasonable | interesting | a great deal | within week before exam | NO txtbk used | moderate | adequately | very clearly | encourages questions & answers them seriously | fairly easy | | very helpful | Good |
| 2.50–2.99 | all but 1–2 | most readings | 85–94% | A | B | mod. paced&mostly accurate | difficult but reasonable | of no interest | moderate amt | within 2 weeks | NO txtbk used | reasonably fast | adequately | | encourages questions & answers them seriously | difficult but reasonable | | very helpful | Good |
| 3.50–4.00 | most classes | most readings | <70% | A | A | quick & reasonably accurate | fairly simple | interesting | moderate amt | within 2 weeks | | good | moderate | very clearly | accessible for questions | difficult but reasonable | | very helpful | Excellent |
| 3.50–4.00 | few classes | most readings | <70% | C | A | quick & reasonably accurate | fairly simple | of no interest | moderate amt | within week before exam | | somewhat slow | adequately | very clearly | encourages questions & answers them seriously | moderate | | very helpful | Excellent |
| 3.50–4.00 | all but 3–4 | all readings | 70–84% | A | A | very quick&accurate | difficult but reasonable | interesting | a great deal | within week before exam | NO txtbk used | moderate | adequately | very clearly | accessible for questions | fairly easy | | very helpful | Excellent |
| 3.50–4.00 | all but 1–2 | most readings | 85–94% | A | A | very quick&accurate | difficult but reasonable | interesting | moderate amt | within 2–3 days | adequate | moderate | adequately | very clearly | accessible for questions | difficult but reasonable | | somewhat helpful | Excellent |
| 3.50–4.00 | all but 1–2 | most readings | 95–100% | A | A | very quick&accurate | difficult but reasonable | interesting | moderate amt | within 2–3 days | adequate | reasonably fast | adequately | very clearly | accessible for questions | difficult but reasonable | | very helpful | Excellent |
| 3.50–4.00 | most classes | most readings | 95–100% | A | A | quick & reasonably accurate | difficult but reasonable | interesting | moderate amt | within 2–3 days | poor | moderate | poorly | | I did not seek help | moderate | | very helpful | Excellent |

| GPA | ATTENDANCE | RDGS COMPLETED | HW COMPLETED | GRD DESERVED | GRD EXPECTED | MATH ABILITY | MATERIAL COVERED | MATERIAL COVERED | I LEARNED | HW STARTED | TEXTBK | PACE OF LECTS | DIFF/SUBTLE PTS | PROFESSOR | PROFESSOR | EXAMS WERE | BLKBD SUPPLEMENTS | MAST.PHYSICS | COURSE OVERALL |
|---|---|---|---|---|---|---|---|---|---|---|---|---|---|---|---|---|---|---|---|
| 2.00–2.49 | all but 1–2 | most readings | <70% | B | B | mod.paced&mostly accurate | difficult but reasonable | interesting | moderate amt | within week before exam | adequate | moderate | adequately | accessible for questions | encourages questions & answers them seriously | difficult but reasonable | somewhat helpful | | Excellent |
| 3.50–4.00 | most classes | most readings | <70% | A | A | very quick&accurate | fairly simple | of some interest | | within week before exam | adequate | somewhat slow | very clearly | I did not seek extra help | encourages questions & answers them seriously | fairly easy | I never consulted Blkbd supplemts | haven't tried any Mastering HW | Excellent |
| 3.50–4.00 | most classes | some readings | <70% | A | A | quick & reasonably accurate | difficult but reasonable | interesting | moderate amt | within week before exam | good | somewhat slow | very clearly | accessible for questions | encourages questions & answers them seriously | moderate | very helpful | haven't tried any Mastering HW | Good |
| 3.50–4.00 | most classes | most readings | <70% | A | A | very quick&accurate | fairly simple | interesting | D | within 2 weeks | adequate | moderate | very clearly | accessible for questions | encourages questions & answers them seriously | fairly easy | not at all helpful | | Excellent |
| 2.50–2.99 | all classes | all readings | <70% | B | B | quick & reasonably accurate | difficult but reasonable | of some interest | a great deal | within 2 weeks | good | reasonably fast | very clearly | accessible for questions | encourages questions & answers them seriously | moderate | somewhat helpful | | Good |
| 3.00–3.49 | most classes | all readings | 95–100% | A | A | very quick&accurate | difficult but reasonable | interesting | moderate amt | within 2 weeks | NO txtbk used | too rapid for proper understanding | very clearly | accessible for questions | encourages questions & answers them seriously | fairly easy | very helpful | not at all helpful | Good |
| 3.00–3.49 | few classes | none of rdgs | 95–100% | A | A | mod.paced&mostly accurate | fairly simple | interesting | moderate amt | within 2 weeks | NO txtbk used | somewhat slow | adequately | I did not seek extra help | encourages questions & answers them seriously | moderate | very helpful | very helpful | |
| 3.50–4.00 | all but 3–4 | | 95–100% | A | A | mod.paced&mostly accurate | difficult but reasonable | very exciting | moderate amt | within 2–3 days | adequate | moderate | adequately | accessible for questions | encourages questions & answers them seriously | difficult but reasonable | very helpful | somewhat helpful | Excellent |
| 3.00–3.49 | most classes | some readings | 70–84% | A | A | mod.paced&mostly accurate | difficult but reasonable | of some interest | moderate amt | within 2–3 days | NO txtbk used | moderate | adequately | accessible for questions | encourages questions & answers them seriously | difficult but reasonable | very helpful | not at all helpful | The best or very nearly the best |
| 3.50–4.00 | most classes | | 95–100% | A | A | very quick&accurate | of no interest | almost nothing | within 1 day of assignment | NO txtbk used | much too slow | poorly | accessible for questions | encourages questions & answers them seriously | too difficult | | not at all helpful | Poor |
| 2.00–2.49 | all but 1–2 | most readings | NO HW assigned | B | A | very quick&accurate | difficult but reasonable | of no interest | moderate amt | NO HW assigned | adequate | moderate | adequately | accessible for questions | encourages questions & answers them seriously | moderate | | haven't tried any Mastering HW | |
| 2.50–2.99 | all but 1–2 | all readings | 85–94% | A | A | very quick&accurate | difficult but reasonable | interesting | a great deal | within 2 weeks | adequate | moderate | adequately | could rarely be found | encourages questions & answers them seriously | difficult but reasonable | very helpful | not at all helpful | Good |
| 3.00–3.49 | all classes | all readings | <70% | B | A | very quick&accurate | difficult but reasonable | of some interest | moderate amt | within 2 weeks | good | moderate | adequately | accessible for questions | encourages questions & answers them seriously | difficult but reasonable | very helpful | very helpful | Excellent |
| 3.50–4.00 | all classes | all readings | 95–100% | A | A | very quick&accurate | difficult but reasonable | of some interest | moderate amt | within 2 weeks | good | moderate | adequately | accessible for questions | encourages questions & answers them seriously | moderate | too difficult | not at all helpful | Excellent |
| 3.00–3.49 | all classes | all readings | 95–100% | A | A | very quick&accurate | fairly simple | interesting | a great deal | within 2 weeks | NO txtbk used | moderate | very clearly | accessible for questions | encourages questions & answers them seriously | moderate | very helpful | not at all helpful | The best or very nearly the best |
| 3.50–4.00 | all classes | all readings | 95–100% | A | A | quick & reasonably accurate | difficult but reasonable | interesting | a great deal | within 2–3 days | NO txtbk used | moderate | adequately | I did not seek extra help | encourages questions & answers them seriously | difficult but reasonable | very helpful | not at all helpful | Good |
| 3.00–3.49 | all but 1–2 | most readings | 95–100% | B | C | mod.paced&mostly accurate | difficult but reasonable | of no interest | moderate amt | within 2–3 days | adequate | reasonably fast | poorly | accessible for questions | encourages questions & answers them seriously | difficult but reasonable | somewhat helpful | very helpful | Fair |
| 3.50–4.00 | all classes | some readings | 95–100% | A | A | mod.paced&mostly accurate | fairly simple | interesting | a great deal | within 2 weeks | good | reasonably fast | very clearly | accessible for questions | encourages questions & answers them seriously | difficult but reasonable | somewhat helpful | somewhat helpful | Excellent |
| 3.50–4.00 | all but 1–2 | none of rdgs | NO HW assigned | B | B | quick & reasonably accurate | difficult but reasonable | interesting | moderate amt | within 2 weeks | NO txtbk used | moderate | very clearly | I did not seek extra help | encourages questions & answers them seriously | difficult but reasonable | very helpful | not at all helpful | Good |
| 3.50–4.00 | most classes | most readings | <70% | A | A | mod.paced&mostly accurate | too difficult | interesting | moderate amt | within week before exam | NO txtbk used | moderate | adequately | accessible for questions | encourages questions & answers them seriously | too difficult | very helpful | haven't tried much Mastering HW | Excellent |
| 3.50–4.00 | all classes | some readings | 95–100% | A | A | mod.paced&mostly accurate | too difficult | of no interest | almost nothing | within week before exam | NO txtbk used | too rapid for proper understanding | poorly | could rarely be found | does not encourage questions but answers them seriously | difficult but reasonable | very helpful | somewhat helpful | Poor |
| 2.00–2.49 | most classes | most readings | <70% | C | B | mod.paced&mostly accurate | fairly simple | of some interest | moderate amt | within 2 weeks | good | moderate | very clearly | accessible for questions | encourages questions & answers them seriously | moderate | very helpful | not at all helpful | Excellent |
| 3.50–4.00 | all but 1–2 | most readings | 70–84% | A | A | very quick&accurate | difficult but reasonable | of some interest | moderate amt | within 1 day of assignment | NO txtbk used | moderate | adequately | I did not seek extra help | encourages questions & answers them seriously | difficult but reasonable | very helpful | haven't tried any Mastering HW | Excellent |
| 2.50–2.99 | most classes | all readings | 95–100% | A | A | mod.paced&mostly accurate | difficult but reasonable | of no interest | moderate amt | within 1 day of assignment | adequate | moderate | adequately | could rarely be found | encourages questions & answers them seriously | difficult but reasonable | somewhat helpful | not at all helpful | Good |
| 3.00–3.49 | most classes | most readings | 95–100% | B | C | mod.paced&mostly accurate | difficult but reasonable | of no interest | moderate amt | within 2–3 days | adequate | reasonably fast | adequately | accessible for questions | encourages questions & answers them seriously | difficult but reasonable | somewhat helpful | not at all helpful | Good |
| 3.50–4.00 | all but 1–2 | none of rdgs | 95–100% | A | A | very quick&accurate | fairly simple | interesting | a great deal | within 2 weeks | adequate | reasonably fast | very clearly | accessible for questions | encourages questions & answers them seriously | moderate | very helpful | somewhat helpful | Excellent |
| 3.50–4.00 | all classes | all readings | 95–100% | A | A | very quick&accurate | fairly simple | interesting | a great deal | within 1 day of assignment | NO txtbk used | moderate | very clearly | accessible for questions | encourages questions & answers them seriously | moderate | very helpful | somewhat helpful | Excellent |
| 3.50–4.00 | most classes | some readings | <70% | B | B | very quick&accurate | fairly simple | interesting | a great deal | within 1 day of assignment | NO txtbk used | moderate | very clearly | accessible for questions | encourages questions & answers them seriously | moderate | somewhat helpful | haven't tried much Mastering HW | Excellent |
| 3.00–3.49 | most classes | most readings | 85–94% | A | A | very quick&accurate | difficult but reasonable | interesting | moderate amt | within 2 weeks | NO txtbk used | moderate | very clearly | I did not seek extra help | encourages questions & answers them seriously | moderate | very helpful | not at all helpful | Excellent |
| 3.50–4.00 | all classes | some readings | <70% | A | A | very quick&accurate | too simple | interesting | almost nothing | within 1 day of assignment | NO txtbk used | much too slow | very clearly | accessible for questions | encourages questions & answers them seriously | much too easy | I never consulted Blkbd supplemts | somewhat helpful | Good |
| 3.50–4.00 | all classes | most readings | 95–100% | A | B | mod.paced&mostly accurate | difficult but reasonable | interesting | moderate amt | within 2–3 days | NO txtbk used | moderate | adequately | accessible for questions | encourages questions & answers them seriously | moderate | somewhat helpful | not at all helpful | Excellent |
| 3.50–4.00 | all classes | all readings | 95–100% | A | A | very quick&accurate | fairly simple | interesting | moderate amt | within 1 day of assignment | adequate | moderate | very clearly | accessible for questions | encourages questions & answers them seriously | fairly easy | very helpful | somewhat helpful | The best or very nearly the best |
| 3.00–3.49 | all classes | some readings | <70% | A | A | quick & reasonably accurate | fairly simple | of no interest | moderate amt | within week before exam | NO txtbk used | somewhat slow | adequately | accessible for questions | encourages questions & answers them seriously | fairly easy | somewhat helpful | not at all helpful | Fair |
| 3.00–3.49 | all classes | most readings | 95–100% | B | C | less quick&/or accurate than I'd like | fairly simple | of no interest | almost nothing | within week before exam | NO txtbk used | somewhat slow | very clearly | accessible for questions | encourages questions & answers them seriously | difficult but reasonable | very helpful | not at all helpful | Fair |
| 2.50–2.99 | all classes | all readings | 95–100% | B | B | very quick&accurate | difficult but reasonable | interesting | a great deal | within 1 day of assignment | good | moderate | adequately | accessible for questions | encourages questions & answers them seriously | difficult but reasonable | very helpful | | Good |
| 3.00–3.49 | most classes | all readings | 70–84% | B | A | quick & reasonably accurate | fairly simple | interesting | moderate amt | within week before exam | NO txtbk used | reasonably fast | poorly | accessible for questions | encourages questions & answers them seriously | moderate | very helpful | | Good |
| 2.50–2.99 | all classes | all readings | 70–84% | A | A | mod.paced&mostly accurate | difficult but reasonable | of some interest | moderate amt | within week before exam | adequate | moderate | adequately | accessible for questions | encourages questions & answers them seriously | moderate | very helpful | haven't tried any Mastering HW | |
| 3.00–3.49 | most classes | most readings | <70% | C | C | less quick&/or accurate than I'd like | difficult but reasonable | of some interest | moderate amt | within week before exam | NO txtbk used | reasonably fast | adequately | accessible for questions | encourages questions & answers them seriously | difficult but reasonable | somewhat helpful | haven't tried any Mastering HW | Fair |
| 3.00–3.49 | few classes | most readings | <70% | A | A | mod.paced&mostly accurate | difficult but reasonable | interesting | moderate amt | within 2–3 days | NO txtbk used | moderate | very clearly | accessible for questions | encourages questions & answers them seriously | difficult but reasonable | very helpful | haven't tried any Mastering HW | |
| 3.00–3.49 | all but 1–2 | most readings | <70% | A | A | quick & reasonably accurate | fairly simple | interesting | moderate amt | within 2–3 days | NO txtbk used | moderate | very clearly | accessible for questions | encourages questions & answers them seriously | difficult but reasonable | very helpful | haven't tried any Mastering HW | Excellent |
| 3.50–4.00 | all classes | most readings | 70–84% | A | B | mod.paced&mostly accurate | difficult but reasonable | interesting | a great deal | within week before exam | NO txtbk used | much too slow | very clearly | I did not seek extra help | encourages questions & answers them seriously | moderate | somewhat helpful | not at all helpful | Good |
| 3.50–4.00 | few classes | | E | E | A | quick & reasonably accurate | too difficult | very exciting | almost nothing | NO HW assigned | NO txtbk used | too rapid for proper understanding | E | D | encourages questions & answers them seriously | too difficult | I never consulted Blkbd supplemts | very helpful | The best or very nearly the best |
| 3.50–4.00 | few classes | | <70% | A | A | quick & reasonably accurate | very exciting | a great deal | good | somewhat slow | very clearly | accessible for questions | encourages questions & answers them seriously | moderate | very helpful | very helpful | The best or very nearly the best | | |
| 3.00–3.49 | all but 1–2 | all readings | 95–100% | A | A | very quick&accurate | fairly simple | very exciting | a great deal | within 2 weeks | good | moderate | very clearly | I did not seek extra help | encourages questions & answers them seriously | fairly easy | very helpful | very helpful | The best or very nearly the best |
| 3.50–4.00 | all but 3–4 | most readings | 85–94% | A | A | very quick&accurate | too simple | of no interest | almost nothing | within week before exam | adequate | somewhat slow | adequately | accessible for questions | encourages questions & answers them seriously | difficult but reasonable | somewhat helpful | somewhat helpful | The best or very nearly the best |
| 3.00–3.49 | most classes | most readings | NO HW assigned | A | A | mod.paced&mostly accurate | very exciting | a great deal | within week before exam | NO txtbk used | moderate | very clearly | I did not seek extra help | encourages questions & answers them seriously | difficult but reasonable | somewhat helpful | somewhat helpful | Excellent | |
| 3.50–4.00 | all but 1–2 | most readings | 95–100% | A | A | mod.paced&mostly accurate | difficult but reasonable | interesting | moderate amt | within week before exam | good | moderate | very clearly | I did not seek extra help | encourages questions & answers them seriously | moderate | very helpful | somewhat helpful | Excellent |
| 3.00–3.49 | most classes | most readings | 85–94% | A | A | very quick&accurate | fairly simple | interesting | moderate amt | within week before exam | adequate | moderate | very clearly | accessible for questions | encourages questions & answers them seriously | fairly easy | very helpful | very helpful | The best or very nearly the best |
| 3.00–3.49 | most classes | some readings | 70–84% | B | B | quick & reasonably accurate | fairly simple | of no interest | moderate amt | within 2 weeks | NO txtbk used | somewhat slow | very clearly | accessible for questions | encourages questions & answers them seriously | difficult but reasonable | very helpful | not at all helpful | Good |
| 3.50–4.00 | all classes | most readings | 70–84% | A | A | quick & reasonably accurate | fairly simple | interesting | moderate amt | within 2 weeks | adequate | moderate | very clearly | accessible for questions | encourages questions & answers them seriously | moderate | very helpful | very helpful | Excellent |
| 3.50–4.00 | all classes | most readings | 85–94% | A | A | mod.paced&mostly accurate | difficult but reasonable | of no interest | moderate amt | within week before exam | poor | reasonably fast | adequately | accessible for questions | encourages questions & answers them seriously | moderate | very helpful | not at all helpful | Excellent |
| 3.50–4.00 | all but 1–2 | most readings | 70–84% | A | B | very quick&accurate | fairly simple | interesting | moderate amt | within 2 weeks | excellent | reasonably fast | adequately | accessible for questions | encourages questions & answers them seriously | difficult but reasonable | somewhat helpful | not at all helpful | Good |
| 3.50–4.00 | most classes | some readings | <70% | A | A | mod.paced&mostly accurate | fairly simple | interesting | a great deal | within week before exam | NO txtbk used | reasonably fast | adequately | accessible for questions | encourages questions & answers them seriously | too difficult | very helpful | not at all helpful | Excellent |
| 3.00–3.49 | all classes | most readings | NO HW assigned | A | A | quick & reasonably accurate | difficult but reasonable | interesting | a great deal | within 1 day of assignment | adequate | moderate | very clearly | accessible for questions | encourages questions & answers them seriously | moderate | very helpful | haven't tried much Mastering HW | Excellent |
| 3.50–4.00 | all classes | some readings | NO HW assigned | A | A | quick & reasonably accurate | too difficult | of no interest | almost nothing | NO HW assigned | NO txtbk used | too rapid for proper understanding | not at all | could rarely be found | openly discourages questions in class | too difficult | not helpful | not at all helpful | Poor |
| 3.00–3.49 | all classes | | <70% | A | A | of no interest | moderate amt | within week before exam | adequate | reasonably fast | poorly | accessible for questions | encourages questions & answers them seriously | moderate | somewhat helpful | not at all helpful | Excellent | | |
| 3.50–4.00 | all but 1–2 | some readings | 70–84% | B | B | mod.paced&mostly accurate | difficult but reasonable | of no interest | moderate amt | within week before exam | adequate | reasonably fast | not at all | I did not seek extra help | does not encourage questions & is reluctant to answer them | difficult but reasonable | somewhat helpful | not at all helpful | Excellent |
| 3.00–3.49 | all classes | most readings | <70% | A | A | very quick&accurate | fairly simple | of some interest | moderate amt | within week before exam | poor | moderate | adequately | I did not seek extra help | encourages questions & answers them seriously | fairly easy | very helpful | haven't tried much Mastering HW | Good |
| 3.00–3.49 | all but 1–2 | | NO HW assigned | A | A | quick & reasonably accurate | fairly simple | interesting | moderate amt | within 2 weeks | NO txtbk used | somewhat slow | adequately | I did not seek extra help | encourages questions & answers them seriously | difficult but reasonable | somewhat helpful | haven't tried any Mastering HW | Good |
| 3.00–3.49 | most classes | some readings | <70% | A | A | mod.paced&mostly accurate | fairly simple | a great deal | within week before exam | NO txtbk used | moderate | adequately | I did not seek extra help | does not encourage questions but answers them seriously | difficult but reasonable | I never consulted Blkbd supplemts | haven't tried any Mastering HW | Good | |
| 3.50–4.00 | few classes | some readings | <70% | A | A | very quick&accurate | fairly simple | interesting | moderate amt | within week before exam | NO txtbk used | moderate | very clearly | accessible for questions | encourages questions & answers them seriously | moderate | very helpful | haven't tried much Mastering HW | Excellent |
| | all classes | all readings | 95–100% | A | A | quick & reasonably accurate | fairly simple | of some interest | moderate amt | within week before exam | poor | somewhat slow | very clearly | accessible for questions | encourages questions & answers them seriously | moderate | very helpful | very helpful | The best or very nearly the best |

| GPA | ATTENDANCE | RDGS COMPLETED | HW COMPLETED | GRD DESERVED | GRD EXPECTED | MATH ABILITY | MATERIAL COVERED | MATERIAL COVERED | I LEARNED | HW STARTED | TEXTBK | PACE OF LECTS | DIFF/SUBTLE PTS | PROFESSOR | PROFESSOR | EXAMS WERE | BLKBD SUPPLEMENTS | MAST.PHYSICS | COURSE OVERALL |
|---|---|---|---|---|---|---|---|---|---|---|---|---|---|---|---|---|---|---|---|
| 2.50–2.99 | most classes | all readings | 95–100% | A | A | quick & reasonably accurate | difficult but reasonable | interesting | moderate amt | within 2 weeks | good | reasonably fast | adequately | accessible for questions | does not encourage questions but answers them seriously | moderate | somewhat helpful | not at all helpful | The best or very nearly the best |
| 2.50–2.99 | few classes | some readings | 95–100% | B | A | very quick&accurate | fairly simple | interesting | moderate amt | within week before exam | adequate | moderate | adequately | accessible for questions | encourages questions & answers them seriously | fairly easy | very helpful | not at all helpful | Good |
| 3.50–4.00 | all classes | some readings | NO HW assigned | B | A | quick & reasonably accurate | difficult but reasonable | interesting | moderate amt | within week before exam | NO txtbk used | moderate | adequately | I did not seek extra help | encourages questions & answers them seriously | moderate | somewhat helpful | haven't tried any Mastering HW | Good |
| 3.50–4.00 | all classes | all readings | 95–100% | A | A | very quick&accurate | difficult but reasonable | interesting | a great deal | within 2–3 days | NO txtbk used | moderate | adequately | very clearly | accessible for questions | encourages questions & answers them seriously | moderate | very helpful | very helpful | Good |
| 3.50–4.00 | all but 3–4 | some readings | <70% | A | A | very quick&accurate | fairly simple | of some interest | a great deal | within 2–3 days | NO txtbk used | moderate | adequately | very clearly | accessible for questions | encourages questions & answers them seriously | difficult but reasonable | not helpful | somewhat helpful | Good |
| 3.00–3.49 | few classes | most readings | 95–100% | A | A | very quick&accurate | difficult but reasonable | interesting | moderate amt | within 1 day of assignment | adequate | moderate | adequately | very clearly | accessible for questions | encourages questions & answers them seriously | moderate | very helpful | very helpful | The best or very nearly the best |
| 3.00–3.49 | most classes | some readings | 85–94% | A | A | quick & reasonably accurate | difficult but reasonable | interesting | moderate amt | within 2 weeks | adequate | moderate | adequately | very clearly | accessible for questions | encourages questions & answers them seriously | difficult but reasonable | very helpful | very helpful | Good |
| 3.50–4.00 | all but 3–4 | some readings | NO HW assigned | A | A | very quick&accurate | difficult but reasonable | of no interest | moderate amt | NO HW assigned | NO txtbk used | somewhat slow | adequately | I did not seek extra help | encourages questions & answers them seriously | difficult but reasonable | not helpful | haven't tried any Mastering HW | Fair |
| 3.00–3.49 | all classes | most readings | <70% | A | A | very quick&accurate | difficult but reasonable | of some interest | a great deal | NO HW assigned | NO txtbk used | reasonably fast | adequately | I did not seek extra help | does not encourage questions but answers them seriously | too difficult | E | not at all helpful | |
| 3.50–4.00 | all classes | most readings | 85–94% | A | A | very quick&accurate | difficult but reasonable | interesting | a great deal | within 2–3 days | adequate | moderate | adequately | very clearly | accessible for questions | encourages questions & answers them seriously | moderate | very helpful | somewhat helpful | |
| 3.50–4.00 | all but 1–2 | most readings | 95–100% | A | A | very quick&accurate | difficult but reasonable | interesting | moderate amt | within 2–3 days | poor | moderate | adequately | very clearly | accessible for questions | encourages questions & answers them seriously | moderate | very helpful | somewhat helpful | Good |
| 2.00–2.49 | most classes | most readings | 95–100% | B | C | quick & reasonably accurate | difficult but reasonable | interesting | moderate amt | within week before exam | adequate | moderate | adequately | very clearly | accessible for questions | does not encourage questions & is reluctant to answer them | difficult but reasonable | very helpful | not at all helpful | |
| 3.00–3.49 | all classes | most readings | <70% | A | A | quick & reasonably accurate | fairly simple | interesting | moderate amt | within 2–3 days | NO txtbk used | moderate | adequately | very clearly | accessible for questions | encourages questions & answers them seriously | moderate | somewhat helpful | somewhat helpful | Excellent |
| 3.50–4.00 | all but 3–4 | most readings | 85–94% | A | A | quick & reasonably accurate | difficult but reasonable | interesting | a great deal | within week before exam | adequate | somewhat slow | adequately | very clearly | accessible for questions | encourages questions & answers them seriously | fairly easy | very helpful | not at all helpful | Excellent |
| 3.00–3.49 | all but 3–4 | most readings | 95–100% | B | B | quick & reasonably accurate | too simple | interesting | moderate amt | within week before exam | poor | moderate | adequately | very clearly | accessible for questions | encourages questions & answers them seriously | difficult but reasonable | very helpful | not at all helpful | Good |
| 3.00–3.49 | all classes | most readings | <70% | A | A | quick & reasonably accurate | fairly simple | interesting | moderate amt | within 2–3 days | good | moderate | adequately | very clearly | accessible for questions | encourages questions & answers them seriously | moderate | somewhat helpful | somewhat helpful | |
| 3.00–3.49 | all but 1–2 | all readings | | A | A | quick & reasonably accurate | fairly simple | of some interest | moderate amt | | NO txtbk used | moderate | adequately | I did not seek extra help | encourages questions & answers them seriously | difficult but reasonable | somewhat helpful | haven't tried any Mastering HW | Good |
| 3.50–4.00 | all classes | most readings | <70% | A | A | quick & reasonably accurate | fairly simple | of no interest | moderate amt | within 2 weeks | NO txtbk used | moderate | not at all | could rarely be found | encourages questions & answers them seriously | moderate | very helpful | not at all helpful | Good |
| 3.50–4.00 | all classes | most readings | 85–94% | A | A | very quick&accurate | difficult but reasonable | very exciting | moderate amt | within 2–3 days | NO txtbk used | moderate | adequately | very clearly | accessible for questions | encourages questions & answers them seriously | fairly easy | very helpful | somewhat helpful | The best or very nearly the best |
| 3.00–3.49 | all but 1–2 | some readings | 70–84% | A | A | very quick&accurate | difficult but reasonable | of some interest | moderate amt | within 2–3 days | adequate | moderate | poorly | accessible for questions | encourages questions & answers them seriously | difficult but reasonable | I never consulted Blkbd supplmts | not at all helpful | Fair |
| 3.00–3.49 | all but 1–2 | most readings | 70–84% | A | A | very quick&accurate | fairly simple | interesting | moderate amt | within 2 weeks | NO txtbk used | somewhat slow | adequately | very clearly | accessible for questions | encourages questions & answers them seriously | moderate | I never consulted Blkbd supplmts | not at all helpful | Fair |
| 2.50–2.99 | few classes | most readings | 70–84% | B | A | mod. paced&mostly accurate | difficult but reasonable | interesting | moderate amt | within week before exam | adequate | moderate | adequately | very clearly | accessible for questions | encourages questions & answers them seriously | moderate | somewhat helpful | somewhat helpful | Excellent |
| 3.00–3.49 | all but 1–2 | all readings | <70% | A | A | quick & reasonably accurate | fairly simple | interesting | a great deal | within week before exam | poor | moderate | adequately | very clearly | accessible for questions | encourages questions & answers them seriously | moderate | not helpful | somewhat helpful | The best or very nearly the best |
| 3.00–3.49 | most classes | most readings | <70% | A | A | quick & reasonably accurate | difficult but reasonable | of some interest | a great deal | within 2–3 days | NO txtbk used | reasonably fast | adequately | very clearly | accessible for questions | encourages questions & answers them seriously | moderate | somewhat helpful | somewhat helpful | Excellent |
| 3.00–3.49 | all but 1–2 | some readings | <70% | B | B | mod. paced&mostly accurate | difficult but reasonable | interesting | moderate amt | within 2 weeks | adequate | somewhat slow | adequately | very clearly | accessible for questions | encourages questions & answers them seriously | moderate | somewhat helpful | somewhat helpful | Excellent |
| 3.50–4.00 | all classes | all readings | 95–100% | A | A | very quick&accurate | fairly simple | interesting | moderate amt | within 2–3 days | adequate | moderate | adequately | very clearly | accessible for questions | encourages questions & answers them seriously | fairly easy | very helpful | very helpful | Excellent |
| 3.50–4.00 | most classes | most readings | <70% | B | B | mod. paced&mostly accurate | difficult but reasonable | interesting | a great deal | within week before exam | adequate | moderate | adequately | very clearly | accessible for questions | encourages questions & answers them seriously | moderate | very helpful | not at all helpful | Excellent |
| 3.00–3.49 | all but 3–4 | all readings | <70% | B | A | quick & reasonably accurate | difficult but reasonable | interesting | a great deal | within week before exam | adequate | reasonably fast | adequately | very clearly | accessible for questions | encourages questions & answers them seriously | difficult but reasonable | very helpful | haven't tried much Mastering HW | Excellent |
| 3.00–3.49 | all but 1–2 | most readings | 95–100% | A | A | quick & reasonably accurate | difficult but reasonable | very exciting | a great deal | within 2–3 days | good | reasonably fast | adequately | very clearly | accessible for questions | encourages questions & answers them seriously | difficult but reasonable | very helpful | very helpful | The best or very nearly the best |
| 3.50–4.00 | most classes | most readings | 95–100% | A | A | very quick&accurate | fairly simple | very exciting | a great deal | within 2–3 days | excellent | too rapid for proper understanding | very clearly | accessible for questions | does not encourage questions but answers them seriously | moderate | I never consulted Blkbd supplmts | very helpful | |
| 3.50–4.00 | all classes | all readings | 95–100% | A | A | very quick&accurate | fairly simple | of some interest | a great deal | within 2–3 days | poor | moderate | adequately | very clearly | accessible for questions | encourages questions & answers them seriously | moderate | I never consulted Blkbd supplmts | somewhat helpful | Excellent |
| 3.00–3.49 | most classes | most readings | 85–94% | A | A | very quick&accurate | fairly simple | of some interest | moderate amt | within 2 weeks | adequate | reasonably fast | adequately | very clearly | could rarely be found | encourages questions & answers them seriously | moderate | somewhat helpful | somewhat helpful | Good |
| 3.00–3.49 | all classes | most readings | 70–84% | A | A | very quick&accurate | difficult but reasonable | of no interest | moderate amt | within 2–3 days | NO txtbk used | moderate | adequately | very clearly | accessible for questions | encourages questions & answers them seriously | moderate | very helpful | not at all helpful | Good |
| 3.50–4.00 | all but 3–4 | most readings | <70% | B | B | mod. paced&mostly accurate | difficult but reasonable | of some interest | moderate amt | | | moderate | adequately | very clearly | accessible for questions | encourages questions & answers them seriously | moderate | somewhat helpful | somewhat helpful | |
| 2.00–2.49 | few classes | some readings | NO HW assigned | C | B | mod. paced&mostly accurate | difficult but reasonable | interesting | moderate amt | NO HW assigned | NO txtbk used | moderate | adequately | very clearly | accessible for questions | encourages questions & answers them seriously | difficult but reasonable | very helpful | haven't tried much Mastering HW | The best or very nearly the best |
| 3.50–4.00 | all classes | some readings | <70% | A | A | quick & reasonably accurate | difficult but reasonable | of some interest | a great deal | within week before exam | NO txtbk used | moderate | adequately | very clearly | accessible for questions | encourages questions & answers them seriously | difficult but reasonable | very helpful | not at all helpful | Excellent |
| 3.50–4.00 | all classes | most readings | 85–94% | A | A | very quick&accurate | difficult but reasonable | very exciting | a great deal | within week before exam | adequate | moderate | adequately | very clearly | accessible for questions | encourages questions & answers them seriously | moderate | very helpful | somewhat helpful | Excellent |
| 3.50–4.00 | all but 1–2 | all readings | 95–100% | A | A | very quick&accurate | fairly simple | of no interest | moderate amt | within 1 day of assignment | adequate | somewhat slow | adequately | very clearly | accessible for questions | encourages questions & answers them seriously | moderate | very helpful | not at all helpful | Good |
| 2.00–2.49 | all but 1–2 | some readings | <70% | C | D | less quick&/or accurate than I'd like | fairly simple | interesting | moderate amt | within week before exam | adequate | moderate | adequately | very clearly | encourages questions & answers them seriously | | difficult but reasonable | very helpful | somewhat helpful | The best or very nearly the best |
| 3.00–3.49 | most classes | most readings | 95–100% | A | A | quick & reasonably accurate | fairly simple | of some interest | moderate amt | within 2–3 days | NO txtbk used | moderate | adequately | very clearly | accessible for questions | encourages questions & answers them seriously | moderate | somewhat helpful | somewhat helpful | Fair |
| 2.50–2.99 | all but 1–2 | all readings | <70% | A | A | very quick&accurate | fairly simple | interesting | a great deal | within 2–3 days | NO txtbk used | moderate | adequately | very clearly | accessible for questions | encourages questions & answers them seriously | moderate | very helpful | not at all helpful | |
| 3.50–4.00 | few classes | some readings | <70% | A | A | mod. paced&mostly accurate | fairly simple | interesting | moderate amt | within week before exam | NO txtbk used | much too slow | poorly | accessible for questions | encourages questions & answers them seriously | | moderate | somewhat helpful | haven't tried any Mastering HW | Excellent |
| 2.50–2.99 | all but 1–2 | some readings | NO HW assigned | A | A | very quick&accurate | too simple | of some interest | moderate amt | within week before exam | NO txtbk used | much too slow | very clearly | I did not seek extra help | encourages questions & answers them seriously | | fairly easy | I never consulted Blkbd supplmts | haven't tried any Mastering HW | |
| 3.50–4.00 | all classes | all readings | <70% | A | A | quick & reasonably accurate | fairly simple | interesting | moderate amt | within week before exam | NO txtbk used | too rapid for proper understanding | poorly | could rarely be found | encourages questions & answers them seriously | | fairly easy | somewhat helpful | haven't tried any Mastering HW | Good |
| 3.00–3.49 | all but 1–2 | most readings | 95–100% | A | A | quick & reasonably accurate | fairly simple | interesting | moderate amt | within 2 weeks | adequate | moderate | adequately | very clearly | accessible for questions | encourages questions & answers them seriously | moderate | very helpful | not at all helpful | The best or very nearly the best |
| 2.50–2.99 | most classes | most readings | <70% | A | A | mod. paced&mostly accurate | fairly simple | interesting | moderate amt | within week before exam | NO txtbk used | moderate | adequately | very clearly | accessible for questions | encourages questions & answers them seriously | fairly easy | somewhat helpful | haven't tried any Mastering HW | The best or very nearly the best |
| 3.00–3.49 | all classes | all readings | 95–100% | A | A | quick & reasonably accurate | difficult but reasonable | interesting | a great deal | within 2–3 days | excellent | moderate | adequately | very clearly | accessible for questions | encourages questions & answers them seriously | difficult but reasonable | very helpful | very helpful | Excellent |
| 3.00–3.49 | few classes | most readings | 95–100% | A | A | mod. paced&mostly accurate | difficult but reasonable | interesting | moderate amt | within week before exam | adequate | moderate | adequately | very clearly | accessible for questions | encourages questions & answers them seriously | moderate | very helpful | haven't tried any Mastering HW | The best or very nearly the best |
| 3.50–4.00 | all but 1–2 | most readings | <70% | A | A | very quick&accurate | difficult but reasonable | of some interest | moderate amt | within week before exam | adequate | reasonably fast | adequately | very clearly | accessible for questions | encourages questions & answers them seriously | moderate | very helpful | haven't tried any Mastering HW | Excellent |
| 3.00–3.49 | all but 1–2 | all readings | 95–100% | A | A | mod. paced&mostly accurate | difficult but reasonable | interesting | moderate amt | within 2–3 days | adequate | moderate | adequately | very clearly | accessible for questions | encourages questions & answers them seriously | difficult but reasonable | somewhat helpful | not at all helpful | Excellent |
| 3.50–4.00 | all classes | all readings | 95–100% | A | A | very quick&accurate | difficult but reasonable | interesting | moderate amt | within 1 day of assignment | adequate | moderate | adequately | very clearly | accessible for questions | encourages questions & answers them seriously | moderate | somewhat helpful | somewhat helpful | Excellent |
| 3.00–3.49 | all but 3–4 | all readings | 95–100% | A | A | quick & reasonably accurate | fairly simple | interesting | moderate amt | within 2 weeks | adequate | moderate | adequately | very clearly | accessible for questions | encourages questions & answers them seriously | fairly easy | very helpful | very helpful | Excellent |
| 3.00–3.49 | most classes | most readings | <70% | A | A | mod. paced&mostly accurate | difficult but reasonable | interesting | a great deal | within week before exam | NO txtbk used | moderate | adequately | very clearly | accessible for questions | encourages questions & answers them seriously | moderate | somewhat helpful | not at all helpful | Excellent |
| 3.50–4.00 | all classes | some readings | 95–100% | C | A | mod. paced&mostly accurate | difficult but reasonable | of some interest | moderate amt | within week before exam | NO txtbk used | reasonably fast | adequately | very clearly | accessible for questions | encourages questions & answers them seriously | difficult but reasonable | very helpful | not at all helpful | Good |
| 3.50–4.00 | all but 1–2 | most readings | 95–100% | A | A | mod. paced&mostly accurate | difficult but reasonable | of some interest | almost nothing | within 2–3 days | NO txtbk used | much too slow | poorly | E | encourages questions & answers them seriously | | moderate | somewhat helpful | not at all helpful | Fair |
| 3.50–4.00 | all classes | all readings | 95–100% | A | B | quick & reasonably accurate | fairly simple | interesting | moderate amt | within week before exam | NO txtbk used | moderate | adequately | very clearly | accessible for questions | encourages questions & answers them seriously | moderate | somewhat helpful | not at all helpful | Excellent |
| 3.00–3.49 | all but 1–2 | all readings | NO HW assigned | A | A | very quick&accurate | fairly simple | of no interest | almost nothing | within week before exam | adequate | much too slow | adequately | I did not seek extra help | encourages questions & answers them seriously | difficult but reasonable | somewhat helpful | not at all helpful | Good |
| 3.50–4.00 | all but 1–2 | none of rdgs | 95–100% | A | A | quick & reasonably accurate | fairly simple | of no interest | almost nothing | within week before exam | poor | reasonably fast | poorly | accessible for questions | openly discourages questions in class | too difficult | not helpful | not at all helpful | FairPoor |
| 3.50–4.00 | all but 1–2 | most readings | <70% | A | A | very quick&accurate | fairly simple | interesting | moderate amt | within week before exam | NO txtbk used | moderate | adequately | very clearly | accessible for questions | encourages questions & answers them seriously | moderate | I never consulted Blkbd supplmts | haven't tried any Mastering HW | Excellent |
| 3.50–4.00 | most classes | all readings | 95–100% | A | A | quick & reasonably accurate | difficult but reasonable | interesting | a great deal | within 2 weeks | NO txtbk used | moderate | adequately | very clearly | accessible for questions | encourages questions & answers them seriously | difficult but reasonable | very helpful | very helpful | Excellent |
| 3.50–4.00 | all but 1–2 | most readings | 85–94% | A | A | quick & reasonably accurate | difficult but reasonable | of some interest | moderate amt | within 2 weeks | poor | moderate | adequately | very clearly | accessible for questions | encourages questions & answers them seriously | moderate | very helpful | not at all helpful | Excellent |
| 3.50–4.00 | all but 3–4 | some readings | 95–100% | B | A | quick & reasonably accurate | difficult but reasonable | interesting | a great deal | within 2–3 days | good | reasonably fast | adequately | very clearly | accessible for questions | encourages questions & answers them seriously | difficult but reasonable | very helpful | | Excellent |

| GPA | ATTENDANCE | RDGS COMPLETED | HW COMPLETED | GRD DESERVED | GRD EXPECTED | MATH ABILITY | MATERIAL COVERED | MATERIAL COVERED | I LEARNED | HW STARTED | TEXTBK | PACE OF LECTS | DIFF/SUBTLE PTS | PROFESSOR | PROFESSOR | EXAMS WERE | BLKBD SUPPLEMENTS | MAST.PHYSICS | COURSE OVERALL |
|---|---|---|---|---|---|---|---|---|---|---|---|---|---|---|---|---|---|---|---|
| 2.50–2.99 | all but 1–2 | most readings | <70% | A | A | quick & reasonably accurate | fairly simple | | moderate amt | within week before exam | adequate | moderate | very clearly | accessible for questions | encourages questions & answers them seriously | fairly easy | very helpful | not at all helpful | Good |
| 3.50–4.00 | most classes | some readings | 70–84% | A | A | mod. paced & mostly accurate | | interesting | moderate amt | within week before exam | good | moderate | very clearly | accessible for questions | encourages questions & answers them seriously | fairly easy | somewhat helpful | not at all helpful | Good |
| 3.00–3.49 | all but 3–4 | all readings | 95–100% | A | A | quick & reasonably accurate | fairly simple | of some interest | a great deal | within week before exam | good | moderate | very clearly | accessible for questions | encourages questions & answers them seriously | moderate | very helpful | somewhat helpful | The best or very nearly the best |
| | most classes | | | A | A | | | | moderate amt | | | | very clearly | | encourages questions & answers them seriously | | somewhat helpful | | |
| 3.50–4.00 | all classes | all readings | NO HW assigned | A | A | very quick & accurate | fairly simple | interesting | a great deal | NO HW assigned | NO txtbk used | somewhat slow | very clearly | I did not seek extra help | encourages questions & answers them seriously | fairly easy | very helpful | haven't tried any Mastering HW | Excellent |
| 3.50–4.00 | all but 3–4 | some readings | <70% | B | A | very quick & accurate | difficult but reasonable | of no interest | moderate amt | within week before exam | poor | moderate | adequately | I did not seek extra help | does not encourage questions but answers them seriously | moderate | somewhat helpful | | Good |
| 3.50–4.00 | most classes | some readings | <70% | A | A | very quick & accurate | fairly simple | almost nothing | a great deal | within week before exam | adequate | somewhat slow | poorly | accessible for questions | encourages questions & answers them seriously | fairly easy | very helpful | | |
| 3.50–4.00 | most classes | some readings | 95–100% | A | A | very quick & accurate | difficult but reasonable | of some interest | moderate amt | within 2 weeks | poor | moderate | adequately | accessible for questions | encourages questions & answers them seriously | moderate | somewhat helpful | not at all helpful | Good |
| 2.50–2.99 | all classes | all readings | 95–100% | A | A | mod. paced & mostly accurate | difficult but reasonable | interesting | moderate amt | within week before exam | good | moderate | adequately | accessible for questions | does not encourage questions & is reluctant to answer them | difficult but reasonable | somewhat helpful | | Excellent |
| 3.50–4.00 | all classes | most readings | 70–84% | A | A | quick & reasonably accurate | fairly simple | interesting | a great deal | within 1 day of assignment | NO txtbk used | reasonably fast | adequately | I did not seek extra help | encourages questions & answers them seriously | fairly easy | 70–84% | not at all helpful | Excellent |
| 3.50–4.00 | all but 3–4 | most readings | 70–84% | A | A | quick & reasonably accurate | fairly simple | of no interest | almost nothing | within 1 day of assignment | poor | somewhat slow | adequately | accessible for questions | encourages questions & answers them seriously | moderate | not at all helpful | | Fair |
| | all classes | all readings | 95–100% | A | A | very quick & accurate | difficult but reasonable | of some interest | a great deal | within 2–3 days | adequate | moderate | adequately | could rarely be found | encourages questions & answers them seriously | moderate | somewhat helpful | not at all helpful | Good |
| 3.00–3.49 | all classes | most readings | 70–84% | A | A | mod. paced & mostly accurate | fairly simple | interesting | a great deal | within week before exam | NO txtbk used | reasonably fast | very clearly | accessible for questions | encourages questions & answers them seriously | difficult but reasonable | very helpful | not at all helpful | Excellent |
| 2.50–2.99 | all but 3–4 | most readings | 95–100% | A | B | quick & reasonably accurate | fairly simple | of some interest | moderate amt | | NO txtbk used | reasonably fast | adequately | accessible for questions | encourages questions & answers them seriously | moderate | somewhat helpful | | Excellent |
| 3.50–4.00 | all but 3–4 | some readings | NO HW assigned | A | A | very quick & accurate | fairly simple | very exciting | a great deal | NO HW assigned | NO txtbk used | moderate | very clearly | E | encourages questions & answers them seriously | moderate | very helpful | not at all helpful | The best or very nearly the best |
| 3.00–3.49 | most classes | most readings | 95–100% | A | A | mod. paced & mostly accurate | difficult but reasonable | interesting | moderate amt | within 2 weeks | NO txtbk used | moderate | adequately | I did not seek extra help | encourages questions & answers them seriously | difficult but reasonable | very helpful | | Excellent |
| 3.00–3.49 | all but 1–2 | most readings | <70% | A | B | mod. paced & mostly accurate | fairly simple | of some interest | moderate amt | within 2–3 days | NO txtbk used | moderate | adequately | accessible for questions | encourages questions & answers them seriously | difficult but reasonable | | | |
| 2.00–2.49 | all classes | all readings | 95–100% | A | A | very quick & accurate | difficult but reasonable | interesting | moderate amt | within 2–3 days | good | reasonably fast | very clearly | accessible for questions | encourages questions & answers them seriously | difficult but reasonable | very helpful | not at all helpful | Excellent |
| 2.50–2.99 | all but 3–4 | most readings | <70% | B | A | mod. paced & mostly accurate | fairly simple | of some interest | moderate amt | within 1 day of assignment | NO txtbk used | somewhat slow | adequately | accessible for questions | encourages questions & answers them seriously | difficult but reasonable | somewhat helpful | not at all helpful | Fair |
| 3.00–3.49 | most classes | all readings | 95–100% | A | A | mod. paced & mostly accurate | difficult but reasonable | interesting | moderate amt | within 2 weeks | adequate | reasonably fast | adequately | accessible for questions | encourages questions & answers them seriously | difficult but reasonable | somewhat helpful | | |
| 3.00–3.49 | all classes | none of rdgs | <70% | A | A | quick & reasonably accurate | difficult but reasonable | interesting | moderate amt | within week before exam | adequate | moderate | adequately | accessible for questions | encourages questions & answers them seriously | moderate | very helpful | haven't tried any Mastering HW | Good |
| 3.50–4.00 | all classes | none of rdgs | <70% | A | A | very quick & accurate | | interesting | a great deal | within 2 weeks | poor | moderate | very clearly | accessible for questions | encourages questions & answers them seriously | difficult but reasonable | somewhat helpful | | Excellent |
| 3.00–3.49 | all but 3–4 | most readings | <70% | A | A | mod. paced & mostly accurate | difficult but reasonable | of some interest | moderate amt | | adequate | reasonably fast | too difficult | accessible for questions | encourages questions & answers them seriously | moderate | very helpful | haven't tried any Mastering HW | |
| 2.50–2.99 | all classes | most readings | <70% | A | A | quick & reasonably accurate | difficult but reasonable | interesting | a great deal | within week before exam | NO txtbk used | moderate | adequately | accessible for questions | encourages questions & answers them seriously | moderate | haven't tried any Mastering HW | | The best or very nearly the best |
| 2.00–2.49 | all but 3–4 | all readings | <70% | B | B | very quick & accurate | | | moderate amt | within 2 weeks | NO txtbk used | moderate | adequately | accessible for questions | encourages questions & answers them seriously | moderate | very helpful | not at all helpful | Good |
| 3.50–4.00 | all classes | all readings | 95–100% | A | A | quick & reasonably accurate | difficult but reasonable | interesting | a great deal | within 2–3 days | NO txtbk used | moderate | adequately | accessible for questions | encourages questions & answers them seriously | fairly easy | very helpful | somewhat helpful | The best or very nearly the best |
| 2.50–2.99 | all but 1–2 | some readings | 95–100% | B | C | less quick &/or accurate than I'd like | difficult but reasonable | of some interest | almost nothing | | adequate | moderate | adequately | accessible for questions | encourages questions & answers them seriously | moderate | somewhat helpful | not at all helpful | Fair |
| 2.00–2.49 | all classes | all readings | <70% | A | A | mod. paced & mostly accurate | | interesting | a great deal | within 1 day of assignment | NO txtbk used | reasonably fast | adequately | accessible for questions | encourages questions & answers them seriously | difficult but reasonable | very helpful | not at all helpful | |
| 3.50–4.00 | all but 1–2 | all readings | 85–94% | A | A | very quick & accurate | difficult but reasonable | interesting | a great deal | within 2–3 days | NO txtbk used | reasonably fast | adequately | accessible for questions | encourages questions & answers them seriously | difficult but reasonable | very helpful | not at all helpful | Excellent |
| 3.00–3.49 | all but 1–2 | most readings | 95–100% | B | B | mod. paced & mostly accurate | | interesting | a great deal | within 2 weeks | poor | moderate | very clearly | accessible for questions | encourages questions & answers them seriously | fairly easy | somewhat helpful | | Excellent |
| 3.00–3.49 | all classes | all readings | 95–100% | A | A | mod. paced & mostly accurate | fairly simple | interesting | moderate amt | within 1 day of assignment | NO txtbk used | moderate | adequately | I did not seek extra help | encourages questions & answers them seriously | moderate | | | Good |
| 3.50–4.00 | all but 1–2 | most readings | 85–94% | A | A | | difficult but reasonable | interesting | a great deal | within 2 weeks | poor | moderate | adequately | accessible for questions | encourages questions & answers them seriously | moderate | somewhat helpful | | |
| 2.50–2.99 | most classes | all readings | 85–94% | A | A | very quick & accurate | difficult but reasonable | interesting | a great deal | within 2–3 days | good | reasonably fast | very clearly | accessible for questions | encourages questions & answers them seriously | difficult but reasonable | very helpful | very helpful | The best or very nearly the best |
| 3.50–4.00 | all classes | most readings | 85–94% | B | C | quick & reasonably accurate | difficult but reasonable | interesting | moderate amt | within 2 weeks | NO txtbk used | moderate | adequately | accessible for questions | encourages questions & answers them seriously | difficult but reasonable | very helpful | somewhat helpful | The best or very nearly the best |
| 3.50–4.00 | all but 1–2 | all readings | <70% | A | A | mod. paced & mostly accurate | difficult but reasonable | of no interest | almost nothing | within week before exam | NO txtbk used | moderate | adequately | I did not seek extra help | encourages questions & answers them seriously | too difficult | very helpful | haven't tried much Mastering HW | Fair |
| 3.50–4.00 | all classes | NO HW assigned | | A | A | quick & reasonably accurate | difficult but reasonable | moderate amt | | NO HW assigned | NO txtbk used | moderate | adequately | accessible for questions | encourages questions & answers them seriously | difficult but reasonable | very helpful | haven't tried any Mastering HW | Excellent |
| 3.00–3.49 | all but 3–4 | most readings | 95–100% | A | B | quick & reasonably accurate | fairly simple | of no interest | moderate amt | | poor | moderate | poorly | I did not seek extra help | encourages questions & answers them seriously | moderate | not at all helpful | | Fair |
| 3.00–3.49 | few classes | all readings | <70% | A | A | very quick & accurate | fairly simple | | moderate amt | within 2–3 days | adequate | reasonably fast | adequately | accessible for questions | encourages questions & answers them seriously | fairly easy | very helpful | not at all helpful | Excellent |
| 3.00–3.49 | all classes | all readings | 95–100% | A | A | quick & reasonably accurate | fairly simple | interesting | moderate amt | within 2 weeks | NO txtbk used | somewhat slow | adequately | accessible for questions | encourages questions & answers them seriously | moderate | very helpful | not at all helpful | The best or very nearly the best |
| 3.00–3.49 | all classes | most readings | 85–94% | C | C | very quick & accurate | difficult but reasonable | interesting | moderate amt | within 2–3 days | excellent | reasonably fast | very clearly | accessible for questions | encourages questions & answers them seriously | moderate | very helpful | very helpful | Good |
| 3.00–3.49 | most classes | | 85–94% | A | B | very quick & accurate | fairly simple | of some interest | moderate amt | within week before exam | | moderate | | | | | | | |
| 3.50–4.00 | all but 1–2 | most readings | <70% | A | A | very quick & accurate | | interesting | a great deal | within 1 day of assignment | adequate | moderate | adequately | accessible for questions | encourages questions & answers them seriously | difficult but reasonable | very helpful | | |
| 3.50–4.00 | all classes | most readings | 95–100% | A | B | quick & reasonably accurate | difficult but reasonable | moderate amt | | within week before exam | NO txtbk used | reasonably fast | poorly | I did not seek extra help | encourages questions & answers them seriously | difficult but reasonable | somewhat helpful | not at all helpful | Excellent |
| 3.50–3.49 | all classes | all readings | 95–100% | A | A | mod. paced & mostly accurate | too difficult | of some interest | moderate amt | within 2–3 days | poor | moderate | very clearly | accessible for questions | encourages questions & answers them seriously | too difficult | somewhat helpful | somewhat helpful | Good |
| 3.00–3.49 | most classes | most readings | <70% | A | B | very quick & accurate | fairly simple | very exciting | a great deal | within week before exam | good | moderate | adequately | accessible for questions | encourages questions & answers them seriously | moderate | very helpful | very helpful | Excellent |
| 3.50–4.00 | all but 1–2 | some readings | <70% | A | A | very quick & accurate | difficult but reasonable | interesting | a great deal | NO HW assigned | NO txtbk used | reasonably fast | adequately | I did not seek extra help | encourages questions & answers them seriously | difficult but reasonable | very helpful | haven't tried any Mastering HW | Excellent |
| 3.00–3.49 | most classes | <70% | | B | B | mod. paced & mostly accurate | difficult but reasonable | of some interest | moderate amt | within week before exam | adequate | too rapid for proper understanding | adequately | accessible for questions | encourages questions & answers them seriously | difficult but reasonable | somewhat helpful | haven't tried any Mastering HW | Good |
| 3.00–3.49 | most classes | most readings | 95–100% | B | B | mod. paced & mostly accurate | difficult but reasonable | of some interest | moderate amt | | NO txtbk used | reasonably fast | adequately | accessible for questions | does not encourage questions & is reluctant to answer them | difficult but reasonable | not helpful | not at all helpful | |
| 3.00–3.49 | all classes | all readings | 95–100% | A | A | very quick & accurate | difficult but reasonable | interesting | moderate amt | within 1 day of assignment | adequate | moderate | adequately | accessible for questions | encourages questions & answers them seriously | moderate | very helpful | not at all helpful | Excellent |
| 3.00–3.49 | all but 3–4 | most readings | 85–94% | A | A | very quick & accurate | fairly simple | very exciting | a great deal | within 2 weeks | NO txtbk used | moderate | very clearly | accessible for questions | encourages questions & answers them seriously | difficult but reasonable | very helpful | not at all helpful | Excellent |
| 2.50–2.99 | all classes | most readings | 95–100% | A | A | mod. paced & mostly accurate | difficult but reasonable | of some interest | moderate amt | within 2 weeks | poor | moderate | adequately | accessible for questions | encourages questions & answers them seriously | difficult but reasonable | very helpful | | Good |
| 3.00–3.49 | all but 1–2 | most readings | 95–100% | A | A | mod. paced & mostly accurate | difficult but reasonable | of some interest | moderate amt | within 2–3 days | adequate | moderate | adequately | accessible for questions | encourages questions & answers them seriously | moderate | somewhat helpful | very helpful | Good |
| 3.50–4.00 | most classes | most readings | 95–100% | A | A | very quick & accurate | fairly simple | of some interest | a great deal | within 1 day of assignment | excellent | too rapid for proper understanding | very clearly | accessible for questions | encourages questions & answers them seriously | too difficult | very helpful | very helpful | The best or very nearly the best |
| 3.50–4.00 | all but 3–4 | none of rdgs | <70% | A | A | very quick & accurate | too simple | of no interest | almost nothing | NO HW assigned | NO txtbk used | much too slow | very clearly | I did not seek extra help | encourages questions & answers them seriously | fairly easy | very helpful | haven't tried much Mastering HW | Fair |
| 3.50–4.00 | all but 1–2 | most readings | 95–100% | A | A | quick & reasonably accurate | difficult but reasonable | of some interest | moderate amt | | adequate | moderate | adequately | I did not seek extra help | encourages questions & answers them seriously | moderate | very helpful | | |
| 3.00–3.49 | all but 1–2 | most readings | <70% | B | B | quick & reasonably accurate | difficult but reasonable | of some interest | moderate amt | NO HW assigned | NO txtbk used | moderate | very clearly | I did not seek extra help | encourages questions & answers them seriously | fairly easy | very helpful | haven't tried any Mastering HW | The best or very nearly the best |
| 3.00–3.49 | all but 3–4 | most readings | <70% | B | B | quick & reasonably accurate | difficult but reasonable | of some interest | moderate amt | within week before exam | NO txtbk used | moderate | adequately | accessible for questions | encourages questions & answers them seriously | moderate | very helpful | haven't tried much Mastering HW | |
| 3.00–3.49 | all but 1–2 | most readings | 70–84% | B | B | quick & reasonably accurate | difficult but reasonable | interesting | a great deal | within 2 weeks | adequate | moderate | adequately | accessible for questions | encourages questions & answers them seriously | difficult but reasonable | somewhat helpful | haven't tried any Mastering HW | Excellent |
| 2.50–2.99 | most classes | most readings | <70% | B | B | quick & reasonably accurate | difficult but reasonable | of some interest | a great deal | within week before exam | NO txtbk used | moderate | adequately | accessible for questions | encourages questions & answers them seriously | moderate | very helpful | not at all helpful | Good |
| 3.50–4.00 | all but 1–2 | most readings | 95–100% | A | A | quick & reasonably accurate | difficult but reasonable | interesting | a great deal | within week before exam | good | moderate | adequately | I did not seek extra help | encourages questions & answers them seriously | moderate | somewhat helpful | | The best or very nearly the best |
| 2.50–2.99 | all but 3–4 | most readings | 70–84% | B | C | mod. paced & mostly accurate | difficult but reasonable | of no interest | a great deal | within 1 day of assignment | adequate | moderate | adequately | accessible for questions | encourages questions & answers them seriously | moderate | very helpful | | Good |
| 3.50–4.00 | all classes | most readings | <70% | A | A | mod. paced & mostly accurate | difficult but reasonable | of some interest | a great deal | within week before exam | NO txtbk used | moderate | adequately | accessible for questions | encourages questions & answers them seriously | moderate | very helpful | not at all helpful | Good |
| 3.00–3.49 | few classes | some readings | NO HW assigned | B | A | quick & reasonably accurate | too difficult | of some interest | moderate amt | NO HW assigned | | poor | moderate | adequately | encourages questions & answers them seriously | difficult but reasonable | very helpful | haven't tried any Mastering HW | Fair |

| GPA | ATTENDANCE | RDGS COMPLETED | HW COMPLETED | GRD DESERVED | GRD EXPECTED | MATH ABILITY | MATERIAL COVERED | MATERIAL COVERED | I LEARNED | HW STARTED | TEXTBK | PACE OF LECTS | DIFF/SUBTLE PTS | PROFESSOR | PROFESSOR | EXAMS WERE | BLKBD SUPPLEMENTS | MAST.PHYSICS | COURSE OVERALL |
|---|---|---|---|---|---|---|---|---|---|---|---|---|---|---|---|---|---|---|---|
| 3.00–3.49 | all classes | most readings | <70% | B | B | quick & reasonably accurate | difficult but reasonable | of no interest | moderate amt | within 2 weeks | adequate | reasonably fast | adequately | accessible for questions | encourages questions & answers them seriously | somewhat helpful | difficult but reasonable | not at all helpful | Fair |
| 3.00–3.49 | all but 1–2 | most readings | 95–100% | A | A | quick & reasonably accurate | fairly simple | interesting | moderate amt | within 2 weeks | NO txtbk used | moderate | very clearly | accessible for questions | encourages questions & answers them seriously | difficult but reasonable | very helpful | not at all helpful | Good |
| 3.50–4.00 | all but 1–2 | most readings | <70% | A | A | mod. paced&mostly accurate | fairly simple | of some interest | moderate amt | within 2 weeks | adequate | moderate | very clearly | accessible for questions | encourages questions & answers them seriously | fairly easy | very helpful | not at all helpful | Good |
| 2.50–2.99 | all but 1–2 | some readings | 95–100% | C | C | less quick&/or accurate than I'd like | difficult but reasonable | interesting | a great deal | within 2 weeks | adequate | moderate | very clearly | accessible for questions | encourages questions & answers them seriously | moderate | very helpful | somewhat helpful | The best or very nearly the best |
| 2.50–2.99 | all classes | most readings | 95–100% | A | A | less quick&/or accurate than I'd like | difficult but reasonable | interesting | a moderate deal | within 1 day of assignment | NO txtbk used | moderate | adequately | accessible for questions | encourages questions & answers them seriously | moderate | very helpful | somewhat helpful | Excellent |
| 3.00–3.49 | all classes | most readings | 85–94% | B | D | quick & reasonably accurate | difficult but reasonable | interesting | a great deal | within 2–3 days | NO txtbk used | moderate | adequately | accessible for questions | encourages questions & answers them seriously | difficult but reasonable | moderate | somewhat helpful | Excellent |
| 2.50–2.99 | all classes | most readings | 95–100% | A | A | mod. paced&mostly accurate | difficult but reasonable | of some interest | a great deal | within 2–3 days | NO txtbk used | moderate | adequately | accessible for questions | encourages questions & answers them seriously | difficult but reasonable | very helpful | somewhat helpful | Excellent |
| 3.00–3.49 | few classes | some readings | <70% | B | B | less quick&/or accurate than I'd like | difficult but reasonable | of no interest | moderate amt | within 2 weeks | good | moderate | very clearly | accessible for questions | encourages questions & answers them seriously | difficult but reasonable | somewhat helpful | not at all helpful | Fair |
| 2.00–2.49 | all but 1–2 | most readings | 70–84% | B | B | quick & reasonably accurate | difficult but reasonable | of no interest | a great deal | within week before exam | adequate | reasonably fast | adequately | accessible for questions | encourages questions & answers them seriously | moderate | very helpful | haven't tried any Mastering HW | The best or very nearly the best |
| 3.50–4.00 | most classes | some readings | <70% | A | A | quick & reasonably accurate | difficult but reasonable | of some interest | moderate amt | within week before exam | NO txtbk used | somewhat slow | adequately | accessible for questions | encourages questions & answers them seriously | difficult but reasonable | very helpful | haven't tried any Mastering HW | Good |
| 3.00–3.49 | all but 1–2 | most readings | 95–100% | A | A | quick & reasonably accurate | difficult but reasonable | interesting | moderate amt | within 2 weeks | adequate | moderate | adequately | accessible for questions | encourages questions & answers them seriously | difficult but reasonable | E | haven't tried any Mastering HW | Good |
| 3.00–3.49 | all but 1–2 | most readings | <70% | C | C | mod. paced&mostly accurate | fairly simple | of no interest | moderate amt | within 2 weeks | good | reasonably fast | adequately | accessible for questions | encourages questions & answers them seriously | fairly easy | very helpful | haven't tried any Mastering HW | Good |
| 3.00–3.49 | all but 1–2 | all readings | NO HW assigned | A | A | quick & reasonably accurate | difficult but reasonable | of no interest | moderate amt | NO HW assigned | NO txtbk used | somewhat slow | adequatelyE | I did not seek extra help | encourages questions & answers them seriously | fairly easy | very helpful | haven't tried any Mastering HW | Good |
| 3.50–4.00 | all but 1–2 | none of rdgs | NO HW assigned | A | A | very quick&accurate | fairly simple | very exciting | a great deal | NO HW before exam | NO txtbk used | moderate | very clearly | accessible for questions | encourages questions & answers them seriously | fairly easy | somewhat helpful | haven't tried much Mastering HW | Good |
| 3.00–3.49 | all but 3–4 | most readings | NO HW assigned | B | B | quick & reasonably accurate | difficult but reasonable | of some interest | moderate amt | NO HW assigned | NO txtbk used | moderate | adequately | accessible for questions | encourages questions & answers them seriously | difficult but reasonable | very helpful | haven't tried any Mastering HW | Excellent |
| | all classes | all readings | <70% | A | A | mod. paced&mostly accurate | difficult but reasonable | interesting | a great deal | within 2–3 days | good | too rapid for proper understanding | very clearly | could rarely be found | encourages questions & answers them seriously | difficult but reasonable | somewhat helpful | | |
| 2.50–2.99 | all classes | most readings | 95–100% | A | B | quick & reasonably accurate | fairly simple | of some interest | a great deal | within 2–3 days | NO txtbk used | moderate | very clearly | encourages questions & answers them seriously | moderate | I never consulted Blkbd supplemts | somewhat helpful | | Good |
| 3.00–3.49 | all classes | most readings | 70–84% | B | B | mod. paced&mostly accurate | difficult but reasonable | of some interest | moderate amt | within 2 weeks | NO txtbk used | reasonably fast | adequately | accessible for questions | I did not seek extra helpE | encourages questions & answers them seriously | difficult but reasonable | very helpful | not at all helpful | Excellent |
| 3.00–3.49 | all but 1–2 | some readings | <70% | B | B | quick & reasonably accurate | difficult but reasonable | of some interest | moderate amt | within week before exam | poor | somewhat slow | adequately | accessible for questions | encourages questions & answers them seriously | fairly easy | very helpful | not at all helpful | Fair |
| 2.50–2.99 | all classes | all readings | 95–100% | B | B | mod. paced&mostly accurate | difficult but reasonable | of some interest | moderate amt | within 2–3 days | adequate | too rapid for proper understanding | poorly | accessible for questions | encourages questions & answers them seriously | moderate | very helpful | somewhat helpful | Good |
| 2.50–2.99 | all but 1–2 | most readings | 95–100% | A | B | less quick&/or accurate than I'd like | difficult but reasonable | interesting | moderate amt | within week before exam | NO txtbk used | moderate | adequately | accessible for questions | encourages questions & answers them seriously | difficult but reasonable | somewhat helpful | not at all helpful | Excellent |
| 3.50–4.00 | all classes | all readings | 95–100% | A | A | very quick&accurate | fairly simple | interesting | a great deal | within 1 day of assignment | good | somewhat slow | adequately | accessible for questions | encourages questions & answers them seriously | moderate | very helpful | somewhat helpful | Excellent |
| 3.50–4.00 | all classes | most readings | 85–94% | A | A | quick & reasonably accurate | fairly simple | interesting | moderate amt | within 2–3 days | NO txtbk used | reasonably fast | adequately | accessible for questions | encourages questions & answers them seriously | moderate | very helpful | not at all helpful | Excellent |
| 3.00–3.49 | all classes | most readings | 95–100% | A | A | very quick&accurate | difficult but reasonable | of some interest | moderate amt | within 1 day of assignment | excellent | reasonably fast | very clearly | accessible for questions | encourages questions & answers them seriously | too difficult | very helpful | somewhat helpful | Excellent |
| 3.50–4.00 | all classes | some readings | <70% | A | A | mod. paced&mostly accurate | difficult but reasonable | of some interest | moderate amt | within week before exam | NO txtbk used | moderate | adequately | accessible for questions | encourages questions & answers them seriously | difficult but reasonable | somewhat helpful | not at all helpful | Good |
| 3.00–3.49 | all classes | most readings | NO HW assigned | B | A | quick & reasonably accurate | difficult but reasonable | interesting | a great deal | within 2–3 daysNO HW assigned | good | too rapid for proper understanding | adequatelyE | could rarely be found | encourages questions & answers them seriously | difficult but reasonablemuch too easy | somewhat helpful | very helpful | |
| 3.00–3.49 | most classes | all readings | 95–100% | A | A | very quick&accurate | fairly simple | very exciting | a great deal | within 2 weeks | poor | reasonably fast | adequately | accessible for questions | encourages questions & answers them seriously | moderate | very helpful | very helpful | The best or very nearly the best |
| 2.50–2.99 | all but 3–4 | all readings | 95–100% | A | A | very quick&accurate | too difficult | of no interest | almost nothing | NO HW assigned | NO txtbk used | much too slow | not at all | I did not seek extra help | openly discourages questions in class | fairly easy | very helpful | haven't tried any Mastering HW | The best or very nearly the best |
| 2.50–2.99 | all but 3–4 | all readings | 95–100% | A | A | very quick&accurate | difficult but reasonable | interesting | moderate amt | within 1 day of assignment | excellent | moderate | adequately | accessible for questions | encourages questions & answers them seriously | difficult but reasonable | very helpful | very helpful | Excellent |
| 3.00–3.49 | all classes | all readings | 95–100% | A | A | very quick&accurate | fairly simple | very exciting | a great deal | within 2 weeks | excellent | too rapid for proper understanding | very clearly | accessible for questions | encourages questions & answers them seriously | moderate | very helpful | not at all helpful | Excellent |
| 3.00–3.49 | all but 1–2 | most readings | 95–100% | A | A | quick & reasonably accurate | difficult but reasonable | interesting | moderate amt | within 2–3 days | excellent | moderate | very clearly | accessible for questions | encourages questions & answers them seriously | moderate | very helpful | not at all helpful | Excellent |
| 2.50–2.99 | most classes | most readings | <70% | C | C | quick & reasonably accurate | difficult but reasonable | of some interest | moderate amt | within week before exam | adequate | moderate | adequately | accessible for questions | encourages questions & answers them seriously | moderate | I never consulted Blkbd supplemts | somewhat helpful | Fair |
| 3.00–3.49 | all classes | most readings | 70–84% | B | B | quick & reasonably accurate | fairly simple | interesting | moderate amt | within 2–3 days | good | moderate | adequately | accessible for questions | encourages questions & answers them seriously | moderate | not helpful | somewhat helpful | Good |
| | all but 1–2 | some readings | 95–100% | B | B | very quick&accurate | difficult but reasonable | interesting | a great deal | | | | | | | | | | |
| 3.00–3.49 | all but 1–2 | some readings | 95–100% | A | A | quick & reasonably accurate | difficult but reasonable | interesting | moderate amt | within 2–3 days | NO txtbk used | too rapid for proper understanding | adequately | could rarely be found | does not encourage questions but answers them seriously | difficult but reasonable | somewhat helpful | somewhat helpful | Excellent |
| 3.50–4.00 | all but 1–2 | most readings | 70–84% | B | C | mod. paced&mostly accurate | difficult but reasonable | of some interest | moderate amt | within week before exam | poor | moderate | adequately | accessible for questions | encourages questions & answers them seriously | difficult but reasonable | very helpful | somewhat helpful | Fair |
| 3.50–4.00 | all classes | most readings | 95–100% | A | A | quick & reasonably accurate | difficult but reasonable | interesting | moderate amt | within week before exam | NO txtbk used | moderate | very clearly | accessible for questions | encourages questions & answers them seriously | difficult but reasonable | very helpful | somewhat helpful | Excellent |
| 3.00–3.49 | all but 3–4 | most readings | 70–84% | B | B | mod. paced&mostly accurate | too simple | of no interest | almost nothing | within week before exam | NO txtbk used | somewhat slow | poorly | I did not seek extra help | encourages questions & answers them seriously | difficult but reasonable | somewhat helpful | not at all helpful | Fair |
| 3.50–4.00 | all but 1–2 | some readings | 95–100% | A | A | mod. paced&mostly accurate | difficult but reasonable | of some interest | a great deal | within 2–3 days | good | reasonably fast | very clearly | accessible for questions | encourages questions & answers them seriously | difficult but reasonable | very helpful | somewhat helpful | The best or very nearly the best |
| 3.00–3.49 | most classes | most readings | 95–100% | A | A | quick & reasonably accurate | difficult but reasonable | interesting | moderate amt | within 2–3 days | poor | moderate | adequately | accessible for questions | encourages questions & answers them seriously | moderate | very helpful | not at all helpful | Good |
| 3.00–3.49 | all classes | all readings | <70% | A | B | quick & reasonably accurate | difficult but reasonable | interesting | moderate amt | within week before exam | poor | moderate | poorly | accessible for questions | encourages questions & answers them seriously | too difficult | very helpful | not at all helpful | Good |
| 3.50–4.00 | all but 3–4 | most readings | <70% | A | A | quick & reasonably accurate | difficult but reasonable | of some interest | moderate amt | within week before exam | adequate | reasonably fast | adequately | accessible for questions | encourages questions & answers them seriously | difficult but reasonable | very helpful | haven't tried much Mastering HW | Good |
| 3.50–4.00 | all classes | some readings | 95–100% | A | A | very quick&accurate | fairly simple | of some interest | moderate amt | within week before exam | adequate | much too slow | very clearly | accessible for questions | encourages questions & answers them seriously | moderate | very helpful | haven't tried much Mastering HW | Excellent |
| 3.50–4.00 | all classes | some readings | <70% | B | A | quick & reasonably accurate | fairly simple | interesting | moderate amt | within week before exam | adequate | much too slow | very clearly | accessible for questions | encourages questions & answers them seriously | moderate | very helpful | haven't tried much Mastering HW | Good |
| 2.00–2.49 | most classes | most readings | 70–84% | C | C | mod. paced&mostly accurate | difficult but reasonable | of no interest | moderate amt | within week before exam | NO txtbk used | moderate | adequately | accessible for questions | encourages questions & answers them seriously | difficult but reasonable | very helpful | not at all helpful | Fair |
| 3.50–4.00 | all but 3–4 | none of rdgs | 70–84% | A | A | mod. paced&mostly accurate | difficult but reasonable | of no interest | moderate amt | within week before exam | poor | somewhat slow | adequately | accessible for questions | encourages questions & answers them seriously | too difficult | not helpful | not at all helpful | Good |
| 3.00–3.49 | all but 3–4 | some readings | <70% | A | A | very quick&accurate | fairly simple | of some interest | moderate amt | within week before exam | NO txtbk used | moderate | adequately | accessible for questions | encourages questions & answers them seriously | moderate | very helpful | haven't tried much Mastering HW | Good |
| 3.50–4.00 | all classes | most readings | <70% | A | A | very quick&accurate | fairly simple | interesting | moderate amt | within 2 weeks | good | moderate | very clearly | I did not seek extra help | encourages questions & answers them seriously | moderate | very helpful | very helpful | The best or very nearly the best |
| 3.50–4.00 | all classes | all readings | 95–100% | A | A | quick & reasonably accurate | difficult but reasonable | of no interest | almost nothing | within 1 day of assignment | NO txtbk used | too rapid for proper understanding | poorly | could rarely be found | does not encourage questions & is reluctant to answer them | difficult but reasonable | somewhat helpful | not at all helpful | Poor |
| 3.50–4.00 | all but 1–2 | some readings | 85–94% | A | A | very quick&accurate | difficult but reasonable | of some interest | moderate amt | within 2–3 days | adequate | moderate | very clearly | accessible for questions | encourages questions & answers them seriously | difficult but reasonable | very helpful | somewhat helpful | Excellent |
| 3.50–4.00 | all but 1–2 | all readings | 95–100% | A | A | mod. paced&mostly accurate | difficult but reasonable | of no interest | a great deal | within 2–3 days | adequate | moderate | adequately | accessible for questions | encourages questions & answers them seriously | moderate | very helpful | somewhat helpful | Excellent |
| 2.50–2.99 | all but 1–2 | all readings | 95–100% | A | B | quick & reasonably accurate | difficult but reasonable | interesting | a great deal | within 2–3 days | adequate | moderate | adequately | accessible for questions | encourages questions & answers them seriously | difficult but reasonable | very helpful | somewhat helpful | Excellent |
| 3.00–3.49 | all classes | all readings | 85–94% | A | A | quick & reasonably accurate | fairly simple | very exciting | a great deal | within week before exam | adequate | moderate | adequately | accessible for questions | encourages questions & answers them seriously | fairly easy | very helpful | very helpful | The best or very nearly the best |
| 3.00–3.49 | most classes | most readings | 95–100% | B | B | quick & reasonably accurate | difficult but reasonable | interesting | moderate amt | within 2–3 days | adequate | moderate | adequately | accessible for questions | encourages questions & answers them seriously | difficult but reasonable | very helpful | not at all helpful | The best or very nearly the best |
| 3.50–4.00 | all classes | all readings | 95–100% | A | A | very quick&accurate | fairly simple | very exciting | a great deal | within 2 weeks | good | moderate | very clearly | I did not seek extra help | encourages questions & answers them seriously | moderate | very helpful | not at all helpful | Excellent |
| 3.00–3.49 | all but 1–2 | | 95–100% | A | A | very quick&accurate | fairly simple | of some interest | a great deal | within 2 weeks | adequate | moderate | adequately | accessible for questions | encourages questions & answers them seriously | difficult but reasonable | very helpful | not at all helpful | Good |
| 3.00–3.49 | all but 1–2 | most readings | <70% | A | B | quick & reasonably accurate | fairly simple | of some interest | moderate amt | within 2–3 days | NO txtbk used | moderate | adequately | accessible for questions | encourages questions & answers them seriously | fairly easy | somewhat helpful | not at all helpful | Excellent |
| 2.00–2.49 | all classes | most readings | 85–94% | C | C | mod. paced&mostly accurate | difficult but reasonable | of no interest | almost nothing | within week before exam | NO txtbk used | reasonably fast | adequately | accessible for questions | encourages questions & answers them seriously | moderate | somewhat helpful | not at all helpful | Good |
| 3.00–3.49 | all classes | some readings | <70% | A | A | very quick&accurate | difficult but reasonable | of some interest | moderate amt | within week before exam | NO txtbk used | moderate | adequately | accessible for questions | I did not seek extra help | encourages questions & answers them seriously | moderate | somewhat helpful | not at all helpful | Good |
| 3.50–4.00 | all classes | most readings | <70% | A | A | mod. paced&mostly accurate | difficult but reasonable | interesting | moderate amt | within 2 weeks | NO txtbk used | moderate | poorly | accessible for questions | encourages questions & answers them seriously | moderate | very helpful | haven't tried any Mastering HW | Excellent |
| 3.50–4.00 | all but 1–2 | most readings | <70% | A | A | very quick&accurate | difficult but reasonable | interesting | moderate amt | within 2 weeks | NO txtbk used | moderate | adequately | accessible for questions | encourages questions & answers them seriously | moderate | very helpful | haven't tried any Mastering HW | Good |
| 3.00–3.49 | all but 3–4 | most readings | <70% | A | A | mod. paced&mostly accurate | fairly simple | of no interest | moderate amt | within week before exam | NO txtbk used | moderate | adequately | accessible for questions | encourages questions & answers them seriously | fairly easy | very helpful | not at all helpful | Excellent |
| 2.50–2.99 | all but 1–2 | all readings | 95–100% | A | A | mod. paced&mostly accurate | difficult but reasonable | of no interest | almost nothing | within 2–3 days | NO txtbk used | too rapid for proper understanding | adequately | could rarely be found | does not encourage questions & is reluctant to answer them | difficult but reasonable | very helpful | not at all helpful | |

| GPA | ATTENDANCE | RDGS COMPLETED | HW COMPLETED | GRD DESERVED | GRD EXPECTED | MATH ABILITY | MATERIAL COVERED | MATERIAL COVERED | I LEARNED | HW STARTED | TEXTBK | PACE OF LECTS | DIFF/SUBTLE PTS | PROFESSOR | PROFESSOR | EXAMS WERE | BLKBD SUPPLEMENTS | MAST.PHYSICS | COURSE OVERALL |
|---|---|---|---|---|---|---|---|---|---|---|---|---|---|---|---|---|---|---|---|
| 2.00–2.49 | most classes | none of rdgs | <70% | A | A | mod.paced&mostly accurate | difficult but reasonable | of some interest | a great deal | | | good | moderate | adequately | encourages questions & answers them seriously | moderate | very helpful | not at all helpful | Good |
| 3.00–3.49 | all but 1–2 | most readings | 85–94% | A | B | | | | of no interest | within 2 weeks | NO txtbk used | reasonably fast | adequately | could rarely be found | openly discourages questions in class | difficult but reasonable | somewhat helpful | somewhat helpful | Fair |
| 3.50–4.00 | | some readings | 95–100% | A | A | mod.paced&mostly accurate | difficult but reasonable | interesting | | within week before exam | | adequate | reasonably fast | adequately | I did not seek extra help | | difficult but reasonable | very helpful | somewhat helpful | Good |
| 3.00–3.49 | | all but 1–2 | all readings | 95–100% | A | A | quick & reasonably accurate | | interesting | a great deal | within 2 weeks | NO txtbk used | moderate | adequately | accessible for questions | encourages questions & answers them seriously | difficult but reasonable | somewhat helpful | haven't tried any Mastering HW | Good |
| 3.50–4.00 | all classes | all readings | 95–100% | A | A | very quick&accurate | too difficult | very exciting | a great deal | within 1 day of assignment | | excellent | too rapid for proper understanding | very clearly | accessible for questions | encourages questions & answers them seriously | too difficult | very helpful | very helpful | The best or very nearly the best |
| 3.00–3.49 | all but 1–2 | all readings | 95–100% | B | B | quick & reasonably accurate | fairly simple | of some interest | moderate amt | within 2 weeks | | good | moderate | adequately | accessible for questions | encourages questions & answers them seriously | fairly easy | somewhat helpful | very helpful | Good |
| 3.00–3.49 | all but 1–2 | all readings | <70% | A | A | mod.paced&mostly accurate | fairly simple | interesting | a great deal | | NO txtbk used | moderate | adequately | accessible for questions | encourages questions & answers them seriously | moderate | I never consulted Blkbd supplemts | haven't tried any Mastering HW | |
| 3.50–4.00 | all but 3–4 | most readings | <70% | A | A | mod.paced&mostly accurate | fairly simple | interesting | | within week before exam | | adequate | somewhat slow | adequately | I did not seek extra help | accessible for questions | fairly easy | very helpful | haven't tried any Mastering HW | The best or very nearly the best |
| 3.50–4.00 | all but 3–4 | some readings | 95–100% | B | B | mod.paced&mostly accurate | fairly simple | of some interest | moderate amt | within week before exam | | adequate | moderate | adequately | I did not seek extra help | encourages questions & answers them seriously | difficult but reasonable | very helpful | haven't tried any Mastering HW | Fair |
| 3.00–3.49 | most classes | all readings | <70% | A | A | quick & reasonably accurate | difficult but reasonable | interesting | a great deal | | | | | | | | | | | |
| 3.00–3.49 | all classes | some readings | <70% | D | E | mod.paced&mostly accurate | | interesting | a great deal | within 2 weeks | | moderate | adequately | could rarely be found | encourages questions & answers them seriously | difficult but reasonable | very helpful | haven't tried any Mastering HW | Fair |
| 3.50–4.00 | few classes | all readings | NO HW assigned | A | A | quick & reasonably accurate | | of some interest | moderate amt | NO HW assigned | NO txtbk used | moderate | adequately | accessible for questions | encourages questions & answers them seriously | moderate | very helpful | haven't tried any Mastering HW | Good |
| 3.50–4.00 | all but 3–4 | none of rdgs | NO HW assigned | A | A | very quick&accurate | difficult but reasonable | of some interest | moderate amt | NO HW assigned | | moderate | adequately | accessible for questions | encourages questions & answers them seriously | moderate | very helpful | haven't tried any Mastering HW | Good |
| 3.50–4.00 | all but 1–2 | none of rdgs | 95–100% | A | A | very quick&accurate | fairly simple | interesting | moderate amt | within 2–3 days | NO txtbk used | somewhat slow | adequately | I did not seek extra help | encourages questions & answers them seriously | moderate | very helpful | very helpful | Good |
| 3.50–4.00 | all readings | NO HW assigned | A | A | very quick&accurate | difficult but reasonable | interesting | moderate amt | | | moderate | adequately | very clearly | accessible for questions | encourages questions & answers them seriously | | somewhat helpful | | |
| 2.50–2.99 | all but 3–4 | most readings | 95–100% | B | B | quick & reasonably accurate | difficult but reasonable | interesting | a great deal | within week before exam | NO txtbk used | moderate | adequately | accessible for questions | encourages questions & answers them seriously | difficult but reasonable | very helpful | very helpful | Excellent |
| 3.50–4.00 | all but 3–4 | most readings | 95–100% | A | A | very quick&accurate | difficult but reasonable | interesting | moderate amt | within 2 weeks | | adequate | moderate | adequately | I did not seek extra help | encourages questions & answers them seriously | difficult but reasonable | very helpful | somewhat helpful | Excellent |
| 2.50–2.99 | all but 1–2 | most readings | 95–100% | A | A | very quick&accurate | fairly simple | interesting | a great deal | within 2–3 days | | good | reasonably fast | adequately | very clearly | accessible for questions | moderate | very helpful | very helpful | Excellent |
| 3.50–4.00 | all classes | all readings | 95–100% | A | A | quick & reasonably accurate | fairly simple | of some interest | moderate amt | within 2–3 days | | adequate | moderate | adequately | accessible for questions | encourages questions & answers them seriously | moderate | somewhat helpful | somewhat helpful | Good |
| 3.50–4.00 | all classes | most readings | 95–100% | A | A | very quick&accurate | difficult but reasonable | of some interest | a great deal | within 1 day of assignment | | poor | reasonably fast | adequately | could rarely be found | does not encourage questions & is reluctant to answer them | difficult but reasonable | somewhat helpful | not at all helpful | Good |
| 3.00–3.49 | all but 1–2 | | <70% | A | B | quick & reasonably accurate | too difficult | of no interest | almost nothing | within week before exam | | poor | too rapid for proper understanding | poorly | could rarely be found | does not encourage questions but answers them seriously | too difficult | E | haven't tried any Mastering HW | Fair |
| 3.00–3.49 | all but 1–2 | some readings | 70–84% | A | A | mod.paced&mostly accurate | difficult but reasonable | interesting | moderate amt | within week before exam | NO txtbk used | moderate | adequately | very clearly | accessible for questions | encourages questions & answers them seriously | moderate | very helpful | not at all helpful | The best or very nearly the best |
| 3.50–4.00 | all classes | all readings | <70% | A | A | quick & reasonably accurate | fairly simple | interesting | a great deal | within week before exam | | poor | reasonably fast | very clearly | accessible for questions | encourages questions & answers them seriously | moderate | very helpful | very helpfulhaven't tried much Mastering HW | Excellent |
| 2.00–2.49 | all classes | none of rdgs | 70–84% | D | C | less quick&/or accurate than I'd like | too simple | of no interest | almost nothing | within week before exam | | poor | somewhat slow | E | E | openly discourages questions in class | moderate | somewhat helpful | somewhat helpful | Excellent |
| 3.50–4.00 | all but 1–2 | most readings | 70–84% | B | B | mod.paced&mostly accurate | fairly simple | interesting | moderate amt | within 1 day of assignment | | poor | moderate | adequately | accessible for questions | does not encourage questions but answers them seriously | difficult but reasonable | very helpful | not at all helpful | Excellent |
| 2.50–2.99 | all but 1–2 | most readings | 70–84% | B | B | mod.paced&mostly accurate | difficult but reasonable | interesting | moderate amt | within 2 weeks | | poor | adequately | | I did not seek extra help | encourages questions & answers them seriously | difficult but reasonable | very helpful | haven't tried any Mastering HW | Excellent |
| 3.00–3.49 | all but 1–2 | most readings | <70% | A | A | mod.paced&mostly accurate | difficult but reasonable | of some interest | moderate amt | NO HW assigned | NO txtbk used | moderate | adequately | very clearly | accessible for questions | difficult but reasonable | very helpful | haven't tried any Mastering HW | Fair |
| 3.00–3.49 | most classes | most readings | 95–100% | A | A | mod.paced&mostly accurate | difficult but reasonable | of no interest | almost nothing | within 1 day of assignment | NO txtbk used | moderate | adequately | very clearly | does not encourage questions & is reluctant to answer them | difficult but reasonable | very helpful | not at all helpful | |
| 3.00–3.49 | all but 3–4 | most readings | 95–100% | A | B | very quick&accurate | difficult but reasonable | of some interest | moderate amt | within 2 weeks | | poor | moderate | very clearly | I did not seek extra help | encourages questions & answers them seriously | difficult but reasonable | very helpful | very helpful | Good |
| 3.00–3.49 | all but 1–2 | | 95–100% | A | A | very quick&accurate | difficult but reasonable | interesting | moderate amt | within 2 weeks | NO txtbk used | excellent | moderate | adequately | could rarely be found | does not encourage questions but answers them seriously | difficult but reasonable | moderate | I never consulted Blkbd supplemts | haven't tried any Mastering HW | Fair |
| 3.00–3.49 | few classes | some readings | 70–84% | B | B | mod.paced&mostly accurate | difficult but reasonable | of some interest | moderate amt | within week before exam | | poor | somewhat slow | poorly | accessible for questions | does not encourage questions but answers them seriously | moderate | very helpful | somewhat helpful | Good |
| 3.50–4.00 | all classes | all readings | 85–94% | A | A | quick & reasonably accurate | fairly simple | of some interest | moderate amt | within 1 day of assignment | | somewhat slow | adequately | | encourages questions & answers them seriously | much too easy | somewhat helpful | | | |
| 3.50–4.00 | all but 3–4 | none of rdgs | <70% | A | A | quick & reasonably accurate | fairly simple | of some interest | moderate amt | within week before exam | NO txtbk used | somewhat slow | adequately | I did not seek extra help | encourages questions & answers them seriously | moderate | very helpful | haven't tried any Mastering HW | Fair |
| 2.50–2.99 | all but 1–2 | some readings | 95–100% | A | A | very quick&accurate | difficult but reasonable | interesting | a great deal | within week before exam | | poor | | adequately | accessible for questions | encourages questions & answers them seriously | somewhat helpful | | | |
| 3.50–4.00 | most classes | some readings | <70% | A | A | mod.paced&mostly accurate | difficult but reasonable | of some interest | moderate amt | within 2 weeks | NO txtbk used | moderate | poorly | accessible for questions | does not encourage questions but answers them seriously | difficult but reasonable | not helpful | Fair | |
| 3.00–3.49 | all classes | most readings | 70–84% | A | B | quick & reasonably accurate | | interesting | moderate amt | within 2–3 days | | good | moderate | adequately | accessible for questions | encourages questions & answers them seriously | moderate | very helpful | haven't tried any Mastering HW | Excellent |
| 3.50–4.00 | most classes | most readings | 70–84% | A | A | quick & reasonably accurate | fairly simple | interesting | moderate amt | | | | | | | | | | | |
| 3.50–4.00 | few classes | some readings | 95–100% | A | A | mod.paced&mostly accurate | difficult but reasonable | of no interest | almost nothing | within 2 weeks | NO txtbk used | moderate | adequately | accessible for questions | encourages questions & answers them seriously | moderate | somewhat helpful | not at all helpful | Fair |
| 3.50–4.00 | all classes | all readings | 95–100% | A | B | mod.paced&mostly accurate | difficult but reasonable | very exciting | a great deal | within 2 weeks | | adequate | moderate | very clearly | accessible for questions | encourages questions & answers them seriously | fairly easy | very helpful | very helpful | Excellent |
| 3.50–4.00 | all but 1–2 | most readings | 95–100% | B | B | less quick&/or accurate than I'd like | difficult but reasonable | very exciting | a great deal | within week before exam | | excellent | moderate | very clearly | accessible for questions | encourages questions & answers them seriously | difficult but reasonable | very helpful | somewhat helpful | The best or very nearly the best |
| 3.50–4.00 | most classes | all readings | | A | B | less quick&/or accurate than I'd like | difficult but reasonable | of no interest | almost nothing | within 2–3 days | NO txtbk used | reasonably fast | adequately | | | | difficult but reasonable | somewhat helpful | haven't tried any Mastering HW | Fair |
| 3.00–3.49 | few classes | none of rdgs | NO HW assigned | A | A | very quick&accurate | too difficult | of no interest | almost nothing | within week before exam | NO txtbk used | somewhat slow | | could rarely be found | | moderate | E | haven't tried any Mastering HW | Fair | |
| 3.50–4.00 | all classes | | NO HW assigned | A | B | mod.paced&mostly accurate | too difficult | of no interest | almost nothing | within week before exam | | much too slow | poorly | encourages questions & answers them seriously | too difficult | very helpful | too difficult | not at all helpful | Fair | |
| 3.50–4.00 | few classes | some readings | 95–100% | A | A | very quick&accurate | too difficult | of no interest | almost nothing | within week before exam | | somewhat slow | adequately | I did not seek extra help | encourages questions & answers them seriously | moderate | very helpful | not at all helpful | Good |
| 3.00–3.49 | all classes | most readings | NO HW assigned | A | A | mod.paced&mostly accurate | difficult but reasonable | fairly simple | interesting | moderate amt | within week before exam | NO txtbk used | much too slow | adequately | I did not seek extra help | accessible for questions | encourages questions & answers them seriously | fairly easy | I never consulted Blkbd supplemts | haven't tried any Mastering HW | Excellent |
| 2.50–2.99 | most classes | none of rdgs | NO HW assigned | A | A | very quick&accurate | fairly simple | interesting | moderate amt | NO HW assigned | NO txtbk used | much too slow | adequately | I did not seek extra help | accessible for questions | encourages questions & answers them seriously | fairly easy | I never consulted Blkbd supplemts | haven't tried any Mastering HW | Excellent |
| 3.00–3.49 | most classes | all readings | <70% | D | D | mod.paced&mostly accurate | difficult but reasonable | of no interest | almost nothing | within week before exam | | adequate | moderate | poorly | I did not seek extra help | encourages questions & answers them seriously | moderate | very helpful | not at all helpful | Poor |
| 3.00–3.49 | most classes | some readings | <70% | B | B | mod.paced&mostly accurate | difficult but reasonable | of no interest | moderate amt | within week before exam | | adequate | moderate | poorly | I did not seek extra help | encourages questions & answers them seriously | difficult but reasonable | moderate | very helpful | haven't tried much Mastering HW | Good |
| 3.00–3.49 | most classes | most readings | <70% | B | E | mod.paced&mostly accurate | fairly simple | of no interest | almost nothing | within week before exam | NO txtbk used | much too slow | poorly | I did not seek extra help | does not encourage questions but answers them seriously | moderate | very helpful | haven't tried any Mastering HW | Poor |
| 3.50–4.00 | few classes | some readings | <70% | A | A | very quick&accurate | too simple | interesting | moderate amt | within week before exam | NO txtbk used | much too slow | | I did not seek extra help | | much too easy | very helpful | haven't tried any Mastering HW | Good | |
| 2.50–2.99 | all but 3–4 | all readings | <70% | A | B | quick & reasonably accurate | too difficult | of no interest | moderate amt | within 1 day of assignment | | poor | too rapid for proper understanding | poorly | could rarely be found | openly discourages questions in class | too difficult | very helpful | not at all helpful | Poor |
| 3.00–3.49 | all but 1–2 | some readings | NO HW assigned | A | B | mod.paced&mostly accurate | too simple | very exciting | a great deal | within 1 day of assignment | | excellent | very clearly | accessible for questions | moderate | very helpful | very helpful | The best or very nearly the best | | |
| 3.00–3.49 | all but 1–2 | most readings | 95–100% | B | C | mod.paced&mostly accurate | difficult but reasonable | of some interest | moderate amt | within week before exam | NO txtbk used | reasonably fast | adequately | too difficult | very helpful | very helpful | | | | |
| 3.00–3.49 | all but 1–2 | some readings | <70% | B | A | quick & reasonably accurate | fairly simple | interesting | moderate amt | within week before exam | NO txtbk used | moderate | adequately | I did not seek extra help | encourages questions & answers them seriously | moderate | not at all helpful | very helpful | | |
| 3.00–3.49 | all but 3–4 | most readings | 95–100% | A | B | quick & reasonably accurate | difficult but reasonable | interesting | a great deal | within week before exam | | reasonably fast | adequately | I did not seek extra help | encourages questions & answers them seriously | difficult but reasonable | not helpful | very helpful | Good |
| 3.50–4.00 | all but 1–2 | most readings | 95–100% | A | A | very quick&accurate | difficult but reasonable | interesting | a great deal | within week before exam | | somewhat slow | adequately | very clearly | accessible for questions | encourages questions & answers them seriously | moderate | very helpful | very helpful | Excellent |
| 3.50–4.00 | most classes | all readings | 95–100% | A | A | quick & reasonably accurate | difficult but reasonable | interesting | a great deal | within week before exam | NO txtbk used | moderate | adequately | very clearly | accessible for questions | encourages questions & answers them seriously | moderate | very helpful | haven't tried any Mastering HW | Excellent |
| 3.50–4.00 | most classes | most readings | <70% | A | A | quick & reasonably accurate | fairly simple | interesting | a great deal | within 2 weeks | | good | moderate | very clearly | accessible for questions | encourages questions & answers them seriously | moderate | very helpful | very helpful | Good |
| 3.50–4.00 | all but 1–2 | most readings | 95–100% | A | A | quick & reasonably accurate | difficult but reasonable | of some interest | moderate amt | within 2 weeks | NO txtbk used | too rapid for proper understanding | not at all | accessible for questions | encourages questions & answers them seriously | moderate | not helpful | not at all helpful | Fair | |
| 3.00–3.49 | most classes | some readings | 70–84% | A | A | less quick&/or accurate than I'd like | difficult but reasonable | very exciting | moderate amt | within 2–3 days | NO txtbk used | reasonably fast | adequately | accessible for questions | encourages questions & answers them seriously | difficult but reasonable | somewhat helpful | somewhat helpful | Good |

| GPA | ATTENDANCE | RDGS COMPLETED | HW COMPLETED | GRD DESERVED | GRD EXPECTED | MATH ABILITY | MATERIAL COVERED | MATERIAL COVERED | I LEARNED | HW STARTED | TEXTBK | PACE OF LECTS | DIFF/SUBTLE PTS | PROFESSOR | PROFESSOR | EXAMS WERE | BLKBD SUPPLEMENTS | MAST.PHYSICS | COURSE OVERALL |
|---|---|---|---|---|---|---|---|---|---|---|---|---|---|---|---|---|---|---|---|
| 3.50–4.00 | few classes | some readings | 95-100% | C | B | mod.paced&mostly accurate | fairly simple | interesting | moderate | within week before exam | NO txtbk used | moderate | adequately | very clearly | encourages questions & answers them seriously | difficult but reasonable | very helpful | very helpful | Excellent |
| 3.00-3.49 | few classes | some readings | <70% | A | A | mod. paced&mostly accurate | too difficult | interesting | within 2 weeks | poor | too rapid for proper understanding | adequately | accessible for questions | encourages questions & answers them seriously | too difficult | not at all helpful | Good |
| 3.50-4.00 | all but 1–2 | most readings | 95-100% | A | B | very quick&accurate | difficult but reasonable | of some interest | a great deal | within 2-3 days | NO txtbk used | moderate | adequately | accessible for questions | encourages questions & answers them seriously | difficult but reasonable | somewhat helpful | somewhat helpful | Good |
| 3.00-3.49 | all but 1–2 | most readings | 95-100% | A | B | mod. paced&mostly accurate | difficult but reasonable | interesting | a great deal | within week before exam | NO txtbk used | reasonably fast | adequately | very clearly | accessible for questions | too difficult | very helpful | very helpful | Good |
| 3.50-4.00 | all but 1–2 | all readings | <70% | A | A | very quick&accurate | difficult but reasonable | interesting | a great deal | within week before exam | NO txtbk used | moderate | adequately | very clearly | accessible for questions | difficult but reasonable | somewhat helpful | haven't tried any Mastering HW | Excellent |
| 3.50-4.00 | most classes | 95-100% | A | B | quick & reasonably accurate | difficult but reasonable | of some interest | moderate amt | within 2-3 days | NO txtbk used | moderate | adequately | accessible for questions | encourages questions & answers them seriously | too difficult | very helpful | Good |
| 2.50-2.99 | most classes | all readings | 70-84% | B | B | mod.paced&mostly accurate | difficult but reasonable | interesting | a great deal | within 2-3 days | NO txtbk used | reasonably fast | adequately | accessible for questions | encourages questions & answers them seriously | difficult but reasonable | somewhat helpful | very helpful | Good |
| 3.50-4.00 | all but 1–2 | most readings | 85-94% | A | A | very quick&accurate | too simple | of no interest | almost nothing | within 1 day of assignment | poor | much too slow | adequately | I did not seek extra help | encourages questions & answers them seriously | much too easy | somewhat helpful | not at all helpful | Fair |
| 3.50-4.00 | all but 1–2 | most readings | 85-94% | B | B | very quick&accurate | fairly simple | of some interest | moderate amt | within 1 day of assignment | adequate | moderate | adequately | accessible for questions | encourages questions & answers them seriously | moderate | somewhat helpful | not at all helpful | Good |
| 3.00-3.49 | most classes | most readings | <70% | B | A | very quick&accurate | fairly simple | interesting | moderate amt | within week before exam | adequate | moderate | adequately | I did not seek extra help | encourages questions & answers them seriously | moderate | moderate | Good |
| 3.50-4.00 | few classes | none of rdgs | 95-100% | E | A | very quick&accurate | fairly simple | very exciting | almost nothing | within week before exam | adequate | too rapid for proper understanding | very clearly | accessible for questions | encourages questions & answers them seriously | moderate | haven't tried any Mastering HW | Excellent |
| 3.50-4.00 | most classes | most readings | 70-84% | A | A | less quick&/or accurate than I'd like | difficult but reasonable | of no interest | moderate amt | within week before exam | NO txtbk used | moderate | adequately | I did not seek extra help | encourages questions & answers them seriously | moderate | very helpful | haven't tried any Mastering HW | Good |
| 3.50-4.00 | all classes | 95-100% | A | A | very quick&accurate | too difficult | very exciting | a great deal | NO HW assigned | good | too rapid for proper understanding | very clearly | accessible for questions | encourages questions & answers them seriously | difficult but reasonable | somewhat helpful | haven't tried any Mastering HW | Good |
| 2.50-2.99 | all but 3–4 | some readings | 95-100% | B | B | mod.paced&mostly accurate | difficult but reasonable | of some interest | almost nothing | within week before exam | NO txtbk used | too rapid for proper understanding | adequately | I did not seek extra help | encourages questions & answers them seriously | too difficult | not helpful | Fair |
| 3.50-4.00 | all classes | most readings | 70-84% | A | A | very quick&accurate | fairly simple | interesting | moderate amt | within 2 weeks | NO txtbk used | somewhat slow | adequately | I did not seek extra help | encourages questions & answers them seriously | fairly easy | very helpful | somewhat helpful | The best or very nearly the best |
| 2.50-2.99 | all but 3–4 | most readings | 70-84% | B | B | mod.paced&mostly accurate | difficult but reasonable | interesting | a great deal | within week before exam | good | moderate | adequately | very clearly | encourages questions & answers them seriously | difficult but reasonable | somewhat helpful | very helpful | Good |
| 2.50-2.99 | all classes | some readings | 95-100% | B | A | quick & reasonably accurate | difficult but reasonable | of some interest | a great deal | within week before exam | NO txtbk used | moderate | adequately | very clearly | accessible for questions | difficult but reasonable | somewhat helpful | not at all helpful | The best or very nearly the best |
| 2.50-2.99 | all but 1–2 | most readings | 85-94% | B | B | very quick&accurate | difficult but reasonable | of some interest | a great deal | within 2 weeks | NO txtbk used | moderate | adequately | very clearly | accessible for questions | difficult but reasonable | somewhat helpful | very helpful | Excellent |
| 3.50-4.00 | all but 1–2 | some readings | 95-100% | A | A | very quick&accurate | fairly simple | very exciting | a great deal | within 1 day of assignment | adequate | moderate | adequately | accessible for questions | encourages questions & answers them seriously | moderate | somewhat helpful | The best or very nearly the best |
| 3.50-4.00 | all but 3–4 | some readings | <70% | C | A | very quick&accurate | fairly simple | of no interest | almost nothing | within week before exam | NO txtbk used | much too slow | very clearly | accessible for questions | encourages questions & answers them seriously | fairly easy | somewhat helpful | haven't tried any Mastering HW | Fair |
| 3.00-3.49 | all but 1–2 | most readings | <70% | A | B | quick & reasonably accurate | fairly simple | of no interest | almost nothing | within week before exam | adequate | somewhat slow | adequately | I did not seek extra help | encourages questions & answers them seriously | difficult but reasonable | very helpful | not at all helpful | Good |
| 3.50-4.00 | all but 1–2 | most readings | <70% | A | A | very quick&accurate | fairly simple | interesting | a great deal | within week before exam | excellent | moderate | adequately | accessible for questions | encourages questions & answers them seriously | moderate | very helpful | haven't tried any Mastering HW | Excellent |
| 2.50-2.99 | all classes | most readings | 95-100% | A | A | quick & reasonably accurate | difficult but reasonable | of some interest | a great deal | within week before exam | adequate | reasonably fast | adequately | accessible for questions | encourages questions & answers them seriously | difficult but reasonable | moderate | not at all helpful | Excellent |
| 2.00–2.49 | all classes | all readings | 95-100% | A | B | mod.paced&mostly accurate | difficult but reasonable | interesting | a great deal | within 2 weeks | adequate | moderate | very clearly | accessible for questions | encourages questions & answers them seriously | moderate | very helpful | Excellent |
| 3.00-3.49 | few classes | some readings | NO HW assigned | C | C | less quick&/or accurate than I'd like | difficult but reasonable | of no interestE | almost nothing | NO HW assigned | NO txtbk used | much too slow | poorly | I did not seek extra help | openly discourages questions in class | fairly easy | very helpful | haven't tried any Mastering HW | Poor |
| 3.00-3.49 | most classes | all readings | 70-84% | A | A | mod.paced&mostly accurate | difficult but reasonable | interesting | moderate amt | within 2 weeks | good | moderate | adequately | accessible for questions | encourages questions & answers them seriously | difficult but reasonable | somewhat helpful | haven't tried much Mastering HW | Excellent |
| 3.50-4.00 | all classes | most readings | 70-84% | C | E | quick & reasonably accurate | fairly simple | interesting | moderate amt | within 2 weeks | NO txtbk used | somewhat slow | adequately | accessible for questions | encourages questions & answers them seriously | too difficult | very helpful | haven't tried much Mastering HW | |
| 3.50-4.00 | all but 1–2 | some readings | 70-84% | A | A | very quick&accurate | fairly simple | interesting | a great deal | within 2-3 days | NO txtbk used | moderate | adequately | very clearly | accessible for questions | encourages questions & answers them seriously | moderate | very helpful | haven't tried much Mastering HW | The best or very nearly the best |
| 3.50-4.00 | all but 1–2 | all readings | 95-100% | A | A | very quick&accurate | fairly simple | very exciting | a great deal | within week before exam | excellent | moderate | adequately | very clearly | accessible for questions | encourages questions & answers them seriously | fairly easy | very helpful | haven't tried much Mastering HW | The best or very nearly the best |
| 2.50-2.99 | all but 1–2 | some readings | 95-100% | B | B | very quick&accurate | difficult but reasonable | of some interest | a great deal | within week before exam | adequate | moderate | adequately | very clearly | accessible for questions | encourages questions & answers them seriously | difficult but reasonable | very helpful | not at all helpful | The best or very nearly the best |
| 3.50-4.00 | few classes | some readings | <70% | A | A | very quick&accurate | difficult but reasonable | of some interest | moderate amt | within week before exam | poor | reasonably fast | adequately | accessible for questions | encourages questions & answers them seriously | difficult but reasonable | somewhat helpful | somewhat helpful | Good |
| 3.50-4.00 | all but 1–2 | most readings | <70% | A | A | quick & reasonably accurate | difficult but reasonable | of no interest | moderate amt | within 2 weeks | NO txtbk used | reasonably fast | adequately | very clearly | accessible for questions | encourages questions & answers them seriously | moderate | very helpful | haven't tried any Mastering HW | Excellent |
| 2.50-2.99 | all classes | some readings | 95-100% | B | B | very quick&accurate | difficult but reasonable | interesting | a great deal | within 2 weeks | NO txtbk used | moderate | adequately | very clearly | accessible for questions | encourages questions & answers them seriously | moderate | very helpful | very helpful | The best or very nearly the best |
| 3.00-3.49 | few classes | most readings | 85-94% | A | B | mod.paced&mostly accurate | difficult but reasonable | of some interest | a great deal | within week before exam | NO txtbk used | moderate | adequately | accessible for questions | encourages questions & answers them seriously | difficult but reasonable | very helpful | not at all helpful | Fair |
| 2.50-2.99 | all but 3–4 | some readings | 70-84% | C | C | less quick&/or accurate than I'd like | difficult but reasonable | of some interest | a great deal | within 2 weeks | NO txtbk used | moderate | adequately | accessible for questions | encourages questions & answers them seriously | difficult but reasonable | somewhat helpful | not at all helpful | Good |
| 3.50-4.00 | all classes | most readings | 70-84% | A | A | less quick&/or accurate than I'd like | difficult but reasonable | of some interest | a great deal | within 2 weeks | NO txtbk used | moderate | adequately | accessible for questions | encourages questions & answers them seriously | difficult but reasonable | somewhat helpful | haven't tried any Mastering HW | Fair |
| 2.00–2.49 | all classes | most readings | 95-100% | C | C | quick & reasonably accurate | difficult but reasonable | of some interest | moderate amt | within 2 weeks | adequate | moderate | adequately | accessible for questions | encourages questions & answers them seriously | moderate | moderate | very helpful | |
| 3.00-3.49 | few classes | most readings | 95-100% | A | A | quick & reasonably accurate | difficult but reasonable | interesting | a great deal | within week before exam | NO txtbk used | moderate | very clearly | I did not seek extra help | encourages questions & answers them seriously | moderate | somewhat helpful | Excellent |
| 3.00-3.49 | all classes | all readings | 95-100% | A | B | quick & reasonably accurate | fairly simple | very exciting | a great deal | within 2-3 days | NO txtbk used | moderate | adequately | very clearly | encourages questions & answers them seriously | difficult but reasonable | I did not seek extra help | not at all helpful | The best or very nearly the best |
| 3.00-3.49 | all but 3–4 | most readings | 70-84% | B | B | mod.paced&mostly accurate | difficult but reasonable | of some interest | moderate amt | within week before exam | NO txtbk used | somewhat slow | very clearly | I did not seek extra help | encourages questions & answers them seriously | difficult but reasonable | somewhat helpful | somewhat helpful | |
| 2.50-2.99 | all but 3–4 | most readings | 95-100% | B | B | very quick&accurate | difficult but reasonable | very exciting | a great deal | within 1 day of assignment | excellent | reasonably fast | adequately | very clearly | encourages questions & answers them seriously | difficult but reasonable | very helpful | Excellent |
| 3.00-3.49 | few classes | most readings | NO HW assigned | C | B | mod.paced&mostly accurate | difficult but reasonable | of some interest | moderate amt | NO HW assigned | NO txtbk used | moderate | adequately | accessible for questions | encourages questions & answers them seriously | difficult but reasonable | very helpful | haven't tried any Mastering HW | Excellent |
| 3.00-3.49 | most classes | most readings | 85-94% | A | B | quick & reasonably accurate | difficult but reasonable | interesting | a great deal | within 2 weeks | adequate | moderate | very clearly | accessible for questions | encourages questions & answers them seriously | moderate | very helpful | haven't tried any Mastering HW | Excellent |
| 3.50-4.00 | all classes | some readings | <70% | A | A | very quick&accurate | fairly simple | interesting | a great deal | NO HW assigned | NO txtbk used | reasonably fast | adequately | accessible for questions | encourages questions & answers them seriously | fairly easy | somewhat helpful | very helpful | |
| 3.50-4.00 | all but 3–4 | some readings | <70% | A | A | quick & reasonably accurate | fairly simple | of some interest | moderate amt | within week before exam | NO txtbk used | reasonably fast | adequately | accessible for questions | encourages questions & answers them seriously | fairly easy | somewhat helpful | somewhat helpful | |
| 2.00–2.49 | all classes | most readings | 95-100% | A | B | quick & reasonably accurate | difficult but reasonable | of some interest | moderate amt | within 2-3 days | adequate | moderate | adequately | accessible for questions | encourages questions & answers them seriously | difficult but reasonable | very helpful | fairly easy | Excellent |
| 3.00-3.49 | few classes | 95-100% | A | B | quick & reasonably accurate | too difficult | of no interest | adequate | moderate | adequately | accessible for questions | encourages questions & answers them seriously | too difficult | somewhat helpful | not at all helpful | Fair |
| 3.00-3.49 | all but 3–4 | all readings | <70% | B | B | mod.paced&mostly accurate | too difficult | of no interest | moderate amt | within week before exam | NO txtbk used | too rapid for proper understanding | adequately | accessible for questions | encourages questions & answers them seriously | too difficult | somewhat helpful | not at all helpful | Fair |
| 3.00-3.49 | all but 3–4 | some readings | 95-100% | C | C | mod.paced&mostly accurate | difficult but reasonable | interesting | moderate amt | within week before exam | adequate | somewhat slow | very clearly | I did not seek extra help | encourages questions & answers them seriously | difficult but reasonable | somewhat helpful | very helpful | Excellent |
| 3.50-4.00 | most classes | 95-100% | E | A | very quick&accurate | too difficult | of some interest | a great deal | within 2-3 days | poor | much too slow | very clearly | encourages questions & answers them seriously | fairly easy | very helpful | haven't tried any Mastering HW | The best or very nearly the best |
| 3.50-4.00/3.00–3.49 | 95-100% | A | A | a great deal | within 2 weeks | good | accessible for questions | very helpful | somewhat helpful | The best or very nearly the best |
| | all but 3–4 | most readings | 95-100% | A | A | quick & reasonably accurate | fairly simple | of some interest | a great deal | within week before exam | adequate | moderate | adequately | accessible for questions | encourages questions & answers them seriously | moderate | somewhat helpful | Excellent |
| 3.50-4.00 | all classes | most readings | 85-94% | A | A | quick & reasonably accurate | difficult but reasonable | interesting | a great deal | within week before exam | NO txtbk used | reasonably fast | adequately | very clearly | encourages questions & answers them seriously | difficult but reasonable | not at all helpful | The best or very nearly the best |
| 3.00-3.49 | few classes | all readings | 85-94% | A | B | very quick&accurate | difficult but reasonable | interesting | a great deal | within week before exam | moderate | adequately | very clearly | accessible for questions | encourages questions & answers them seriously | moderate | very helpful | Excellent |
| 3.00-3.49 | all classes | some readings | 70-84% | B | B | quick & reasonably accurate | difficult but reasonable | interesting | moderate amt | within week before exam | NO txtbk used | moderate | adequately | accessible for questions | encourages questions & answers them seriously | difficult but reasonable | very helpful | somewhat helpful | Good |
| 3.00-3.49 | all but 1–2 | some readings | 70-84% | A | A | mod.paced&mostly accurate | difficult but reasonable | interesting | moderate amt | within week before exam | NO txtbk used | moderate | adequately | accessible for questions | encourages questions & answers them seriously | fairly easy | somewhat helpful | somewhat helpful | Good |
| 3.50-4.00 | all classes | most readings | NO HW assigned | A | A | quick & reasonably accurate | difficult but reasonable | of some interest | moderate amt | NO HW assigned | NO txtbk used | moderate | adequately | very helpful | encourages questions & answers them seriously | moderate | moderate | very helpful | |
| 3.50-4.00 | few classes | most readings | <70% | A | A | quick & reasonably accurate | fairly simple | of some interest | moderate amt | NO HW assigned | NO txtbk used | somewhat slow | adequately | I did not seek extra help | encourages questions & answers them seriously | fairly easy | haven't tried any Mastering HW | Good |
| 3.50-4.00 | all classes | most readings | 95-100% | A | A | quick & reasonably accurate | difficult but reasonable | interesting | a great deal | within week before exam | NO txtbk used | moderate | adequately | very clearly | accessible for questions | moderate | very helpful | somewhat helpful | The best or very nearly the best |
| 3.50-4.00 | all classes | most readings | 95-100% | A | A | quick & reasonably accurate | difficult but reasonable | interesting | moderate amt | within 2-3 days | good | moderate | adequately | very clearly | accessible for questions | moderate | very helpful | somewhat helpful | The best or very nearly the best |
| 2.50-2.99 | few classes | most readings | 70-84% | B | B | quick & reasonably accurate | difficult but reasonable | of some interest | moderate amt | within week before exam | good | reasonably fast | adequately | I did not seek extra help | encourages questions & answers them seriously | difficult but reasonable | somewhat helpful | not at all helpful | Good |
| 3.00-3.49 | all classes | most readings | 95-100% | A | A | quick & reasonably accurate | fairly simple | of some interest | D | within 2 weeks | good | reasonably fast | adequately | could rarely be found | does not encourage questions but answers them seriously | difficult but reasonable | somewhat helpful | somewhat helpful | Excellent |
| 2.50-2.99 | all but 3–4 | most readings | 95-100% | A | B | quick & reasonably accurate | difficult but reasonable | of some interest | almost nothing | within week before exam | NO txtbk used | moderate | adequately | could rarely be found | encourages questions & answers them seriously | moderate | not helpful | |

| GPA | ATTENDANCE | RDGS COMPLETED | HW COMPLETED | GRD DESERVED | GRD EXPECTED | MATH ABILITY | MATERIAL COVERED | MATERIAL COVERED | I LEARNED | HW STARTED | TEXTBK | PACE OF LECTS | DIFF/SUBTLE PTS | PROFESSOR | PROFESSOR | EXAMS WERE | BLKBD SUPPLEMENTS | MAST.PHYSICS | COURSE OVERALL |
|---|---|---|---|---|---|---|---|---|---|---|---|---|---|---|---|---|---|---|---|
| 3.50–4.00 | all but 1–2 | some readings | NO HW assigned | A | A | quick & reasonably accurate | moderate amt | of some interest | moderate amt | NO HW assigned | NO txtbk used | reasonably fast | adequately | accessible for questions | encourages questions & answers them seriously | difficult but reasonable | very helpful | haven't tried any Mastering HW | Excellent |
| 3.00–3.49 | all classes | most readings | 70–84% | A | A | mod. paced&mostly accurate | difficult but reasonable | interesting | a great deal | within week before exam | | moderate | very clearly | accessible for questions | encourages questions & answers them seriously | difficult but reasonable | very helpful | haven't tried any Mastering HW | The best or very nearly the best |
| 2.00–2.49 | most classes | most readings | 85–94% | B | B | mod. paced&mostly accurate | difficult but reasonable | interesting | almost nothing | within 1 day of assignment | | adequate | moderate | very clearly | accessible for questions | encourages questions & answers them seriously | moderate | very helpful | not at all helpful | Good |
| 3.50–4.00 | few classes | some readings | NO HW assigned | A | A | quick & reasonably accurate | moderate amt | interesting | moderate amt | NO HW assigned | NO txtbk used | somewhat slow | poorly | accessible for questions | encourages questions & answers them seriously | difficult but reasonable | very helpful | haven't tried any Mastering HW | |
| | all but 3–4 | some readings | <70% | A | B | quick & reasonably accurate | difficult but reasonable | of no interest | almost nothing | within week before exam | NO txtbk used | moderate | adequately | accessible for questions | encourages questions & answers them seriously | | somewhat helpful | not at all helpful | Fair |
| 3.50–4.00 | all but 1–2 | some readings | <70% | B | A | quick & reasonably accurate | fairly simple | of no interest | moderate amt | within week before exam | NO txtbk used | much too slow | very clearly | accessible for questions | encourages questions & answers them seriously | fairly easy | very helpful | haven't tried any Mastering HW | Good |
| 2.50–2.99 | all but 1–2 | most readings | NO HW assigned | A | A | quick & reasonably accurate | | interesting | a great deal | within 2–3 days | NO txtbk used | moderate | adequately | accessible for questions | encourages questions & answers them seriously | fairly easy | | haven't tried any Mastering HW | The best or very nearly the best |
| 3.00–3.49 | all but 1–2 | most readings | 70–84% | B | B | mod. paced&mostly accurate | difficult but reasonable | of some interest | a great deal | within 2 weeks | | adequate | moderate | accessible for questions | encourages questions & answers them seriously | difficult but reasonable | somewhat helpful | haven't tried any Mastering HW | |
| 2.50–2.99 | all classes | most readings | 95–100% | B | B | quick & reasonably accurate | difficult but reasonable | of some interest | a great deal | within 1 day of assignment | | adequate | moderate | accessible for questions | encourages questions & answers them seriously | difficult but reasonable | very helpful | somewhat helpful | Good |
| 2.50–2.99 | most classes | most readings | 70–84% | C | A | quick & reasonably accurate | moderate amt | interesting | moderate amt | within week before exam | | excellent | moderate | accessible for questions | encourages questions & answers them seriously | | | | |
| 3.00–3.49 | most classes | some readings | 70–84% | B | B | mod. paced&mostly accurate | difficult but reasonable | interesting | moderate amt | within week before exam | | poor | moderate | adequately | accessible for questions | encourages questions & answers them seriously | difficult but reasonable | very helpful | very helpful | Excellent |
| 2.50–2.99 | most classes | some readings | <70% | C | C | mod. paced&mostly accurate | difficult but reasonable | of some interest | moderate amt | within week before exam | NO txtbk used | moderate | | very clearly | accessible for questions | does not encourage questions & is reluctant to answer them | too difficult | somewhat helpful | | |
| 3.00–3.49 | all classes | all readings | 70–84% | A | A | mod. paced&mostly accurate | difficult but reasonable | interesting | a great deal | within 2–3 days | NO txtbk used | moderate | | accessible for questions | encourages questions & answers them seriously | moderate | | haven't tried any Mastering HW | |
| 3.50–4.00 | all but 1–2 | all readings | <70% | A | A | quick & reasonably accurate | fairly simple | of some interest | moderate amt | within week before exam | | somewhat slow | | accessible for questions | encourages questions & answers them seriously | fairly easy | very helpful | very helpful | Fair |
| 3.00–3.49 | all but 3–4 | some readings | 70–84% | C | C | mod. paced&mostly accurate | difficult but reasonable | of no interest | moderate amt | within week before exam | | poor | moderate | | accessible for questions | encourages questions & answers them seriously | difficult but reasonable | | very helpful | Good |
| 3.50–4.00 | few classes | some readings | 85–94% | B | B | | too simple | very exciting | almost nothing | NO HW assigned | | adequate | moderate | very clearly | accessible for questions | encourages questions & answers them seriously | difficult but reasonable | very helpful | very helpful | Excellent |
| 2.50–2.99 | most classes | some readings | 85–94% | B | B | mod. paced&mostly accurate | difficult but reasonable | of no interest | moderate amt | within 2–3 days | | poor | reasonably fast | adequately | accessible for questions | encourages questions & answers them seriously | | too difficult | very helpful | somewhat helpful | Good |
| 3.50–4.00 | most classes | some readings | <70% | B | C | quick & reasonably accurate | difficult but reasonable | interesting | moderate amt | within week before exam | | | | | | | | | |
| | all but 3–4 | most readings | <70% | A | A | quick & reasonably accurate | difficult but reasonable | of no interest | moderate amt | NO HW assigned | NO txtbk used | reasonably fast | poorly | D | E | difficult but reasonable | somewhat helpful | haven't tried any Mastering HW | Fair |
| 3.00–3.49 | few classes | all readings | 95–100% | A | A | very quick&accurate | difficult but reasonable | of some interest | a great deal | within week before exam | NO txtbk used | moderate | | accessible for questions | encourages questions & answers them seriously | moderate | very helpful | very helpful | Good |
| 3.00–3.49 | all but 1–2 | most readings | 85–94% | A | A | quick & reasonably accurate | difficult but reasonable | of some interest | moderate amt | within 2–3 days | | adequate | moderate | accessible for questions | encourages questions & answers them seriously | moderate | somewhat helpful | somewhat helpful | |
| 3.00–3.49 | all but 3–4 | most readings | 70–84% | A | B | mod. paced&mostly accurate | difficult but reasonable | of no interest | moderate amt | within 2 weeks | | adequate | moderate | accessible for questions | encourages questions & answers them seriously | | | not at all helpful | |
| 3.50–4.00 | all but 3–4 | most readings | <70% | A | A | quick & reasonably accurate | difficult but reasonable | of no interest | moderate amt | within week before exam | | poor | reasonably fast | very clearly | accessible for questions | encourages questions & answers them seriously | | somewhat helpful | | Good |
| 2.50–2.99 | most classes | most readings | 70–84% | B | B | less quick&/or accurate than I'd like | difficult but reasonable | of some interest | moderate amt | within 1 day of assignment | | adequate | reasonably fast | adequately | accessible for questions | encourages questions & answers them seriously | difficult but reasonable | very helpful | not at all helpful | Excellent |
| 3.00–3.49 | all but 3–4 | most readings | 70–84% | B | A | mod. paced&mostly accurate | difficult but reasonable | interesting | a great deal | within 2–3 days | | adequate | reasonably fast | adequately | I did not seek extra help | encourages questions & answers them seriously | | | not at all helpful | Good |
| 3.00–3.49 | most classes | most readings | <70% | B | B | mod. paced&mostly accurate | difficult but reasonable | interesting | a great deal | within week before exam | | good | moderate | adequately | accessible for questions | encourages questions & answers them seriously | | somewhat helpful | haven't tried any Mastering HW | Excellent |
| 3.00–3.49 | all but 1–2 | some readings | <70% | C | C | mod. paced&mostly accurate | difficult but reasonable | of no interest | moderate amt | within week before exam | | adequate | reasonably fast | adequately | accessible for questions | encourages questions & answers them seriously | difficult but reasonable | | somewhat helpful | Fair |
| 2.50–2.99 | most classes | some readings | <70% | A | A | quick & reasonably accurate | fairly simple | of some interest | moderate amt | within week before exam | NO txtbk used | much too slow | | adequately | I did not seek extra help | does not encourage questions & is reluctant to answer them | moderate | very helpful | haven't tried any Mastering HW | Fair |
| 2.50–2.99 | all but 1–2 | most readings | <70% | B | B | quick & reasonably accurate | difficult but reasonable | of some interest | a great deal | within week before exam | | good | somewhat slow | adequately | accessible for questions | encourages questions & answers them seriously | | | very helpful | |
| 2.50–2.99 | all but 1–2 | all readings | 85–94% | A | B | quick & reasonably accurate | moderate amt | interesting | moderate amt | within 2 weeks | | adequate | reasonably fast | adequately | accessible for questions | encourages questions & answers them seriously | moderate | not helpful | moderate | Excellent |
| 3.00–3.49 | most classes | most readings | 85–94% | A | A | mod. paced&mostly accurate | difficult but reasonable | of some interest | moderate amt | within 2 weeks | | adequate | moderate | adequately | accessible for questions | encourages questions & answers them seriously | moderate | | not at all helpful | |
| 3.50–4.00 | all but 1–2 | most readings | 85–94% | A | A | quick & reasonably accurate | | interesting | moderate amt | within 2 weeks | | good | moderate | very clearly | accessible for questions | encourages questions & answers them seriously | difficult but reasonable | very helpful | haven't tried any Mastering HW | Excellent |
| 3.50–4.00 | few classes | some readings | <70% | A | A | very quick&accurate | fairly simple | of some interest | moderate amt | within 2–3 days | NO txtbk used | moderate | moderate | I did not seek extra help | encourages questions & answers them seriously | fairly easy | very helpful | haven't tried much Mastering HW | Good |
| 3.50–4.00 | most classes | most readings | <70% | B | B | quick & reasonably accurate | difficult but reasonable | interesting | moderate amt | within week before exam | NO txtbk used | | | accessible for questions | encourages questions & answers them seriously | | very helpful | somewhat helpful | |
| 3.00–3.49 | all classes | all readings | 85–94% | A | A | | difficult but reasonable | very exciting | a great deal | within 2–3 days | | good | too rapid for proper understanding | very clearly | accessible for questions | encourages questions & answers them seriously | difficult but reasonable | moderate | haven't tried any Mastering HW | Excellent |
| 3.50–4.00 | all but 1–2 | some readings | NO HW assigned | A | A | very quick&accurate | fairly simple | of some interest | moderate amt | within 2–3 days | | adequate | somewhat slow | very clearly | accessible for questions | | | somewhat helpful | haven't tried any Mastering HW | |
| 3.50–4.00 | most classes | most readings | 70–84% | A | A | quick & reasonably accurate | difficult but reasonable | interesting | moderate amt | within week before exam | | adequate | reasonably fast | very clearly | accessible for questions | encourages questions & answers them seriously | difficult but reasonable | very helpful | haven't tried any Mastering HW | Excellent |
| 2.00–2.49 | all classes | most readings | <70% | A | B | mod. paced&mostly accurate | difficult but reasonable | interesting | moderate amt | within week before exam | NO txtbk used | | | | | | | | |
| | all classes | most readings | 70–84% | | | quick & reasonably accurate | difficult but reasonable | interesting | a great deal | within week before exam | | adequate | too rapid for proper understanding | very clearly | accessible for questions | does not encourage questions but answers them seriously | too difficult | somewhat helpful | | |
| 3.00–3.49 | all but 3–4 | some readings | NO HW assigned | A | B | mod. paced&mostly accurate | fairly simple | of no interest | almost nothing | within week before exam | NO txtbk used | much too slow | poorly | accessible for questions | encourages questions & answers them seriously | | somewhat helpful | haven't tried any Mastering HW | Fair |
| 3.50–4.00 | most classes | most readings | NO HW assigned | A | A | very quick&accurate | fairly simple | interesting | moderate amt | NO HW assigned | | adequate | moderate | | accessible for questions | | | very helpful | haven't tried any Mastering HW | Excellent |
| 3.50–4.00 | most classes | some readings | <70% | B | B | quick & reasonably accurate | fairly simple | of some interest | moderate amt | within week before exam | NO txtbk used | adequate | moderate | very clearly | accessible for questions | encourages questions & answers them seriously | difficult but reasonable | very helpful | | Fair |
| 3.00–3.49 | all but 1–2 | all readings | <70% | B | A | very quick&accurate | difficult but reasonable | very exciting | a great deal | | NO txtbk used | moderate | very clearly | I did not seek extra help | encourages questions & answers them seriously | | very helpful | haven't tried any Mastering HW | The best or very nearly the best |
| 3.00–3.49 | most classes | most readings | <70% | A | A | very quick&accurate | fairly simple | interesting | moderate amt | NO HW assigned | | adequate | moderate | very clearly | accessible for questions | encourages questions & answers them seriously | | very helpful | | |
| 3.00–3.49 | all classes | most readings | 70–84% | A | A | very quick&accurate | fairly simple | interesting | moderate amt | within 2 weeks | | moderate | adequately | accessible for questions | encourages questions & answers them seriously | moderate | | not at all helpful | Excellent |
| 2.50–2.99 | all classes | most readings | <70% | B | C | mod. paced&mostly accurate | difficult but reasonable | of some interest | a great deal | within week before exam | NO txtbk used | moderate | | adequately | I did not seek extra help | encourages questions & answers them seriously | difficult but reasonable | somewhat helpful | haven't tried any Mastering HW | Good |
| 3.50–4.00 | all classes | all readings | NO HW assigned | A | A | very quick&accurate | difficult but reasonable | interesting | moderate amt | NO HW assigned | | poor | somewhat slow | | accessible for questions | encourages questions & answers them seriously | fairly easy | | not at all helpful | Good |
| 3.50–4.00 | most classes | some readings | NO HW assigned | A | A | very quick&accurate | difficult but reasonable | interesting | moderate amt | NO HW assigned | NO txtbk used | moderate | adequately | accessible for questions | encourages questions & answers them seriously | moderate | very helpful | haven't tried any Mastering HW | The best or very nearly the best |
| 3.00–3.49 | all but 3–4 | most readings | NO HW assigned | A | B | quick & reasonably accurate | difficult but reasonable | of some interest | a great deal | NO HW assigned | NO txtbk used | | somewhat slow | very clearly | accessible for questions | encourages questions & answers them seriously | | very helpful | haven't tried much Mastering HW | Excellent |
| 2.50–2.99 | all classes | some readings | <70% | B | C | less quick&/or accurate than I'd like | difficult but reasonable | interesting | moderate amt | within week before exam | | moderate | adequately | accessible for questions | encourages questions & answers them seriously | difficult but reasonable | very helpful | | |
| 3.50–4.00 | all but 1–2 | most readings | <70% | A | A | very quick&accurate | fairly simple | of some interest | a great deal | within week before exam | NO txtbk used | | much too slow | very clearly | accessible for questions | encourages questions & answers them seriously | | | | Fair |
| 3.50–4.00 | all classes | none of rdgs | <70% | A | A | very quick&accurate | fairly simple | of no interest | a great deal | NO HW assigned | NO txtbk used | moderate | very clearly | I did not seek extra help | encourages questions & answers them seriously | fairly easy | | | |
| 3.00–3.49 | all but 3–4 | most readings | NO HW assigned | A | B | mod. paced&mostly accurate | difficult but reasonable | of some interest | moderate amt | NO HW assigned | | adequate | moderate | adequately | accessible for questions | encourages questions & answers them seriously | | somewhat helpful | haven't tried any Mastering HW | Fair |
| 3.50–4.00 | all classes | all readings | 70–84% | A | A | quick & reasonably accurate | difficult but reasonable | of some interest | a great deal | within week before exam | | adequate | reasonably fast | very clearly | accessible for questions | encourages questions & answers them seriously | | very helpful | haven't tried any Mastering HW | Good |
| | all but 3–4 | most readings | 70–84% | A | A | quick & reasonably accurate | too simple | interesting | moderate amt | within week before exam | | adequate | too rapid for proper understanding | very clearly | accessible for questions | encourages questions & answers them seriously | moderate | I never consulted Blkbd supplmnts | haven't tried any Mastering HW | The best or very nearly the best |
| 3.00–3.49 | all but 1–2 | all readings | NO HW assigned | A | A | very quick&accurate | | interesting | moderate amt | NO HW assigned | NO txtbk used | reasonably fast | adequately | accessible for questions | encourages questions & answers them seriously | difficult but reasonable | | not at all helpful | Poor |
| 2.50–2.99 | all but 3–4 | most readings | NO HW assigned | B | C | very quick&accurate | too difficult | of some interest | moderate amt | NO HW assigned | NO txtbk used | | | | | | | | |
| 3.00–3.49 | most classes | all readings | NO HW assigned | A | B | less quick&/or accurate than I'd like | difficult but reasonable | of no interest | moderate amt | within week before exam | | poor | moderate | | accessible for questions | encourages questions & answers them seriously | moderate | very helpful | haven't tried any Mastering HW | Good |
| 3.50–4.00 | most classes | all readings | <70% | A | A | very quick&accurate | difficult but reasonable | interesting | moderate amt | within week before exam | NO txtbk used | moderate | adequately | I did not seek extra help | encourages questions & answers them seriously | | somewhat helpful | haven't tried any Mastering HW | Excellent |
| 3.50–4.00 | most classes | some readings | <70% | B | B | very quick&accurate | difficult but reasonable | of no interest | moderate amt | within week before exam | NO txtbk used | poor | reasonably fast | adequately | I did not seek extra help | encourages questions & answers them seriously | moderate | somewhat helpful | haven't tried any Mastering HW | Excellent |
| | most classes | none of rdgs | NO HW assigned | | | mod. paced&mostly accurate | difficult but reasonable | of some interest | moderate amt | within week before exam | NO txtbk used | moderate | adequately | accessible for questions | encourages questions & answers them seriously | difficult but reasonable | very helpful | haven't tried any Mastering HW | |

| GPA | ATTENDANCE | RDGS COMPLETED | HW COMPLETED | GRD DESERVED | GRD EXPECTED | MATH ABILITY | MATERIAL COVERED | MATERIAL COVERED | I LEARNED | HW STARTED | TEXTBK | PACE OF LECTS | DIFF/SUBTLE PTS | PROFESSOR | PROFESSOR | EXAMS WERE | BLKBD SUPPLEMENTS | MAST.PHYSICS | COURSE OVERALL |
|---|---|---|---|---|---|---|---|---|---|---|---|---|---|---|---|---|---|---|---|
| 3.00–3.49 | all classes | all readings | 70–84% | B | A | mod. paced&mostly accurate | difficult but reasonable | of some interest | moderate amt | within week before exam | good | moderate | very clearly | accessible for questions | encourages questions & answers them seriously | difficult but reasonable | moderate | haven't tried much Mastering HW | Excellent |
| 3.50–4.00 | most classes | <70% | A | A | quick & reasonably accurate | of no interest | a great deal | within 2 weeks | somewhat slow | adequately | accessible for questions | encourages questions & answers them seriously | moderate | E |
| 3.00–3.49 | all but 1–2 | most readings | 70–84% | A | A | quick & reasonably accurate | difficult but reasonable | interesting | a great deal | within 2 weeks | NO txtbk used | reasonably fast | adequately | accessible for questions | encourages questions & answers them seriously | difficult but reasonable | very helpful | haven't tried any Mastering HW | Good |
| 3.00–3.49 | most classes | most readings | <70% | A | A | quick & reasonably accurate | difficult but reasonable | of no interest | a great deal | within 2 weeks | moderate | I did not seek extra help | encourages questions & answers them seriously | very helpful | not at all helpful |
| 3.50–4.00 | all but 3–4 | all readings | NO HW assigned | A | A | mod. paced&mostly accurate | difficult but reasonable | of some interest | moderate amt | NO HW assigned | NO txtbk used | somewhat slow | adequately | accessible for questions | encourages questions & answers them seriously | moderate | very helpful | haven't tried any Mastering HW | The best or very nearly the best |
| 3.50–4.00 | all but 3–4 | all readings | 85–94% | A | A | quick & reasonably accurate | difficult but reasonable | very exciting | a great deal | within 2–3 days | adequate | reasonably fast | very clearly | accessible for questions | encourages questions & answers them seriously | difficult but reasonable | very helpful | very helpful | The best or very nearly the best |
| 3.00–3.49 | all but 1–2 | all readings | <70% | A | A | mod. paced&mostly accurate | difficult but reasonable | interesting | a great deal | NO HW assigned | moderate | very clearly | accessible for questions | encourages questions & answers them seriously | difficult but reasonable | very helpful | haven't tried any Mastering HW |
| 3.50–4.00 | all but 3–4 | some readings | <70% | B | A | mod. paced&mostly accurate | fairly simple | of some interest | moderate amt | within 2–3 days | NO txtbk used | moderate | adequately | very clearly | accessible for questions | encourages questions & answers them seriously | fairly easy | very helpful | haven't tried any Mastering HW | The best or very nearly the best |
| 3.50–4.00 | all but 1–2 | none of rdgs | NO HW assigned | A | A | very quick&accurate | difficult but reasonable | interesting | a great deal | NO HW assigned | NO txtbk used | moderate | very clearly | accessible for questions | encourages questions & answers them seriously | difficult but reasonable | very helpful | haven't tried any Mastering HW | Excellent |
| 3.50–4.00 | few classes | some readings | NO HW assigned | A | A | mod. paced&mostly accurate | fairly simple | interesting | moderate amt | within 2–3 days | moderate | adequately | accessible for questions | encourages questions & answers them seriously | fairly easy | very helpful | haven't tried any Mastering HW |
| 3.50–4.00 | all but 3–4 | most readings | <70% | A | A | very quick&accurate | difficult but reasonable | interesting | a great deal | within week before exam | NO txtbk used | moderate | adequately | accessible for questions | encourages questions & answers them seriously | difficult but reasonable | very helpful | not at all helpful | Excellent |
| 3.50–4.00 | most classes | none of rdgs | <70% | A | A | mod. paced&mostly accurate | difficult but reasonable | interesting | moderate amt | within week before exam | adequate | moderate | adequately | accessible for questions | encourages questions & answers them seriously | moderate | somewhat helpful | haven't tried any Mastering HW | Fair |
| 3.50–4.00 | all but 1–2 | all readings | 85–94% | A | A | very quick&accurate | difficult but reasonable | interesting | a great deal | within 2 weeks | adequate | moderate | adequately | accessible for questions | encourages questions & answers them seriously | difficult but reasonable | somewhat helpful | haven't tried any Mastering HW | The best or very nearly the best |
| 2.50–2.99 | all but 3–4 | most readings | <70% | A | A | quick & reasonably accurate | fairly simple | of some interest | moderate amt | within week before exam | NO txtbk used | moderate | adequately | very clearly | accessible for questions | encourages questions & answers them seriously | fairly easy | very helpful | haven't tried much Mastering HW |
| 3.00–3.49 | all classes | all readings | 95–100% | B | B | quick & reasonably accurate | difficult but reasonable | interesting | a great deal | within 2–3 days | good | moderate | very clearly | accessible for questions | encourages questions & answers them seriously | too difficult | very helpful | somewhat helpful | The best or very nearly the best |
| 3.50–4.00 | few classes | some readings | <70% | B | A | very quick&accurate | fairly simple | of some interest | moderate amt | within week before exam | NO txtbk used | moderate | adequately | very clearly | accessible for questions | encourages questions & answers them seriously | difficult but reasonable | very helpful | haven't tried any Mastering HW | Fair |
| 3.00–3.49 | most classes | most readings | <70% | A | A | very quick&accurate | fairly simple | of no interest | moderate amt | within 2–3 days | NO txtbk used | somewhat slow | adequately | very clearly | accessible for questions | encourages questions & answers them seriously | fairly easy | somewhat helpful | haven't tried any Mastering HW |
| 3.00–3.49 | all but 1–2 | most readings | <70% | B | B | mod. paced&mostly accurate | difficult but reasonable | interesting | a great deal | within week before exam | adequate | moderate | adequately | accessible for questions | encourages questions & answers them seriously | difficult but reasonable | very helpful | not at all helpful | Excellent |
| 3.00–3.49 | all but 1–2 | some readings | <70% | A | A | quick & reasonably accurate | difficult but reasonable | interesting | a great deal | within 2–3 days | adequate | somewhat slow | adequately | accessible for questions | encourages questions & answers them seriously | difficult but reasonable | very helpful | haven't tried any Mastering HW | Excellent |
| 3.50–4.00 | all classes | some readings | <70% | B | B | mod. paced&mostly accurate | difficult but reasonable | interesting | a great deal | within week before exam | NO txtbk used | somewhat slow | adequately | accessible for questions | encourages questions & answers them seriously | difficult but reasonable | very helpful | haven't tried any Mastering HW |
| 3.00–3.49 | most classes | all readings | 70–84% | A | A | quick & reasonably accurate | difficult but reasonable | interesting | a great deal | within week before exam | adequate | reasonably fast | very clearly | accessible for questions | encourages questions & answers them seriously | difficult but reasonable | very helpful | haven't tried any Mastering HW | Excellent |
| 3.50–4.00 | all but 1–2 | most readings | 95–100% | A | A | quick & reasonably accurate | fairly simple | of no interest | moderate amt | within 2–3 days | NO txtbk used | moderate | adequately | very clearly | accessible for questions | encourages questions & answers them seriously | moderate | very helpful | I did not seek extra help | Good |
| 3.50–4.00 | all but 1–2 | most readings | NO HW assigned | A | A | quick & reasonably accurate | fairly simple | interesting | a great deal | within week before exam | somewhat slow | very clearly | very helpful | I did not seek extra help | fairly easy | very helpful |
| 3.00–3.49 | all but 1–2 | most readings | 70–84% | A | B | quick & reasonably accurate | difficult but reasonable | of some interest | moderate amt | within 2 weeks | within 2 weeks | NO txtbk used | moderate | adequately | accessible for questions | encourages questions & answers them seriously | difficult but reasonable | very helpful | haven't tried any Mastering HW | Good |
| 3.50–4.00 | all but 3–4 | some readings | 70–84% | A | A | very quick&accurate | difficult but reasonable | interesting | a great deal | within 2 weeks | NO txtbk used | moderate | adequately | very clearly | accessible for questions | encourages questions & answers them seriously | moderate | very helpful | haven't tried much Mastering HW | The best or very nearly the best |
| 3.00–3.49 | all but 1–2 | none of rdgs | <70% | A | A | quick & reasonably accurate | difficult but reasonable | interesting | a great deal | within week before exam | somewhat slow | very clearly | accessible for questions | somewhat helpful |
| 2.50–2.99 | all but 1–2 | all readings | 70–84% | B | B | mod. paced&mostly accurate | difficult but reasonable | interesting | a great deal | within 2 weeks | adequate | reasonably fast | adequately | accessible for questions | encourages questions & answers them seriously | difficult but reasonable | very helpful | haven't tried any Mastering HW | Good |
| 3.00–3.49 | all classes | all readings | 85–94% | B | B | quick & reasonably accurate | difficult but reasonable | interesting | a great deal | within 2–3 days | NO txtbk used | moderate | adequately | accessible for questions | encourages questions & answers them seriously | difficult but reasonable | somewhat helpful | haven't tried any Mastering HW | Good |
| 3.50–4.00 | most classes | most readings | 70–84% | A | A | very quick&accurate | difficult but reasonable | of some interest | moderate amt | within 2 weeks | NO txtbk used | much too slow | poorly | accessible for questions | encourages questions & answers them seriously | moderate | very helpful | haven't tried any Mastering HW | Fair |
| 3.00–3.49 | all but 3–4 | most readings | 70–84% | A | A | mod. paced&mostly accurate | difficult but reasonable | of some interest | moderate amt | within week before exam | good | reasonably fast | very clearly | accessible for questions | encourages questions & answers them seriously | difficult but reasonable | somewhat helpful | haven't tried any Mastering HW |
| 3.00–3.49 | most classes | most readings | <70% | B | B | mod. paced&mostly accurate | difficult but reasonable | of some interest | moderate amt | within week before exam | adequate | moderate | adequately | accessible for questions | encourages questions & answers them seriously | difficult but reasonable | very helpful | haven't tried any Mastering HW | Excellent |
| 3.00–3.49 | all but 1–2 | all readings | 95–100% | A | A | quick & reasonably accurate | difficult but reasonable | interesting | a great deal | within 1 day of assignment | good | reasonably fast | very clearly | accessible for questions | encourages questions & answers them seriously | difficult but reasonable | very helpful | very helpful | Excellent |
| 3.50–4.00 | most classes | most readings | 70–84% | A | B | very quick&accurate | difficult but reasonable | interesting | moderate amt | within 2 weeks | NO txtbk used | moderate | adequately | accessible for questions | encourages questions & answers them seriously | moderate | somewhat helpful | haven't tried any Mastering HW | Fair |
| 2.50–2.99 | most classes | most readings | 85–94% | B | C | mod. paced&mostly accurate | difficult but reasonable | interesting | moderate amt | within 2–3 days | good | moderate | adequately | accessible for questions | encourages questions & answers them seriously | difficult but reasonable | somewhat helpful | very helpful | Good |
| 3.00–3.49 | all but 1–2 | most readings | <70% | B | C | mod. paced&mostly accurate | difficult but reasonable | of no interest | almost nothing | within week before exam | NO txtbk used | moderate | poorly | accessible for questions | encourages questions & answers them seriously | difficult but reasonable | somewhat helpful | haven't tried any Mastering HW | Fair |
| 3.50–4.00 | all but 1–2 | all readings | 70–84% | A | A | very quick&accurate | difficult but reasonable | interesting | a great deal | within 2 weeks | NO txtbk used | moderate | adequately | accessible for questions | encourages questions & answers them seriously | difficult but reasonable | very helpful | haven't tried any Mastering HW |
| 3.50–4.00 | all but 1–2 | all readings | 85–94% | B | A | very quick&accurate | difficult but reasonable | interesting | a great deal | within 2–3 days | adequate | reasonably fast | very clearly | accessible for questions | encourages questions & answers them seriously | difficult but reasonable | very helpful | haven't tried much Mastering HW | Excellent |
| 3.50–4.00 | all but 3–4 | all readings | 95–100% | A | A | quick & reasonably accurate | difficult but reasonable | interesting | moderate amt | within week before exam | good | moderate | very clearly | accessible for questions | encourages questions & answers them seriously | difficult but reasonable | very helpful | haven't tried any Mastering HW | Excellent |
| 3.00–3.49 | all but 1–2 | all readings | 70–84% | B | B | mod. paced&mostly accurate | difficult but reasonable | interesting | a great deal | within 2–3 days | adequate | moderate | adequately | accessible for questions | encourages questions & answers them seriously | difficult but reasonable | very helpful | haven't tried any Mastering HW |
| 3.50–4.00 | all but 1–2 | all readings | 95–100% | A | A | quick & reasonably accurate | fairly simple | of no interest | almost nothing | within week before exam | reasonably fast | adequately | very helpful | fairly easy | very helpful | haven't tried any Mastering HW | Fair |
| 3.50–4.00 | all classes | most readings | 95–100% | A | A | mod. paced&mostly accurate | fairly simple | of no interest | moderate amt | within 2–3 days | poor | moderate | very clearly | accessible for questions | encourages questions & answers them seriously | very helpful | very helpful | Good |
| 3.50–4.00 | all classes | some readings | <70% | A | A | quick & reasonably accurate | fairly simple | very exciting | a great deal | within week before exam | good | somewhat slow | very clearly | accessible for questions | encourages questions & answers them seriously | fairly easy | very helpful | The best or very nearly the best |
| 3.50–4.00 | all but 1–2 | all readings | 85–94% | A | B | quick & reasonably accurate | difficult but reasonable | interesting | moderate amt | within week before exam | adequate | moderate | adequately | accessible for questions | encourages questions & answers them seriously | difficult but reasonable | very helpful | haven't tried any Mastering HW |
| 3.50–4.00 | all but 1–2 | all readings | 95–100% | A | A | very quick&accurate | fairly simple | of some interest | a great deal | within 1 day of assignment | poor | moderate | very clearly | accessible for questions | encourages questions & answers them seriously | fairly easy | very helpful | haven't tried any Mastering HW | Excellent |
| 3.00–3.49 | all but 1–2 | some readings | 70–84% | A | A | very quick&accurate | difficult but reasonable | very exciting | a great deal | within 2 weeks | NO txtbk used | reasonably fast | very clearly | accessible for questions | encourages questions & answers them seriously | difficult but reasonable | very helpful | haven't tried much Mastering HW | The best or very nearly the best |
| 3.50–4.00 | most classes | most readings | 95–100% | A | A | quick & reasonably accurate | difficult but reasonable | interesting | moderate amt | within week before exam | adequate | moderate | very clearly | accessible for questions | encourages questions & answers them seriously | moderate | very helpful | not at all helpful | Excellent |
| 3.50–4.00 | all classes | all readings | 85–94% | A | A | quick & reasonably accurate | difficult but reasonable | interesting | a great deal | within 2–3 days | adequate | moderate | very clearly | accessible for questions | encourages questions & answers them seriously | moderate | very helpful | somewhat helpful | Excellent |
| 3.50–4.00 | most classes | most readings | 95–100% | A | A | quick & reasonably accurate | difficult but reasonable | very exciting | a great deal | within 2–3 days | adequate | moderate | very clearly | accessible for questions | encourages questions & answers them seriously | moderate | very helpful | haven't tried any Mastering HW | The best or very nearly the best |
| 3.50–4.00 | all but 1–2 | all readings | 95–100% | A | A | quick & reasonably accurate | difficult but reasonable | interesting | a great deal | within 2 weeks | NO txtbk used | moderate | very clearly | accessible for questions | encourages questions & answers them seriously | moderate | very helpful | haven't tried any Mastering HW | The best or very nearly the best |
| 3.00–3.49 | all but 3–4 | most readings | 85–94% | A | B | quick & reasonably accurate | difficult but reasonable | of some interest | moderate amt | within 2–3 days | NO txtbk used | moderate | adequately | accessible for questions | encourages questions & answers them seriously | too difficult | very helpful | somewhat helpful | Fair |
| 3.50–4.00 | all but 1–2 | most readings | <70% | A | A | mod. paced&mostly accurate | fairly simple | interesting | moderate amt | within week before exam | NO txtbk used | moderate | adequately | accessible for questions | encourages questions & answers them seriously | difficult but reasonable | very helpful | haven't tried any Mastering HW |
| 3.00–3.49 | all classes | most readings | 95–100% | A | A | quick & reasonably accurate | difficult but reasonable | interesting | moderate amt | within 2–3 days | adequate | moderate | adequately | accessible for questions | encourages questions & answers them seriously | difficult but reasonable | very helpful | haven't tried much Mastering HW | Good |
| 3.50–4.00 | most classes | all readings | NO HW assigned | A | A | mod. paced&mostly accurate | fairly simple | of some interest | moderate amt | NO HW assigned | NO txtbk used | somewhat slow | adequately | accessible for questions | encourages questions & answers them seriously | moderate | very helpful | haven't tried any Mastering HW | Good |
| 3.50–4.00 | most classes | all readings | NO HW assigned | A | A | mod. paced&mostly accurate | difficult but reasonable | interesting | a great deal | within week before exam | good | moderate | adequately | accessible for questions | encourages questions & answers them seriously | difficult but reasonable | very helpful | haven't tried any Mastering HW | Good |
| 3.50–4.00 | few classes | some readings | <70% | B | B | mod. paced&mostly accurate | difficult but reasonable | interesting | a great deal | within week before exam | adequate | moderate | adequately | accessible for questions | encourages questions & answers them seriously | difficult but reasonable | very helpful | haven't tried any Mastering HW | Excellent |
| 3.00–3.49 | all classes | all readings | 85–94% | A | A | very quick&accurate | difficult but reasonable | interesting | a great deal | within 2–3 days | NO txtbk used | moderate | very clearly | accessible for questions | does not encourage questions but answers them seriously | too difficult | very helpful | not at all helpful | Good |
| 3.00–3.49 | most classes | all readings | 85–94% | A | B | very quick&accurate | difficult but reasonable | of no interest | moderate amt | within week before exam | adequate | too rapid for proper understanding | poorly | accessible for questions | encourages questions & answers them seriously | difficult but reasonable | very helpful | not at all helpful | Fair |
| 3.00–3.49 | all but 3–4 | most readings | 85–94% | A | A | mod. paced&mostly accurate | difficult but reasonable | interesting | moderate amt | within 2 weeks | reasonably fast | very clearly | accessible for questions | encourages questions & answers them seriously |
| 2.50–2.99 | most classes | some readings | <70% | B | B | mod. paced&mostly accurate | difficult but reasonable | of no interest | moderate amt | within week before exam | NO txtbk used | moderate | adequately | I did not seek extra help | encourages questions & answers them seriously | difficult but reasonable | I never consulted Blkbd supplemts | haven't tried any Mastering HW | Good |
| 3.00–3.49 | most classes | most readings | <70% | A | A | quick & reasonably accurate | difficult but reasonable | of some interest | a great deal | within 2 weeks | NO txtbk used | moderate | very clearly | accessible for questions | encourages questions & answers them seriously | difficult but reasonable | very helpful | haven't tried any Mastering HW | The best or very nearly the best |
| 3.00–3.49 | all but 3–4 | most readings | 85–94% | A | A | mod. paced&mostly accurate | difficult but reasonable | interesting | moderate amt | within 2–3 days | excellent | moderate | very clearly | accessible for questions | encourages questions & answers them seriously | difficult but reasonable | very helpful | haven't tried any Mastering HW | Excellent |
| 3.50–4.00 | all classes | most readings | 85–94% | B | B | quick & reasonably accurate | too difficult | of no interest | moderate amt | within 2–3 days | good | too rapid for proper understanding | poorly | accessible for questions | encourages questions & answers them seriously | moderate | somewhat helpful | Fair |
| 3.50–4.00 | most classes | none of rdgs | <70% | A | A | very quick&accurate | difficult but reasonable | of no interest | moderate amt | within week before exam | NO txtbk used | somewhat slow | adequately | accessible for questions | encourages questions & answers them seriously | difficult but reasonable | very helpful | haven't tried any Mastering HW | Good |
| 3.50–4.00 | all classes | some readings | <70% | A | A | very quick&accurate | difficult but reasonable | of no interest | moderate amt | within week before exam | NO txtbk used | somewhat slow | adequately | accessible for questions | encourages questions & answers them seriously | difficult but reasonable | somewhat helpful | haven't tried any Mastering HW | |